# Exploratory Analysis of Cyberattack Patterns on E-commerce Platforms Using Statistical Methods

Submitted in accordance with the requirements for the degree of
Master of Science in Data Science

Adeniya Fatimo Adenike

York St John University

Department of Data Science

August 2025

# Declaration

The candidate confirms that the work submitted is his own and that appropriate credit has been given where reference has been made to the work of others.

| | |
|---|---|
| I have read and understood the Academic Misconduct statement. | Tick to confirm ☒ |
| I have read and understood the Generative Artificial Intelligence use statement. | Tick to confirm ☒ |
| I am satisfied that I have met the Learning Outcomes of this assignment<br>(please check the Assignment Brief if you are unsure) | Met ☒ |

| |
|---|
| **Self-Assessment** – If there are particular aspects of your assignment on which you would like feedback, please indicate below.<br>Optional for students |
| ***Suggested prompt questions-***<br>*How have you developed or progressed your learning in this work?*<br>*What do you feel is the strongest part of this submission?*<br>*What feedback would you give yourself?*<br>*What part(s) of this assignment are you still unsure about?* |

# Acknowledgements

I am deeply grateful to Almighty God for granting me the strength and grace to complete this journey. I dedicate this work to the memory of my late father, Prince Adeniya Bolaji, whose passion for education continues to inspire me.

My heartfelt thanks go to my mother for her prayers and encouragement, and to my siblings, Adeniya Damilola, Razaq, Mosumola, Kazeem, and Hassan, and my cousins, Oyinloluwa Dawodu and Dare Kilani, for their unwavering physical, financial, and emotional support.

I am also indebted to my friends and colleagues, especially Dr. Moji Aderinola, Afolashade Olubode, Ada Orizu, and Omolola Bankole, Idayat Abdulkareem for their steadfast encouragement and shared perseverance.

Finally, I express my sincere appreciation to my supervisor, Shamsuddeen Muhammad, for his guidance, time, and dedication throughout this research.

# Abstract

Cyberattacks on e-commerce platforms have grown in sophistication, threatening consumer trust and operational continuity. This research presents a novel hybrid analytical framework that integrates statistical modelling and machine learning for detecting and forecasting cyberattack patterns in the e-commerce domain. Using the Verizon Community Data Breach (VCDB) dataset, the study applies Auto ARIMA for temporal forecasting and significance testing, including a Mann–Whitney U test ($U = 2{,}579{,}981.5$, $p = 0.0121$), which confirmed that holiday shopping events experienced significantly more severe cyberattacks than non-holiday periods. ANOVA was also used to examine seasonal variation in threat severity, while ensemble machine learning models (XGBoost, LightGBM, and CatBoost) were employed for predictive classification. Results reveal recurrent attack spikes during high-risk periods such as Black Friday and holiday shopping seasons, with breaches involving Personally Identifiable Information (PII) exhibiting significantly elevated threat indicators. Among the machine learning models, CatBoost achieved the highest predictive performance (accuracy = 85.29%, F1 score = 0.2254, ROC AUC = 0.8247), demonstrating the value of non-linear boosting methods in complex cyberattack detection. This approach is unique in combining seasonal statistical forecasting with interpretable ensemble learning, enabling both temporal risk anticipation and breach-type classification. Ethical considerations, including responsible use of sensitive breach data and bias assessment in predictive models, were incorporated into the design process. While the study is limited by dataset class imbalance and reliance on historical breach reporting quality, it provides actionable insights for proactive cybersecurity resource allocation. Future work will extend the framework to streaming threat data and integrate adversarial resilience techniques for robust real-time detection.

## Table of Contents

# Table of Abbreviations

| Abbreviation | Full Term |
|---|---|
| API | Application Programming Interface |
| AWS | Amazon Web Services |
| Azure | Microsoft Azure |
| BDCC | Big Data Cloud Computing |
| CLI | Command Line Interface |
| CVE | Common Vulnerabilities and Exposures |
| EC2 | Elastic Compute Cloud |
| EDA | Exploratory Data Analysis |
| ETL | Extract, Transform, Load |
| FBI IC3 | FBI’s Internet Crime Complaint Center (IC3) |
| GBT | Gradient Boosted Tree |
| GDPR | General Data Protection Regulation |
| IAM | Identity and Access Management |
| IDE | Integrated Development Environment |
| IDS | Intrusion Detection System |
| JSON | JavaScript Object Notation |
| LIME | Local Interpretable Model-agnostic Explanations. |
| ML | Machine Learning |
| MSE | Mean Squared Error |
| MFA | Multi-Factor Authentication |
| SHAP | SHapley Additive exPlanations |
| SOC | Security Operations Center |
| SMEs | Small and Medium-sized Enterprises |
| PCIDSS | Payment Card Industry Data Security Standard |
| PRISMA | Preferred Reporting Items for Systematic Reviews and Meta-Analyses |
| RMSE | Root Mean Squared Error |
| IRB | Institutional Review Board |
| UNCTAD | United Nations Conference on Trade and Development |
| BLACK BOXES | A Model Whose Internal Logic Is Opaque, Despite Observable Inputs and Outputs; Common In Complex ML Models Like Neural Networks and Ensembles. |
| WAF | Web Application Firewall |

# List of Figures

# List of Tables

# 1.0 Research Introduction

The swift growth of e-commerce has revolutionised worldwide retail practices, enhancing consumer accessibility, digital convenience, and international transaction frameworks (Laudon and Traver, 2021, p. 48). This digital development has concurrently introduced new weaknesses, exposing e-commerce systems to cyberattacks. As e-commerce systems increasingly interact with cloud infrastructure, third-party services, and real-time data processing systems, their attack surfaces broaden, rendering them significant targets for hackers (Kshetri, 2021). The escalating complexity of attacks, including credential stuffing, phishing, and JavaScript-based skimming, illustrates the growing dangers confronting online businesses (Zade et al., 2024).

Case studies highlight the gravity of these threats: In 2019, the Macy's data breach, which involved JavaScript injection during a peak shopping season, compromised thousands of client credentials (Romanosky, 2016). The 2018 British Airways hack compromised more than 400,000 customer payment records, resulting in a £20 million fine (Information Commissioner's Office, 2020). These examples underscore the financial and reputational harm inflicted by cyberattacks on e-commerce platforms. Although artificial intelligence (AI)-driven models, such as neural networks and support vector machines, are increasingly employed for anomaly detection, they frequently exhibit limitations, including inadequate interpretability and suboptimal performance on imbalanced datasets (Goodfellow et al., 2014). Conventional statistical techniques, like time-series forecasting and regression analysis, provide enhanced transparency, however, are frequently underexploited in e-commerce cybersecurity scenarios (Zhang et al., 2022).

This research proposes a hybrid approach that integrates traditional statistical models, such as ARIMA and ANOVA, with modern machine learning techniques like XGBoost and LightGBM. The aim is to enhance the accuracy, interpretability, and scalability of cyberattack detection and prediction in the e-commerce sector. By applying this framework to real-world breach data, this study explores its effectiveness in identifying patterns, forecasting incident timing, and improving threat detection in complex cyber environments.

## 1.1 Research Problem

The cybersecurity threat landscape facing e-commerce platforms is increasingly flexible, behaviourally complex, and time sensitive. Retailers are frequently targeted during high-demand periods, such as Black Friday and year-end sales, when system loads increase, monitoring capabilities are challenged, and user habits change substantially (Kumar et al., 2022). The 2024 Neiman Marcus breach illustrated how attackers exploited compromised credentials during a high traffic shopping period, affecting 4.6 million customers (Baker, 2023), while the 2020 Shopify insider threat revealed weaknesses in internal access governance amid operational expansions (ThreatPost, 2020).

Traditional security systems, particularly those relying on rule-matching and signature-based detection, often struggle to recognise new or concealed attack vectors like credential stuffing and supply-chain injection tactics preferred by cybercriminals targeting e-commerce (Forrester, 2022). Although AI-driven anomaly detection solutions have gained popularity, they pose challenges in transparency and interpretability. Many operate as "black boxes," complicating the understanding and justification of their decisions, which is problematic in high-risk industries requiring accountability (Doshi-Velez & Kim, 2017). Moreover, their effectiveness typically decreases on imbalanced datasets where benign user actions outnumber malicious incidents, hindering accurate classification without additional techniques like data resampling (Chawla et al., 2002).

Statistical forecasting techniques such as ARIMA and hypothesis-driven models like ANOVA offer clarity and temporal insight but often fail to account for non-linear or multifaceted relationships among threat components (Bhatt et al., 2020). For example, while ARIMA may detect increased breaches in December, it might overlook compounded risks from phishing and third-party plugin vulnerabilities during promotional events, which ensemble machine learning models like XGBoost or CatBoost identify more effectively (Vähäkainu and Lehto, 2019).

This highlights a key methodological gap: the absence of hybrid systems that combine interpretable, statistically robust methods with advanced machine learning models to improve predictive insights and interpretability. Although longitudinal cyber event data are accessible from sources like the Verizon Community Data Breach (VCDB), existing frameworks rarely use this resource for attack prediction or to examine evolving threats (Nguyen et al., 2023).

In e-commerce cybersecurity, research seldom integrates interpretable statistical methods with ensemble-based machine learning, despite the growing use of AI in threat detection. This study addresses that gap by presenting a hybrid architecture combining time-series forecasting (ARIMA), variance analysis (ANOVA), and tree-based algorithms (Random Forest, XGBoost, LightGBM, CatBoost). While these models are individually recognised for precision and interpretability, their combined use in modelling cyberattack patterns remains underexplored. The study analyses the VCDB dataset, acknowledging challenges like selective reporting, publication delays, and missing dark web incidents. The core issue is evaluating the methodological rigor of this hybrid framework and the dataset's ability to represent the changing threat landscape.

## 1.2 Research Questions

This study explores the evolving dynamics of cybersecurity threats in the e-commerce sector using statistical and machine learning techniques applied to historical breach and threat intelligence data. The investigation is guided by the following research questions:

RQ1: What are the predominant types of cyberattacks impacting e-commerce platforms? This question aims to categorise and quantify threat vectors disproportionately affecting various domains within the e-commerce ecosystem, such as online retail, cloud hosting, auctions, platform services, and application development. Insights from frequency analysis and categorical visualisations help inform sector-specific defence strategies and optimise cybersecurity resource allocation.

RQ2: Do cyberattacks occur more frequently or intensely during specific times of the year, such as holiday seasons? Spikes in e-commerce traffic during peak commercial periods may coincide with increased malicious activity. This question investigates seasonal fluctuations using time-series decomposition and forecasting models, including ARIMA and Prophet. Identifying temporal risk windows supports better timing of security interventions.

RQ3: Is there a correlation between breaches involving Personally Identifiable Information (PII) and elevated threat keyword activity? Incidents exposing sensitive consumer data may be associated with distinct threat behaviours. This question tests statistical associations between the presence of PII and increased threat keyword activity using correlation analysis and ANOVA.

RQ4: Can machine learning models reliably forecast high-risk periods or emerging cyberattack patterns in e-commerce? This question evaluates the predictive power of historical breach data by applying classification algorithms such as Logistic Regression, XGBoost, LightGBM, and CatBoost. The focus includes assessing model performance, interpretability, and real-world applicability for proactive cybersecurity planning.

## 1.3 Research Scope and Limitations

This study investigates cyberattack patterns in the e-commerce sector using structured historical data sourced exclusively from the Verizon Community Data Breach (VCDB) repository. The focus includes identifying prevalent attack types, seasonal variations in breach activity, the correlation between PII and threat indicators, and the use of machine learning models for predictive analysis. The analysis applies statistical techniques and ensemble classifiers XGBoost, LightGBM, and CatBoost to identify high-risk indicators within the constraints of available data.

While the research offers valuable insights into historical cybersecurity patterns, it is constrained by several limitations that affect the generalisability and real-world applicability of its findings:

1. Incomplete and Biased Data

Exclusive reliance on VCDB introduces selection bias, as the dataset contains only publicly disclosed breaches. Many incidents, especially from small and medium-sized enterprises, are underreported or omitted due to reputational and regulatory concerns. This skews the data toward high-profile breaches involving large, well-resourced organisations, limiting the relevance of the findings to broader segments of the e-commerce ecosystem (Böttinger et al., 2022).

2. Absence of Real-Time Threat Telemetry

The use of static, historical data restricts the study's ability to anticipate novel threats such as zero-day exploits or rapidly evolving attack vectors. Without real-time telemetry, the models are inherently retrospective and may underperform during high-risk periods like Black Friday, when attackers adapt quickly to exploit heightened system load and transaction volume (Symonenko and Ivanova, 2021).

3. Lack of Regional and Regulatory Context

VCDB does not include comprehensive metadata on geographic location or applicable regulatory frameworks. This limits the study's ability to account for jurisdictional differences in data protection obligations and enforcement. For instance, breaches under the EU's GDPR may carry more severe consequences than similar events in less regulated regions (Cavusoglu et al., 2022), making cross-context generalisation problematic.

4. No Practical System Evaluation

Although rigorous validation techniques were applied, the models were not deployed in live cybersecurity environments. Their resilience against unstructured, real-time data or adversarial behaviours remains untested. Prior research (Gupta et al., 2022) shows that models performing well under academic conditions may fail in practice, reducing confidence in operational deployment.

5. Limited Model Interpretability

While ensemble models like XGBoost and CatBoost offer strong predictive capabilities, they often lack transparency. In high-stakes cybersecurity operations, interpretability is essential for timely, accountable decision-making. Although SHAP values were used to enhance transparency.

## 1.4 Research Flow

The remaining sections of this research are organised as follows:

Chapter 2 presents a detailed literature review on key cybersecurity threats affecting e-commerce platforms, including a review of machine learning and statistical forecasting techniques used in existing studies.

Chapter 3 outlines the research methodology, including the design framework, data selection criteria, preprocessing techniques, and model development steps used for cyberattack prediction.

Chapter 4 discusses the experimental results and offers reflective insights into the research questions, drawing on statistical trends, attack pattern detection, and model performance evaluations.

Finally, Chapter 5 presents the study's conclusions, summarises key findings, highlights contributions to the field, and suggests directions for future research.

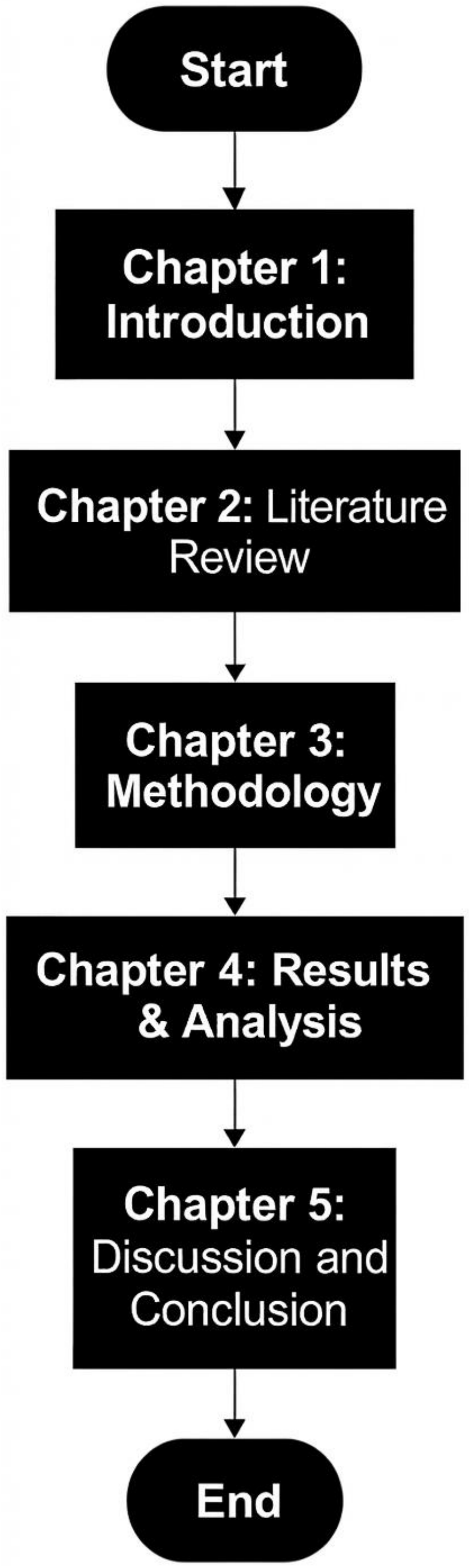

*Figure 1 Research Flow Diagram outlining the sequential structure of the dissertation, from Introduction to Conclusion (Voyant Tools,2025).*

# 2.0 Literature Review

The growth of e-commerce has transformed how businesses engage with consumers, with online retail sales surpassing USD 5.7 trillion globally in 2022 and projected to exceed USD 6.3 trillion by 2025 (UNCTAD, 2023). This expansion, driven by mobile adoption, digital payment infrastructure, and post-pandemic behavioural shifts (Kawa & Maryniak, 2022), has simultaneously increased the exposure of online platforms to cyber threats. As digital transactions multiply, the sector is seeing a surge in attacks such as account takeovers, payment fraud, DDoS events, and phishing (Cui, Ge & Zhang, 2020). These threats are not only more common but increasingly sophisticated, exploiting seasonal peaks and user behaviours (Yu et al., 2021). IBM Security (2023) reports that the average retail data breach now costs over USD 3.3 million, underscoring both financial and reputational risks. Traditional rule-based defences often fall short in this evolving landscape, especially in high-volume transactional environments.

This chapter (beginning with Section 2.2) offers a thematic review of academic and industry literature on cybersecurity risks and detection strategies in online retail. It outlines the methodology for source selection, examines the evolution of cyber risks in e-commerce, and classifies key threat types. It then critically evaluates core detection paradigms rule-based systems, statistical forecasting, machine learning, and hybrid models highlighting their respective strengths and limitations. The chapter concludes by identifying research gaps and explaining this study's contribution to improving cybersecurity in the e-commerce domain.

## 2.1 Process of Selecting Literature for Study

A rigorous and transparent methodology was employed to ensure the literature selected was relevant, credible, and comprehensive. Given the interdisciplinary nature of cybersecurity in e-commerce, the review included both peer-reviewed academic sources and reputable industry reports published between 2010 and 2025. The goal was to capture foundational theories, emerging trends, and current empirical developments related to cyberattack detection, statistical analysis, and machine learning applications in online retail. The search process utilised academic databases such as IEEE Xplore, ACM Digital Library, ScienceDirect, SpringerLink, and Scopus, as well as specialised repositories like arXiv, MPDI, Google Scholar, and PubMed Central for AI and ML-related studies. Keywords were carefully constructed and iteratively refined, including combinations such as "cybersecurity AND e-commerce," "fraud detection in online retail," "machine learning for cyber threat detection," "rule-based intrusion detection systems," and "statistical forecasting in cybersecurity."

The inclusion criteria followed five core principles:

1. Relevance to cybersecurity or fraud detection in online retail/e-commerce.
2. Methodological rigour, favouring empirical studies and reproducibility.
3. Recency, prioritising studies from the past 10 years unless foundational.
4. Peer-reviewed status or origin from reputable organisations (e.g., ENISA, IBM, UNCTAD).

5. Thematic alignment with the study's focus on threat detection via rule-based, statistical, or machine learning approaches.

A PRISMA-inspired filtering process ensured quality control. From an initial set of over 250 articles, duplicates and unrelated titles were excluded, resulting in 176 publications. Screening abstracts and conclusions narrowed the pool to 96 core sources. Following full-text analysis and citation mapping, 53 academic publications and 9 industry whitepapers were retained for in-depth thematic review. Citation tracing was also used to identify highly influential works (e.g., Chawla et al., 2002). Efforts were made to include diverse perspectives across geographical regions and disciplines, such as information systems, cybersecurity, artificial intelligence, and applied statistics.

Overall, the final literature corpus was critically appraised and carefully curated to form a robust evidence base for the thematic analysis that follows, grounded in contemporary, high-impact research.

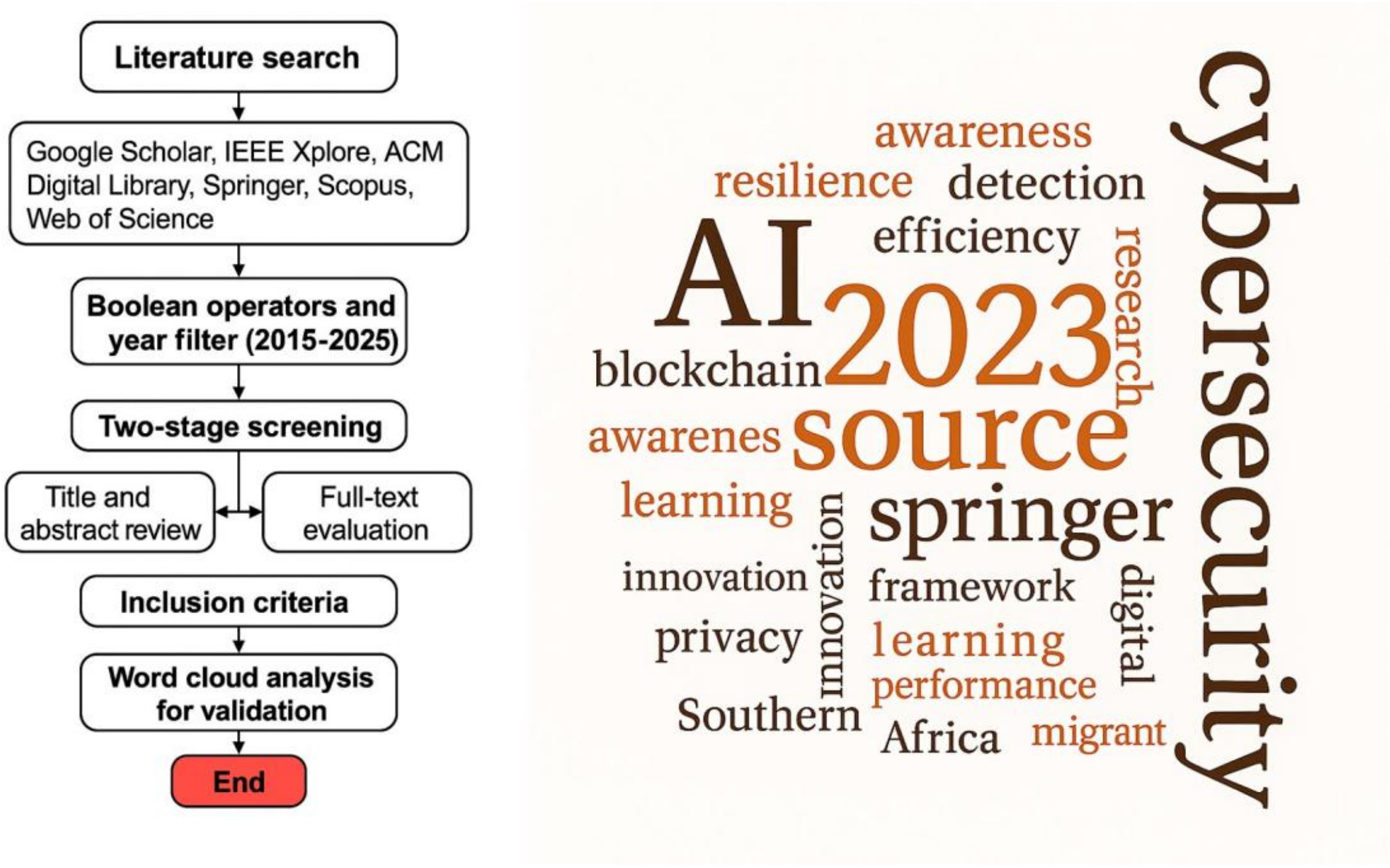

*Figure 2 Literature Selection Process (Voyant Tools,2025).*

## 2.2 The Evolution of E-Commerce

E-commerce has evolved from rigid enterprise systems into a dynamic global marketplace driven by accessibility, scalability, and data intelligence. Its origins trace back to Electronic Data Interchange (EDI) in the 1970s, which enabled machine-readable transactions but suffered from limited adoption due to proprietary formats and high costs (Zwass, 1996). The emergence of the World Wide Web in the 1990s catalysed scalable business-to-consumer (B2C) and business-to-business (B2B) commerce through platforms like Amazon and eBay (Rayport & Jaworski, 2002).

The 2000s saw rapid expansion with mobile technology and internet penetration, leading to real-time, multi-channel shopping, especially across Asia (Chen, Wang & Chen, 2018). Secure payment systems such as PayPal and later blockchain-based solutions improved transaction trust and decentralisation (Zheng et al., 2020). Today's e-commerce models, ranging from direct-to-consumer (D2C) and consumer-to-consumer (C2C) to subscription-based services, leverage cloud computing for real-time analytics, dynamic pricing, and AI-driven fraud detection (Alharthi et al., 2023).

Technologies like blockchain enhance supply chain security (Casino, Dasaklis & Patsakis, 2019), while AI supports personalisation, predictive services, and cybersecurity automation. However, these innovations raise complex concerns around algorithmic transparency, ethics, and governance (Ribeiro et al., 2016), as well as data privacy and compliance in large-scale consumer data processing (Rjoub, Alomari & Yousif, 2023).

Overall, the trajectory from EDI to AI reflects growing platform complexity and data dependence, accompanied by evolving cyber threats that demand adaptive and intelligent security frameworks.

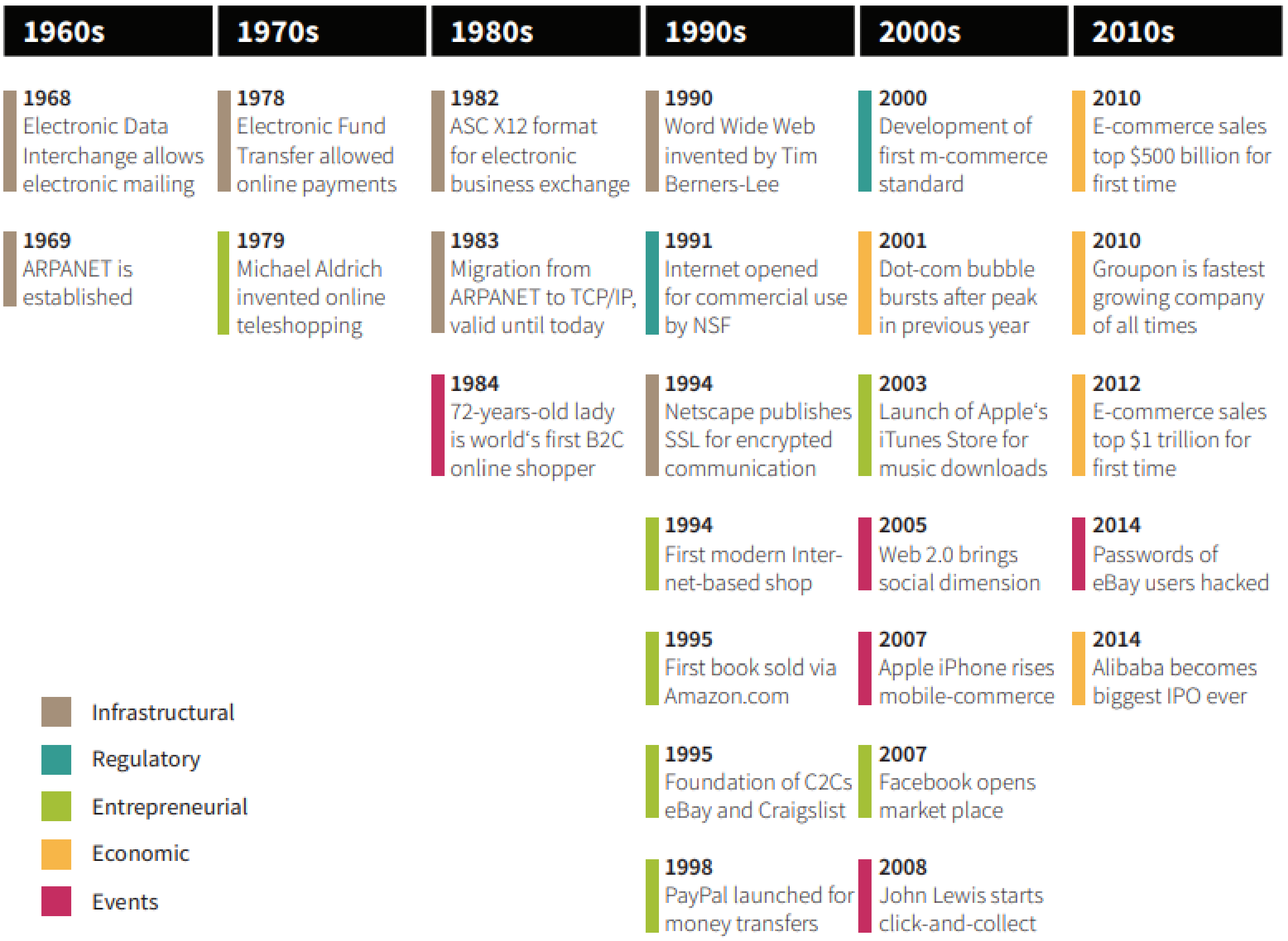

*Figure 3 Categorised milestones in the development of electronic commerce (HENDRIK TERBECK, 2025).*

## 2.3 The Evolving Threat Landscape in E-Commerce

The e-commerce threat landscape has evolved from isolated breaches into complex; multi-phase cyberattack operations. According to FBI IC3 (2023), credential-based intrusions such as credential stuffing and session hijacking now account for over 60% of breach vectors in online retail, up from less than 40% five years ago. This rise is fuelled by password reuse and the underutilisation of multi-factor authentication (MFA). While Microsoft (2019) asserts that MFA could prevent 99.9% of credential attacks, Bada et al. (2019) found that only 27% of UK SMEs had implemented it, often due to resource limitations. Though malware and ransomware remain prevalent, the attack paradigm has shifted. IBM Security (2023) highlights the rise of ransomware-as-a-service (RaaS), which empowers low-skill attackers during lucrative periods like Black Friday. Mirkovic and Reiher (2021) show how attackers increasingly time operations to disrupt businesses during commercial peaks. Liu et al. (2022) report a twofold rise in botnet-fuelled DDoS attacks via IoT devices since 2020. While Radware (2022) warns of evasive techniques like encrypted traffic and adaptive modulation, cloud defences (e.g., AWS Shield, Azure DDoS Protection) offer mitigation, but not immunity.

Human vulnerabilities persist as a critical concern. Hadnagy and Fincher (2015) illustrate how social engineering continues to bypass technical defences. Symantec (2023) notes that AI-generated phishing content increasingly evades both human review and automated filters, underscoring the need for user training and behavioural defences. Larger organisations are adopting layered defences that integrate threat intelligence with real-time analytics (NIST, 2023), but SMEs lag behind. NCCoE (2023) reports that 68% of SMEs still rely on basic antivirus and lack structured penetration testing. This "security maturity gap" creates systemic risk, as breaches in smaller vendors can compromise entire ecosystems through third-party integrations. Collectively, these trends highlight a core challenge for e-commerce: the need to move from fragmented, compliance-driven approaches to holistic, adaptive cybersecurity strategies. These must incorporate behavioural analytics, dynamic threat intelligence, and contextual risk awareness especially for vulnerable SMEs. Without such evolution, the widening gap between attacker innovation and defence readiness will continue to endanger digital commerce.

### 2.3.1 Specific Cyberattack Vectors Targeting E-Commerce

The rapid expansion of e-commerce has increased its appeal to cybercriminals, who target platforms for financial or political gain. This is largely due to the concentration of sensitive data, such as payment credentials and personal identifiers, in online retail systems (Li et al., 2020). Beyond technical vulnerabilities, attackers also exploit consumer behaviours for instance, impulsive clicks during flash sales or discount events (Kshetri, 2021). Modern attacks are no longer limited to single breaches but have become multi-stage operations. According to the FBI IC3 (2023), credential-based attacks like stuffing and session hijacking now account for over 60% of breaches in online retail, up from under 40% five years ago. This increase is linked to password reuse and the low adoption of multi-factor authentication (MFA). While Microsoft (2019) claims that MFA can prevent 99.9% of such breaches, only 27% of UK SMEs have implemented it due to cost and operational barriers (Bada et al., 2019).

The threat landscape has also grown more complex. Malware and ransomware remain prevalent, but IBM Security (2023) reports that ransomware-as-a-service (RaaS) now enables low-skill attackers to launch sophisticated campaigns, especially during high-traffic periods like Black Friday. Temporal targeting further amplifies disruption (Mirkovic & Reiher, 2021). Botnet-driven DDoS attacks have doubled since 2020, fuelled by IoT devices (Liu et al., 2022). Although cloud-based tools like AWS Shield offer mitigation, evolving techniques such as encrypted traffic can evade traditional defences (Radware, 2022). Social engineering remains a persistent risk. Hadnagy and Fincher (2015) emphasise how attackers bypass technical safeguards by exploiting human trust. The emergence of AI-generated phishing content has further complicated detection, often eluding both human review and automated filters. While large enterprises are deploying advanced, layered defences with real-time analytics (NIST, 2023), small and medium-sized enterprises (SMEs) often lag. Many still rely on basic antivirus solutions and lack penetration testing (NCCoE, 2023). This creates sector-wide risk, as compromised SMEs can act as entry points into larger ecosystems through third-party integrations. In summary, the escalating sophistication of cyber threats demands a shift from fragmented, compliance-focused security to holistic, adaptive frameworks. These should integrate behavioural analytics, real-time threat intelligence, and contextual awareness particularly for at-risk SMEs. Without such a shift, the gap between attacker capabilities and defence resilience will continue to grow.

*Table 1 Evolution of Cyberattacks (adapted from Mallick and Nath, 2024))*

| **Period** | **Type of Attack** | **Description** | **Notable Cases/Examples** |
|---|---|---|---|
| 1980s | Early Network Exploits | Initial attacks focused on exploiting network protocols and unpatched systems. | Marcus Hess military system breach (1986) |
| 1990s | Viruses and Worms | Self-replicating malware like Melissa and ILOVEYOU caused widespread damage. | ILOVEYOU virus (2000) |
| 2000s | Phishing and Identity Theft | Growth of social engineering attacks to steal credentials and financial data. | Phishing campaigns targeting banks and e-commerce sites |
| 2010s | Botnets and DDoS Attacks | Use of compromised devices (including IoT) to launch massive DDoS attacks. | Mirai botnet causing major outages (2016) |
| 2010s | Ransomware | Malware encrypting data and demanding ransom, targeting businesses globally. | WannaCry outbreak (2017) |
| 2020s | Supply Chain Attacks | Attacks compromising third-party software to infiltrate target systems. | SolarWinds breach (2020) |
| 2020s | AI-Enhanced Attacks | Use of AI for sophisticated phishing, evasion, and automated exploitation. | Deepfake phishing, automated vulnerability scanning |

#### 2.3.1.1 Payment Card Fraud

Payment card fraud refers to the unauthorised use of credit, debit, or prepaid card information to make fraudulent transactions or gain access to funds or services (Ngai, Hu, Wong, Chen & Sun, 2011). This type of fraud, particularly prevalent in card-not-present (CNP) transactions, remains a critical threat to e-commerce, where fraudsters exploit stolen credentials remotely, bypassing physical verification (Zhai, 2024).

They frequently use automated botnets to carry out credential-stuffing at scale, supplemented by sophisticated social-engineering methods like phishing and deepfake audio to increase success rates (Zhai, 2024). A recent academic study estimates that EU-based merchants suffered over €2 billion in CNP fraud during the first half of 2023; however, inconsistent reporting and limited transparency in emerging markets suggest this figure likely underrepresents the global impact (Singh et al., 202). In response, industry adoption of layered defenses, such as Strong Customer Authentication (e.g., 3-D Secure), tokenisation, and device fingerprinting, has shown strong effectiveness in controlled environments, but many small and medium-sized enterprises (SMEs) struggle to implement them due to resource and technical constraints (Ashby, 2022).

The introduction of behavioral biometrics, including keystroke dynamics, mouse-trajectory monitoring, and touch-pattern recognition, augmented by graph-based machine learning models, has yielded approximately 15% higher detection rates and 20% fewer false positives in real-world pilots, though integration complexity remains a significant barrier (Yin et al., 2022). Beyond financial loss, CNP fraud also causes reputational damage and regulatory repercussions, especially under GDPR, where high-profile cases reveal how non-compliance exacerbates the impact of breaches (Peretti et al., 2022). Peer-reviewed research underscores the importance of a multi-layered defense architecture, combining adaptive authentication, real-time behavioral analytics, regulatory compliance, and organisational cyber-governance, to build resilience against evolving fraud threats and maintain customer trust, particularly for SMEs (Rejeb et al., 2024).

#### 2.3.1.2 Phishing and Social Engineering Attacks

Social engineering attacks pose a significant threat to e-commerce environments by exploiting human psychology rather than technical vulnerabilities (Hadnagy & Fincher, 2015). Among these, phishing remains the most widespread vector, employing deceptive emails, SMS messages, and QR codes to impersonate legitimate entities and extract sensitive information such as login credentials and financial data (Abawajy, Hassan & Alemayehu, 2021). Unlike automated threats like credential stuffing, phishing capitalises on cognitive biases and trust mechanisms, allowing it to bypass conventional technical controls (Krombholz et al., 2015). Recent campaigns have grown increasingly sophisticated, integrating AI-generated content, polymorphic URLs, and domain spoofing to evade detection systems, resulting in a dynamic arms race between attackers and defenders (Nguyen et al., 2023). Although machine learning–based techniques, such as natural language processing and image recognition, have improved detection accuracy (Sahingoz et al., 2019), their effectiveness is often short-lived as attackers rapidly adapt, revealing the limitations of relying solely on technical defences. Beyond external vectors, social engineering also manifests internally through Business Email Compromise (BEC), in which attackers impersonate employees or vendors to conduct fraud or deploy malware. In 2023 alone, BEC attacks accounted for over $2.4 billion in reported global losses, with the e-commerce sector particularly affected due to its transactional volume and trust-based operations (FBI IC3, 2023). These attacks often exploit organisational hierarchies and insufficient internal controls, especially within small and medium enterprises (SMEs) that lack mature cybersecurity infrastructures (Rombaldo Junior et al., 2023).

Given the persistent and adaptive nature of social engineering, a holistic defence strategy is increasingly recognised as essential. While machine learning enhances technical detection, empirical research highlights the importance of complementing these tools with user-focused measures, such as ongoing phishing simulations, real-time browser alerts, and continuous user education, to significantly reduce susceptibility (Bada, Sasse & Nurse, 2019). Larger organisations like Amazon and Shopify have adopted comprehensive socio-technical frameworks that integrate automation, policy, and workforce training (Shopify, 2023). However, many SMEs remain reliant on basic spam filters, leaving them disproportionately vulnerable. As a result, scholars and practitioners increasingly advocate for socio-technical approaches that fuse technological controls with governance mechanisms and cultural transformation to build long-term organisational resilience (Alshaikh et al., 2021).

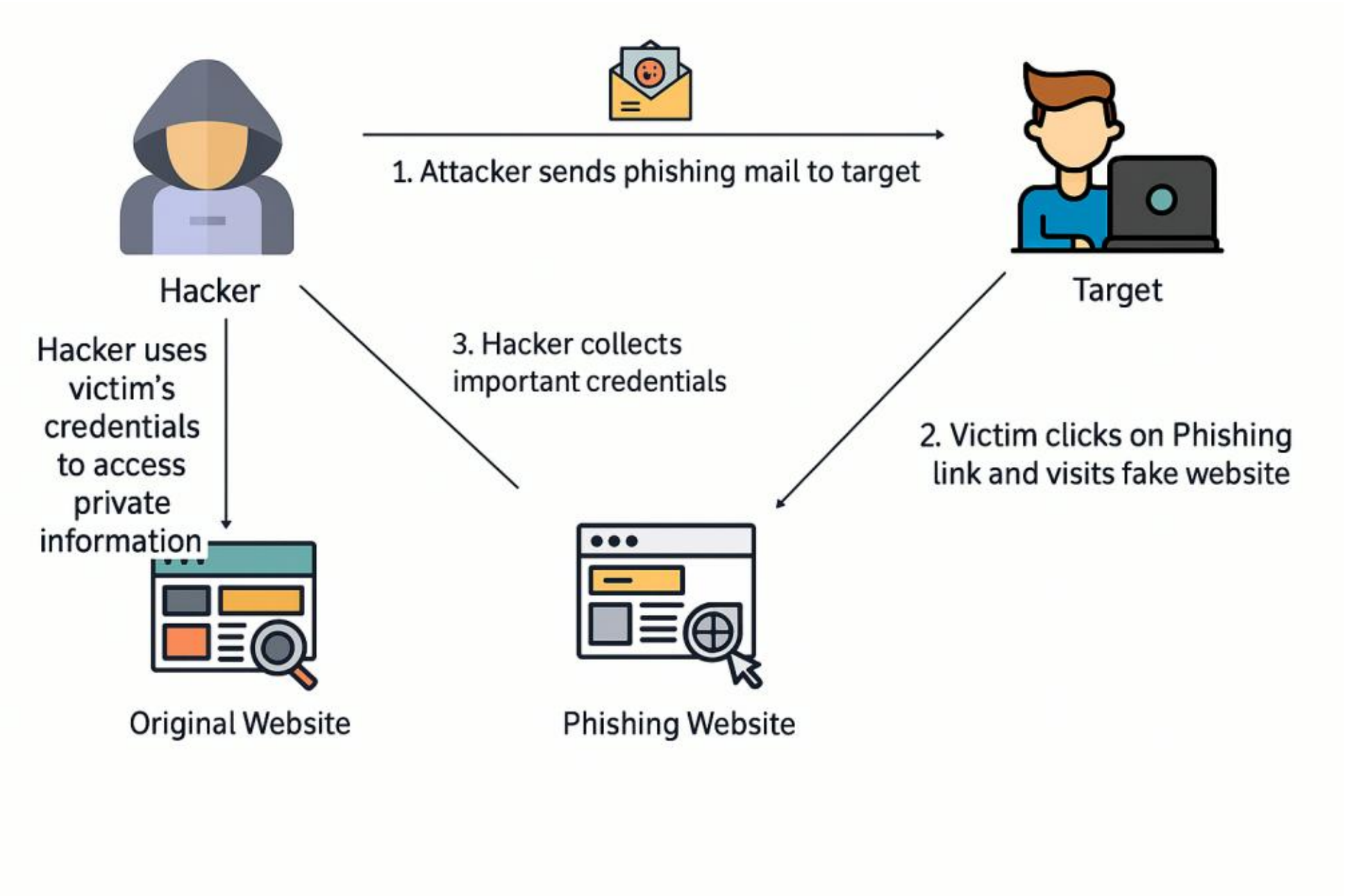

*Figure 4 Phishing Workflow: Exploiting User Credentials for Cyberattacks (Siddiqi, Pak & Siddiqi,2022).*

#### 2.3.1.3 Distributed Denial of Service (DDoS) Attacks

Distributed Denial of Service (DDoS) attacks threaten the availability, performance, and integrity of e-commerce platforms by overwhelming systems, servers, networks, or applications with massive illegal traffic. This renders services inaccessible to legitimate users and causes significant financial and reputational damage (ENISA, 2023). Attacks often coincide with major retail events like Black Friday or Cyber Monday to maximise disruption and cost impact (Mirkovic & Reiher, 2021).

Advances in ensemble and hybrid machine learning methods have improved threat detection on e-commerce platforms, responding to the escalating economic costs of cybercrime, which surpassed $1 trillion globally in 2023 comparable to the GDP of several G20 nations (World Bank, 2024). Beyond immediate financial loss, cyberattacks erode consumer trust, brand loyalty, and market competitiveness (Kannan et al., 2018). Ransomware exemplifies high financial damage, with average costs reaching £3.7 million per attack (Cybersecurity Ventures, 2023). For context, the 2013 Target breach cost over $200 million in direct expenses, excluding reputational and legal fallout (Target Corporation, 2014), while the 2018 British Airways breach led to a £20 million fine and severe brand damage (BBC News, 2020). These incidents highlight the far-reaching consequences of cyberattacks.

Modern DDoS mitigation increasingly employs cloud-based, machine learning-driven solutions. Behavioral analytics help distinguish human from automated traffic, improving defense accuracy (Kshetri, 2021). Services like AWS Shield and Cloudflare combine dynamic scaling, rate limiting, IP reputation scoring, and CAPTCHA challenges to counter threats while preserving user experience (Gartner, 2023). Nonetheless, encrypted traffic and advanced evasion tactics pose ongoing challenges (Sahaf et al., 2020). A notable gap exists between large e-commerce firms and small-to-medium enterprises (SMEs) in DDoS defense capabilities. SMEs often rely on basic firewalls or affordable third-party content delivery networks (CDNs) with minimal protection. The 2020 Magento-related attacks exposed vulnerabilities in smaller shops lacking adequate defences (Cisco Talos, 2020). In contrast, giants like Amazon and eBay deploy multi-tiered mitigation strategies involving specialised response teams and real-time monitoring, resources that SMEs typically cannot access.

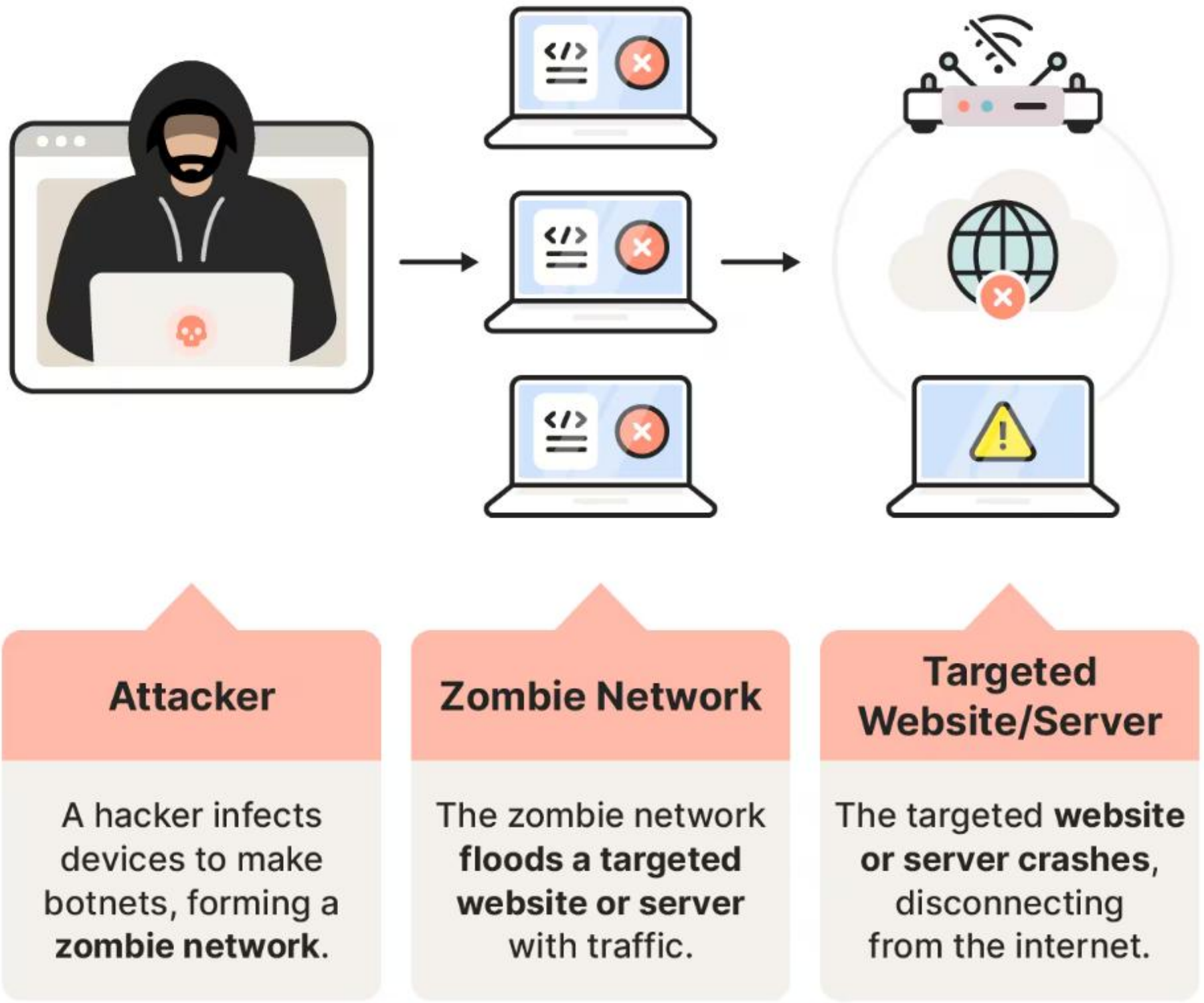

*Figure 5 Visualisation of a DDoS Attack Lifecycle (Norton, no date)*

#### 2.3.1.4 Malware and Ransomware Attacks

Malware and ransomware represent related but distinct threats within e-commerce. Malware broadly refers to malicious software designed to infiltrate systems, exfiltrate data, or maintain persistent access (Alharthi et al., 2023). Ransomware, by contrast, focuses on extortion encrypting critical assets and demanding payment for decryption or to prevent public data exposure (Kshetri, 2021). Both threaten confidentiality, integrity, and availability but require tailored defences.

Malware has become increasingly stealthy, especially with fileless variants that evade signature-based detection. Attackers often exploit third-party components such as CMS platforms and payment plugins to implant keyloggers or JavaScript skimmers. The Magecart campaign, for example, infiltrated over 7,000 e-commerce sites to steal payment data (Europol, 2022). Liu et al. (2022) note that malware's prolonged dwell time makes it especially dangerous, enabling covert data theft compared to the more visible volumetric DDoS attacks. Ransomware has evolved into a highly organised criminal enterprise through ransomware-as-a-service (RaaS) models. Groups like LockBit and BlackCat deploy double extortion tactics encrypting data while threatening public leaks to maximise pressure on victims (IBM Security, 2023). Attack timing often aligns with peak retail periods such as Black Friday, amplifying disruption and financial damage (Tietoevry, 2025).

Organisational readiness is critical. Major platforms like Amazon and Shopify operate advanced Security Operations Centers (SOCs) with threat hunting and behavioural analytics to detect early intrusions (Rombaldo Junior et al., 2023). In contrast, SMEs often lack sophisticated endpoint detection and prolonged threat visibility, increasing their vulnerability (Bada et al., 2019; Alharthi et al., 2023). This highlights a persistent "security maturity gap" in the sector. Compliance standards such as PCIDSS and the NIST Cybersecurity Framework provide foundational guidance (PCI SSC, 2022). However, Schatz and Bashroush (2020) caution that mere compliance is insufficient. The 2020 Magento breach exploiting unpatched CMS flaws on PCI-compliant systems illustrates the danger of complacency without active threat monitoring (Cisco Talos, 2020). Kshetri (2021) advocates embedding compliance within a culture of threat intelligence sharing and adaptive security. Human factors also significantly impact outcomes. Phishing awareness, simulated attacks, and ongoing staff training are crucial to prevent initial malware infections (Cappelli et al., 2012).

While large enterprises institutionalise continuous education, SMEs often lack resources for sustained training, deepening vulnerabilities (Rombaldo Junior et al., 2023). Research differentiates malware and ransomware operationally: malware campaigns like Magecart aim for low-noise, long-term data theft, whereas ransomware seeks rapid disruption and financial gain via high-profile extortion (Europol, 2022). Defences must therefore be multifaceted, combining threat intelligence, anomaly detection, incident response, and employee vigilance. Ultimately, malware and ransomware attacks expose fundamental structural and cultural gaps in cybersecurity preparedness. Effective defense requires integrated, multidimensional strategies that transcend technical fixes to include governance, organisational culture, and equitable resource allocation. Bridging the security maturity gap between SMEs and large firms is essential to protect the broader e-commerce ecosystem against increasingly sophisticated and coordinated cyber threats (Tietoevry, 2025).

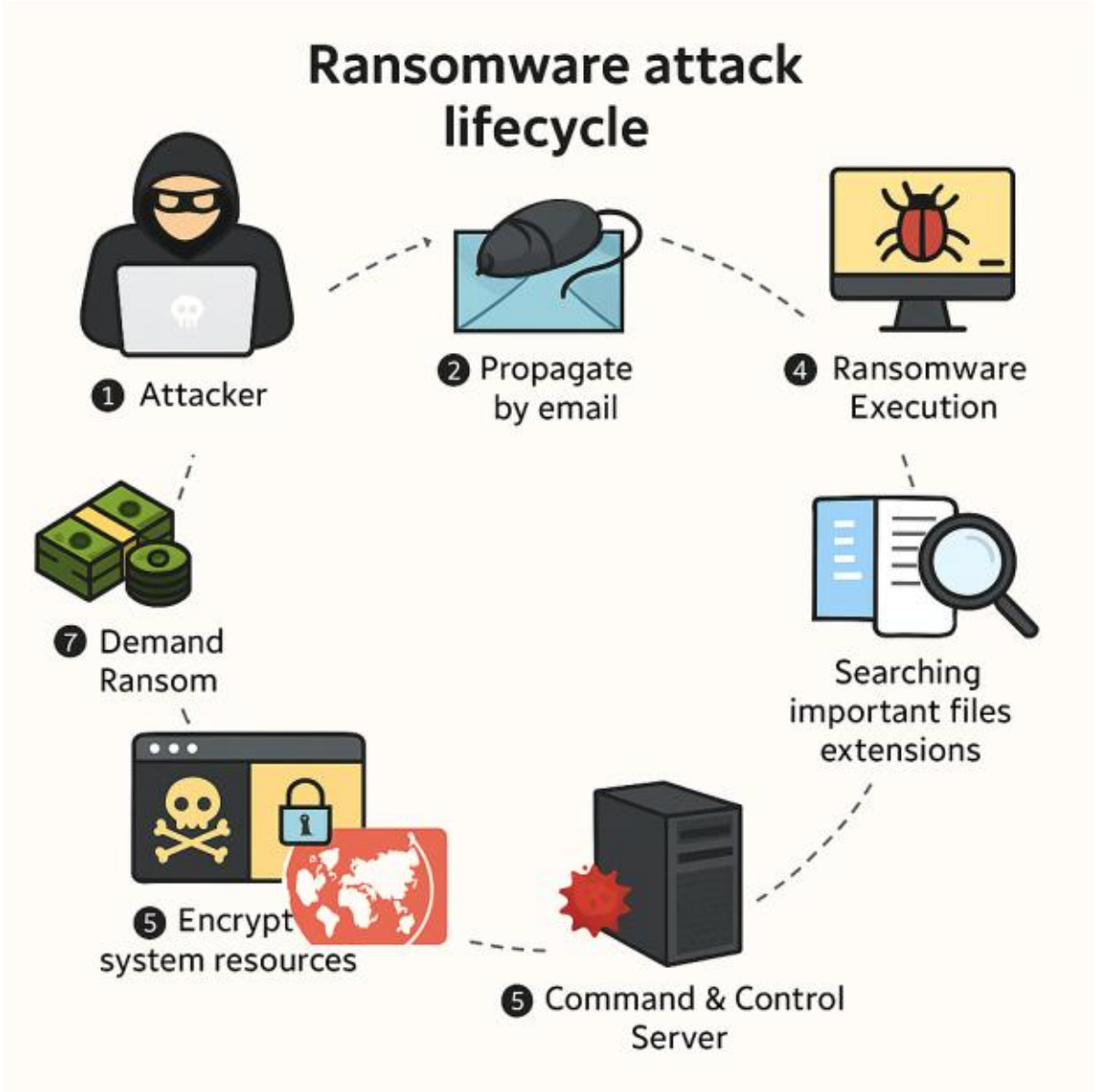

*Figure 6 Ransomware attack lifecycle illustrating key stages from initial infection to ransom demand (Zahoora et al., 2022).*

### 2.3.1.5 SQL Injection and Cross-Site Scripting (XSS) in E-Commerce

SQL Injection (SQLi) and Cross-Site Scripting (XSS) remain among the most persistent and damaging vulnerabilities on e-commerce platforms. Both exploit weaknesses where user input interacts with application logic yet differ in technique and impact. SQLi enables attackers to manipulate database queries via unfiltered inputs, often leading to unauthorised access to sensitive backend data like customer and financial records (Mallick & Nath, 2024). The 2013 Target breach, partially caused by SQLi flaws, exposed over 40 million cardholder details, highlighting risks from poor query sanitisation and legacy architectures (Shu et al., 2017).

XSS attacks compromise the integrity of content rendered in users' browsers by injecting malicious scripts capable of hijacking sessions, redirecting users, or covertly altering interfaces (Gupta et al., 2022). The British Airways breach, where attackers used third-party scripts to redirect customers to fraudulent payment portals illustrates how underestimated XSS threats directly impact transaction security and consumer trust (ICO, 2018). Both vulnerabilities root in inadequate input handling and insufficient secure-by-design development, challenges exacerbated by modern frameworks like Single Page Applications (SPAs) that expand the attack surface (OWASP, 2023).

Mitigation requires layered strategies beyond perimeter defences. SQLi can be prevented via parameterised queries, strict input validation, and role-based access controls, though these are unevenly applied across microservices and cloud environments (Halfond et al., 2006). Web Application Firewalls (WAFs) provide frontline defence but face evasion through payload obfuscation and delayed script execution (Alqahtani et al., 2021). XSS defences include rigorous output encoding, Content Security Policies (CSP), and sanitisation of user-generated content; however, many e-commerce sites misconfigure or over-permit CSP headers, often due to unchecked third-party code (Rjoub et al., 2020). Emerging machine learning techniques offer promising anomaly detection for novel SQLi and XSS exploits by analysing unusual HTTP request patterns and payload entropy, yet high false positives and performance costs hinder widespread use in high-traffic settings (Mallick 2024). These persistent vulnerabilities reflect systemic issues beyond technical flaws. Schatz and Bashroush (2020) argue that compliance-focused security cultures prioritising audit passing over substantive patching worsens long-term risk exposure. SMEs are especially vulnerable due to limited developer training, scarce secure code review resources, and reliance on third-party plugins lacking rigorous scrutiny.

Ultimately, SQL Injection (SQLi) and Cross-Site Scripting (XSS) represent not only technical shortcomings but also governance and process failures within the software development lifecycle (OWASP, 2023). Effective mitigation demands the integration of security principles throughout development workflows, commonly referred to as "secure by design", alongside strict enforcement of contextual input/output validation policies and the deployment of adaptive defences that evolve with emerging attacker techniques (McGraw, 2006). High-profile breaches, such as those affecting Target in 2013 and British Airways in 2018, underscore the profound reputational, regulatory, and financial consequences of neglecting such application-layer vulnerabilities (Arroyabe et al., 2024). These incidents have catalysed industry-wide emphasis on secure coding practices and proactive threat modelling, positioning SQLi and XSS prevention as critical priorities within robust e-commerce cybersecurity strategies.

## 2.4 Vulnerabilities, Exploit Attacks, and Attacker Typologies in E-Commerce Platforms

In 2023 alone, cyberattacks targeting e-commerce and broader online platforms contributed to global economic losses estimated in the hundreds of billions of dollars, based on aggregated data from industry sources such as Norton (USD 172 billion in 2017) up to projected values exceeding USD 10.5 trillion annually by 2025 (Vergara Cobos & Cakir, 2024). These losses often stem from vulnerabilities in web applications and APIs, the foundational technologies underpinning online commerce. According to the 2023 OWASP report, over 90% of e-commerce breaches exploit known vulnerabilities such as injection flaws, broken authentication, and insecure APIs, underscoring the persistent inadequacies in platform security (OWASP, 2023). These vulnerabilities, arising from architectural complexities, rapid deployment cycles, and third-party integrations, provide fertile ground for attackers who exploit them through a variety of sophisticated methods.

This section undertakes a rigorous analysis of these structural weaknesses and the exploit mechanisms leveraged by diverse attacker typologies ranging from opportunistic cybercriminals to advanced persistent threat (APT) actors. Drawing upon recent high-profile breaches such as the Magecart campaigns, the Capital One cloud API compromise, and the Shopify API exploitation, it situates these technical vulnerabilities within the real-world operational context of e-commerce. The discussion further elucidates how internal threats and supply chain dependencies compound risk, creating a multi-dimensional threat environment.

### 2.4.1 Platform Vulnerabilities: The Fragility of Web Applications, APIs, and Third-Party Integrations in E-Commerce

E-commerce ecosystems rely on a complex network of platforms, mainly consisting of web applications, application programming interfaces (APIs), and extensive third-party service integrations that collectively enable customer interactions, transaction processing, and operational workflows (Mallick & Nath, 2024). Although these platforms enhance functionality and user experience, they also present significant vulnerabilities that attackers systematically exploit, ultimately compromising the integrity, confidentiality, and availability of e-commerce services.

**1. Web Application (mobile and desktop) Vulnerabilities**

Web applications are central to e-commerce, but their complexity, including layered architectures, multiple programming languages, and legacy components, makes them prone to security flaws. OWASP highlights common risks like SQL injection and cross-site scripting (XSS), often caused by poor input validation and output encoding (OWASP, 2023). The incident revealed critical failures in authentication and token management, areas often overlooked in API security (OWASP, 2023). Strengthening API defences requires implementing best practices such as OAuth 2.0 for secure authorisation, rate limiting to mitigate abuse, and comprehensive audit logging to enable timely threat detection and forensic analysis ( Lodderstedt, McGloin and Hunt, 2013),

**2. API Vulnerabilities**

APIs are critical for e-commerce scalability and integration, connecting systems like inventory, payments, and customer management (Rjoub, Alomari & Yousif, 2023). However, this reliance expands the attack surface, exposing weaknesses such as poor authentication, missing rate limits, and inadequate monitoring. In the 2021 Shopify breach, attackers exploited insecure API endpoints to access sensitive customer and payment data across merchant accounts ((Uzsunny, 2022). The incident revealed failures in authentication and token management. Strengthening API security requires best practices like OAuth implementation, rate limiting, and detailed audit logging to support timely threat detection and response.

**3. Third-Party Service Integrations**

Third-party services such as payment gateways, marketing platforms, and logistics providers are essential to e-commerce functionality but introduce significant security vulnerabilities. These external systems often lie outside the direct governance of the primary platform and may lack consistent security controls, making them attractive targets for adversaries ( ENISA, 2021). Notably, Magecart-style attacks have exploited weaknesses in third-party JavaScript integrations to inject malicious code into checkout pages, enabling large-scale theft of payment data across thousands of online retailers (Europol, 2022). These supply chain attacks bypass conventional defences by exploiting trusted vendor relationships, rendering perimeter-based protections insufficient (ENISA, 2021). As a result, industry guidelines now emphasise continuous third-party risk monitoring, formal vendor assessment procedures, and adoption of supply chain security frameworks such as NIST SP 800-161r1 to safeguard e-commerce platforms from indirect compromise (Gartner, 2023).

## 2.5 Key Threat Actors in E-Commerce Cybersecurity

E-commerce systems face threats from diverse actors including cybercriminals, insiders, and state-sponsored groups, each exploiting specific vulnerability (Mallick & Nath, 2024). The rise of tools like ransomware-as-a-service and automated attack kits has increased the scale and complexity of these threats (Zetter, 2019). Attacks now extend beyond data theft to include extortion, disruption, and cryptocurrency-based laundering (Anderson et al., 2019).

The subsections below outline the main threat actor categories and their tactics.

1. **Insider Threats**

   Insider threats remain a significant yet often underestimated risk in e-commerce, originating from employees, contractors, or trusted third parties who misuse legitimate access for malicious or negligent purposes (Greitzer & Frincke, 2010). The trusted status of insiders allows them to circumvent traditional perimeter defences, enabling activities such as data theft, fraud, or sabotage. A notable example is the 2014 Home Depot breach, where attackers gained access to the retailer's network via stolen third-party vendor credentials, ultimately compromising over 50 million credit card numbers (Shu et al., 2017). Although the breach involved external actors, the incident highlighted how poorly managed internal access , especially involving third-party insiders , can expose e-commerce systems to serious harm. Research indicates that insiders are responsible for a substantial proportion of data breaches, often driven by motives including financial gain, dissatisfaction, or coercion (McKinsey & Company, 2018). The widespread adoption of remote work and cloud-based services has further expanded the insider threat surface, increasing the complexity of access governance. Unlike overt external attacks, insider threats frequently involve prolonged, discreet misuse of privileges, making them harder to detect. Mitigating such risks requires a combination of strict access control policies, behavioural analytics, continuous monitoring, and comprehensive personnel vetting(Gupta et al., 2021).

2. **Hacktivists and Competitors**

   While most attacks are financially driven, hacktivist groups and rival companies also pose serious threats to e-commerce, motivated by ideology or competition rather than profit (Mallick & Nath, 2024). Hacktivists use tactics like denial-of-service, website defacement, and data leaks to damage reputations or express political messages (Weimann, 2015). Competitive adversaries may engage in espionage, sabotage, or review manipulation to gain market advantage (Holt et al., 2015). Though often less sophisticated than state-sponsored attacks, these threats can severely impact brand trust especially during high-traffic sales events.

3. **State-Sponsored Actors**

   State-sponsored groups pose a rising threat to e-commerce, driven by goals like economic espionage and geopolitical disruption (Li & Liu, 2021). Using advanced persistent threats (APTs), zero-day exploits, and social engineering, they often target cloud providers and supply chains, enabling stealthy, long-term attacks. The 2020 SolarWinds breach demonstrated the risks facing cloud-reliant platforms (FireEye, 2020). These well-resourced actors require robust detection, coordinated response, and global cooperation (Rid & Buchanan, 2015). As the most strategic and persistent threat, they demand multilayered defences combining technology, human oversight, and shared threat intelligence.

E-commerce faces varied threats from cybercriminals, insiders, hacktivists, and state actors. Mitigating these risks requires layered defences, strong internal controls, and ongoing threat monitoring.

## 2.6 Seasonal Forecasting in Cyberattack Prediction

Forecasting seasonal patterns in cyberattacks is essential for proactive threat mitigation in e-commerce, where attack frequency often aligns with high-traffic events such as Black Friday, year-end sales, and promotional campaigns (Hyndman & Athanasopoulos, 2018). These predictable surges in exposure necessitate time series models capable of capturing seasonal trends, enabling cybersecurity teams to allocate resources preemptively and reduce detection latency (Kumar et al., 2021). Among the most employed models are ARIMA (Autoregressive Integrated Moving Average) and its seasonal extension SARIMA, which decompose time series into trend, seasonal, and irregular components (Box, Jenkins & Reinsel, 2015). Empirical studies demonstrate their value in modelling cyclic threat patterns: for instance, Johnson and Wang (2020) applied SARIMA to cyberattack volumes surrounding Cyber Monday, identifying both weekly and annual periodicities. Similarly, Liu et al. (2022) used SARIMA on longitudinal DDoS datasets, revealing seasonal spikes aligned with quarterly retail cycles.

However, the practical utility of ARIMA-based models is constrained by their strong assumptions of linearity and stationarity, along with the need for manual parameter tuning (Lee et al., 2019). These models often underperform in dynamic environments marked by sudden behavioural shifts, such as botnet campaigns or emergent zero-day exploits (Ghahramani et al., 2021). To address these limitations, Auto ARIMA has gained traction for its ability to automate the selection of optimal model parameters using unit root tests and information criteria such as the corrected Akaike Information Criterion (AICc) and the Bayesian Information Criterion (BIC), which evaluate model quality by balancing goodness of fit against model complexity (Hyndman & Khandakar, 2008). This reduces configuration overhead while improving forecast precision, particularly in environments where retraining must be rapid and frequent. For example, Liu et al. (2022) found Auto ARIMA outperformed manually tuned SARIMA in both accuracy and responsiveness in forecasting DDoS surges. Likewise, Yin et al. (2017) highlighted Auto ARIMA's superior robustness when capturing seasonal trends in phishing campaigns around tax season.

Complementing these approaches is Prophet, a decomposable time series model developed by Facebook that is particularly suited for capturing multiple seasonality patterns with irregular intervals (Taylor & Letham, 2018). Prophet handles missing data, outliers, and changepoints more flexibly than classical statistical models, making it well-suited for volatile cyber threat landscapes. Its additive modelling structure accommodates trend shifts and holidays explicitly, making it highly interpretable and effective in contexts like forecasting phishing and fraud attempts during national events or promotional periods. Gupta et al. (2022) demonstrated. Prophet’s utility in predicting transaction anomalies during e-commerce flash sales, noting its advantage in incorporating external regressors such as campaign dates or user activity levels. Hybrid models that integrate time series forecasting with machine learning techniques are also gaining prominence. For instance, Kim and Park (2022) proposed a SARIMA–Random Forest hybrid pipeline in which Auto ARIMA residuals were analysed using supervised classifiers to detect stealthy deviations, enhancing anomaly sensitivity without sacrificing interpretability. These architectures reflect an ongoing shift toward resilient forecasting frameworks that combine statistical rigour with adaptive learning. Although these methods improve seasonal forecasting, challenges remain. Auto ARIMA and Prophet still assume certain regularities in temporal dynamics and may struggle with regime shifts induced by novel attacker tactics or contextual shifts, such as news-driven campaigns. Prophet’s effectiveness also depends on the availability of well-defined changepoints and external features. Moreover, both models may become computationally intensive on high-dimensional or high-frequency data streams (Yin et al., 2017). Deep learning alternatives such as LSTM (Long Short-Term Memory) and GRU (Recurrent Neural Networks) offer solutions by capturing long-range dependencies and nonlinearities without stationarity assumptions (Malhotra et al., 2016), but their opaque nature limits transparency in regulated domains (Doshi-Velez & Kim, 2017).

The COVID-19 pandemic further underscored the need for flexible, real-time forecasting tools. Between 2020 and 2021, researchers observed significant deviations in cyberattack patterns, driven by remote work, digital acceleration, and opportunistic phishing, which rendered seasonal assumptions alone insufficient (Ghahramani et al., 2021). These events accelerated the adoption of hybrid and context-aware forecasting systems capable of real-time anomaly detection and adaptive retraining. In summary, seasonal forecasting is integral to anticipating cyberattack surges within e-commerce ecosystems. While ARIMA, Auto ARIMA, and Prophet offer robust and interpretable forecasting capabilities, their optimal value is realised when embedded within hybrid pipelines that incorporate external context, anomaly detection, and machine learning components. As cyber threats continue to evolve in complexity and frequency, forecasting strategies must integrate statistical depth, operational adaptability, and domain knowledge to remain effective.

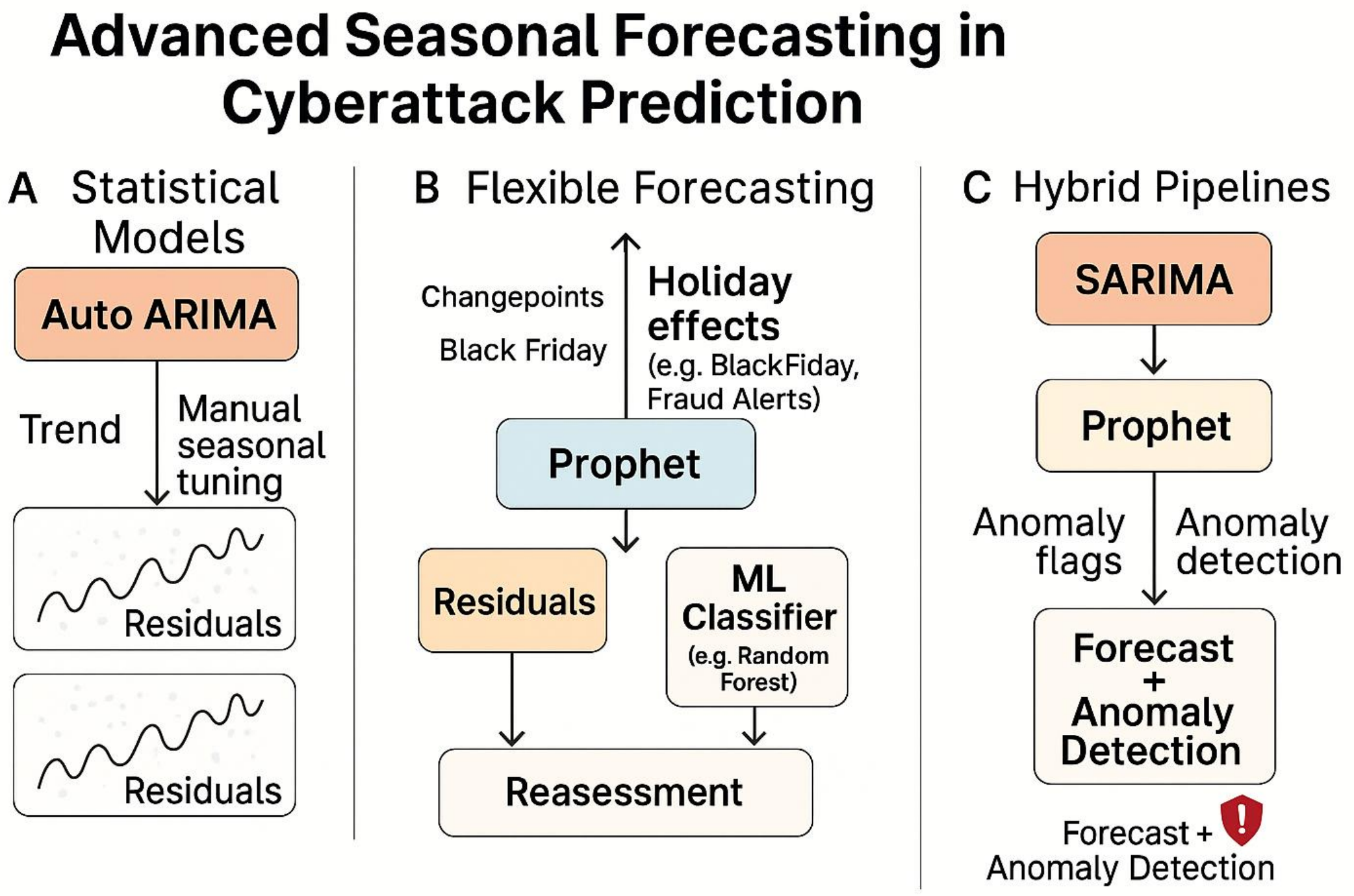

*Figure 7 Comparison of (A) Auto ARIMA, (B) Prophet with ML classifiers, and (C) SARIMA–Prophet hybrids for detecting seasonal and behavioural attack patterns. Adapted from Hyndman and Athanasopoulos (2021).*

## 2.7 Hypothesis Testing for Temporal Impact Analysis

Hypothesis testing methods are foundational in statistically validating the influence of temporal variables such as holidays, promotional campaigns, or policy changes on cyberattack frequency and severity. By establishing statistical significance, these tests help e-commerce platforms prioritise defensive resources during high-risk periods (Nguyen et al., 2023). Nguyen et al. (2023) employed t-tests and ANOVA to demonstrate significant increases in phishing and credential stuffing incidents during major retail events, supporting proactive security postures. The Analysis of Variance (ANOVA) test is particularly effective in determining whether significant differences exist between group means across time windows (e.g., holiday vs non-holiday periods), assuming normally distributed data (Field, 2013). When this assumption is violated, non-parametric alternatives like the Mann–Whitney U test allow robust median comparisons between independent groups (Nachar, 2008). These tests have been applied in cyber risk studies to assess whether transaction anomalies, login attempts, or malware propagation differ significantly across seasonal intervals. For example, Shafiq et al. (2018) used

ANOVA to analyse transaction deviations across quarterly business cycles, while Kumar et al. (2021) applied the Mann–Whitney U test to detect significant spikes in phishing attempts during national holidays.

Conversely, Romanosky (2016) cautions that temporal effects are often confounded by multiple covariates, finding inconsistent seasonal attack patterns across datasets via chi-square tests, underscoring the complexity of attributing causality. Further, Zhao et al. (2022) integrate hypothesis testing with machine learning frameworks to validate anomalies detected in real time, enhancing detection reliability. Yet, the inherent assumptions of independence and homogeneity in classical tests often clash with correlated and heteroskedastic cyber data (Fenz et al., 2014), necessitating more robust statistical frameworks such as generalised linear models (GLMs) or time-varying coefficient models (TVCMs) for nuanced temporal analysis (Zeger & Liang, 1986). While hypothesis testing remains a critical tool for *exploration* data analysis and quantifying initial impacts, many scholars acknowledge that it is insufficient by itself for forecasting or modelling complex temporal dynamics in cyber threats.

## 2.8 Anomaly Detection via Statistical Thresholds

Statistical anomaly detection methods, including Z-score thresholding and time-window analyses, provide interpretable and computationally efficient tools for recognising unusual cyber events relative to historical baselines (Sadreazami et al., 2020). These techniques are particularly prevalent in e-commerce monitoring systems, where they enable immediate alerts and proactive risk management. For instance, Sadreazami et al. (2020) demonstrated effective anomaly detection in e-commerce transaction streams using adaptive Z-score thresholds, successfully pinpointing sudden spikes in fraudulent activities. Similarly, Akbanov et al. (2019) recommended dynamic threshold adjustment to accommodate fluctuating baselines, thereby reducing false positive rates a crucial consideration given the high variability in e-commerce traffic patterns. To further address challenges related to class imbalance in anomaly datasets, Chawla et al. (2002) developed the Synthetic Minority Oversampling Technique (SMOTE), which has since been incorporated into statistical anomaly detection frameworks to improve sensitivity without compromising specificity.

Despite their advantages, statistical anomaly detection methods often lack the granularity needed to categorise different types of attacks or autonomously adapt to evolving threat strategies, necessitating their integration with machine learning classifiers for comprehensive cyber defence (Sarker, 2022). Additionally, statistical forecasting models like ARIMA effectively identify seasonal trends but require augmentation with adaptive methods to manage irregularities in data. Hypothesis testing offers systematic means to evaluate temporal attack impacts but must be refined to account for complex interdependencies within e-commerce datasets. In conclusion, while statistical methods continue to provide vital, interpretable insights into the temporal dynamics of cyberattacks in e-commerce, the literature underscores the ongoing need for hybrid approaches that combine these traditional techniques with machine learning to meet the sophisticated demands of contemporary cybersecurity environments.

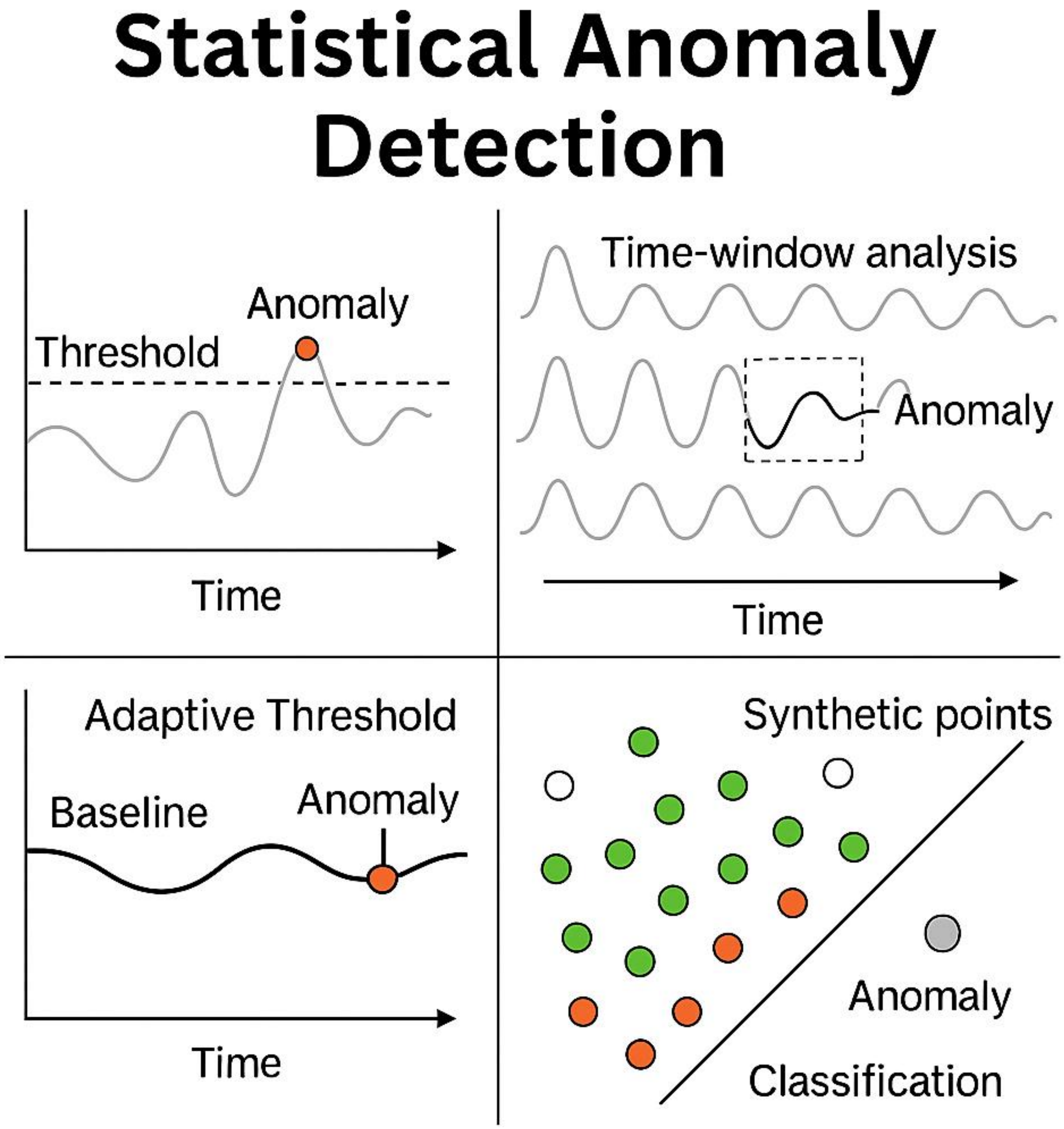

*Figure 8 Anomaly Detection Methods — Z-score, time-windowing, adaptive thresholds, and SMOTE for better anomaly classification (author-created).*

## 2.9 Machine Learning Techniques for Cyberattack Detection in E-Commerce

Machine learning is vital for detecting cyberattacks in e-commerce, with supervised models like Random Forest and XGBoost commonly used to flag fraud based on historical patterns. However, they rely on labelled data and perform poorly against unknown threats. Unsupervised methods, such as clustering and anomaly detection, help identify novel attacks but often generate false positives. Deep learning models like neural networks and LSTMs can capture complex patterns, though they are less interpretable and more resource intensive. To address this, hybrid approaches combining accuracy with explainability using tools like SHAP and LIME are increasingly explored.

### 2.9.1 Unsupervised Learning Approaches

Unsupervised learning algorithms have gained traction in the cybersecurity domain of e-commerce due to their ability to detect novel and unforeseen threats without relying on labelled training data (Chawla et al., 2002). These methods are particularly valuable in identifying zero-day attacks or behavioural anomalies that do not conform to historical patterns, a crucial complement to supervised systems which may fail to generalise to unknown risks. Clustering techniques such as K-Means and DBSCAN are often applied to group transactional records into similar behavioural clusters, flagging outliers as potential fraud instances (Chandola, Banerjee & Kumar, 2009). While effective in low-dimensional contexts, these models exhibit reduced reliability when operating on complex, high-dimensional e-commerce datasets, which can lead to inconsistent cluster boundaries and sensitivity to hyperparameters (Aggarwal & Yu, 2001). Moreover, DBSCAN's density-based assumptions are often violated in dynamic retail environments where customer behaviour shifts seasonally.

Anomaly detection algorithms, including Isolation Forests (Liu, Ting & Zhou, 2008) and Local Outlier Factor (LOF) (Breunig et al., 2000), have shown promise in identifying a typical transaction patterns without prior labelling. These models are particularly useful for detecting rare fraud cases and structural shifts in data distribution. However, high false positive rates remain a major operational challenge, leading to alert fatigue and potential erosion of stakeholder trust (Ahmed, Mahmood & Hu, 2016). Some researchers advocate hybrid solutions that combine unsupervised anomaly scoring with supervised post-classification to reduce false alarms and improve detection accuracy (Fiore et al., 2019).

Despite their adaptability, unsupervised models often lack transparency, which raises concerns under regulatory frameworks like GDPR and PCIDSS that require explainable decision-making. As such, integrating explainable AI (XAI) mechanisms into these systems is increasingly seen as essential for trustworthy deployment in commercial environments. Unsupervised methods offer critical value for detecting emergent threats in e-commerce, especially where labelled data is scarce. However, challenges related to scalability, interpretability, and false positives must be addressed through hybridisation and domain-specific tuning.

### 2.9.2 Deep Learning Techniques

Deep learning (DL) offers powerful capabilities for cyberattack detection in e-commerce, especially in learning patterns directly from raw behavioural data. Unlike traditional models requiring manual feature engineering, DL architectures autonomously extract multi-level abstractions, enabling better detection of sophisticated or evolving threats (LeCun, Bengio & Hinton, 2015). Among DL models, Convolutional Neural Networks (CNNs) have been successfully adapted for security tasks by structuring transactional data into matrices. Initial layers extract localised behaviours, such as repeated access attempts or irregular browsing patterns, using small filters (e.g., 5×5), while subsequent convolutional layers (e.g., 3×3) refine abstract threat features across channels. These outputs are compressed via 1×1 convolutions and passed into fully connected layers to support final decision-making. This layered structure enables the model to transition from low-level anomaly detection to holistic threat classification.

Recurrent Neural Networks (RNNs) particularly LSTM networks excel in modelling sequential data like login histories or session flows. Their ability to capture long-term dependencies makes them effective in detecting fraud and insider threats that unfold over time (Malhotra et al., 2016). LSTMs often outperform static classifiers in time-based anomaly detection tasks.

Autoencoders, trained to reconstruct input data, are commonly used for unsupervised anomaly detection. High reconstruction error flags deviations from learned "normal" patterns, helping identify subtle threats. Variants like VAEs improve generalisation but still face issues with false positives and limited interpretability (Zhou & Paffenroth, 2017).

Despite their power, DL models demand large, labelled datasets, are resource-intensive, and often function as black-box systems, posing challenges for explainability and compliance (Doshi-Velez & Kim, 2017). Hybrid approaches that integrate DL with interpretable models have been proposed to improve auditability (Fiore et al., 2019). Advanced methods like attention mechanisms and transformers further boost real-time detection but increase complexity (Zhang et al., 2022). Given these concerns, and the structured tabular nature of the data, this study did not adopt deep learning models. Instead, tree-based ensembles (e.g., XGBoost, CatBoost) were chosen for their optimal balance between accuracy and interpretability in e-commerce threat detection.

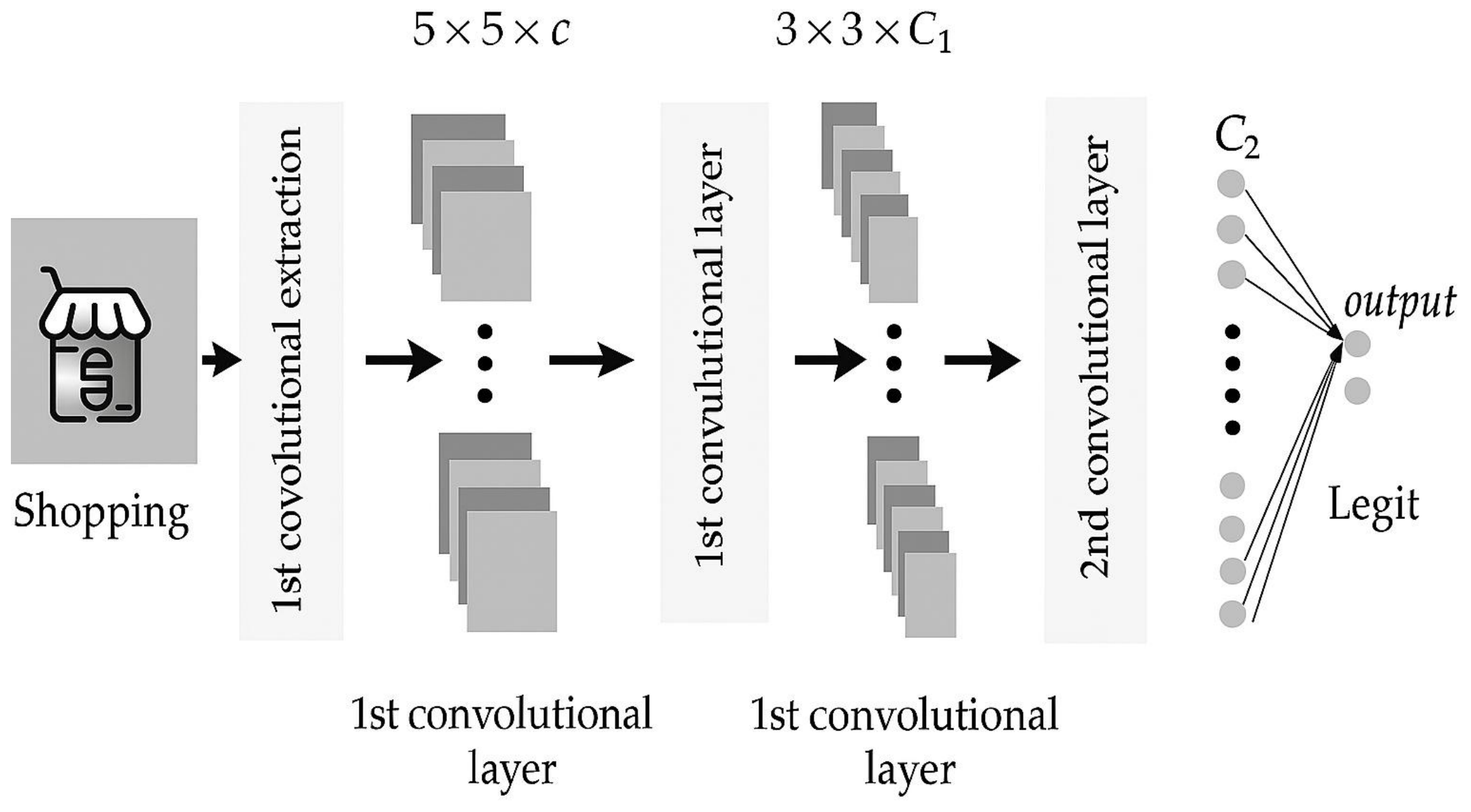

*Figure 9 Two conv CNN-based architecture for classifying e-commerce activity as legitimate. Adapted from Makantasis et al. (2015).*

## 2.10 Robust and Explainable Models for E-Commerce Threat Detection

Ensemble models are widely used in e-commerce cybersecurity for their ability to reduce variance and bias by combining multiple learners, resulting in greater accuracy and robustness (Dietterich, 2000). This section evaluates three leading ensemble methods, XGBoost, CatBoost, and LightGBM, selected for their strong performance on structured cyberattack data. For comparison, logistic regression is also included to illustrate the limitations of simpler linear approaches.

1. **Extreme Gradient Boosting (XGBoost):** This is an advanced gradient boosting framework designed for efficiency, scalability, and accuracy. It builds additive decision trees sequentially using a second-order Taylor approximation of the loss function, capturing subtle feature interactions and correcting previous errors (Chen & Guestrin, 2016). Features like L1/L2 regularisation, sparse data handling, and parallelised learning make it well-suited for high-dimensional cybersecurity datasets. In e-commerce cybersecurity, XGBoost has been widely applied to fraud and intrusion detection. For example, Jain and Rathore (2022) achieved higher precision and recall than neural networks when classifying fraudulent credit card transactions, while Al-Mnayyis et al. (2021) successfully detected phishing domains in retail platforms, effectively managing large, imbalanced datasets. Sharma and Saini (2023) incorporated XGBoost into a hybrid framework to identify abnormal API access behaviours, improving accuracy by 17% over Support Vector Machines. These successes are partly due to XGBoost's ability to handle missing values and regularised learning, enhancing stability with noisy inputs. However, XGBoost's computational cost rises with deeper trees and larger datasets, challenging real-time scalability without distributed systems (Kumar et al., 2020). It also requires careful categorical feature encoding to avoid performance drops, as noted by Zhang et al. (2020). Lastly, its predictions are often opaque without explainability tools like SHAP (Lundberg & Lee, 2017), complicating transparency in regulatory contexts such as GDPR.

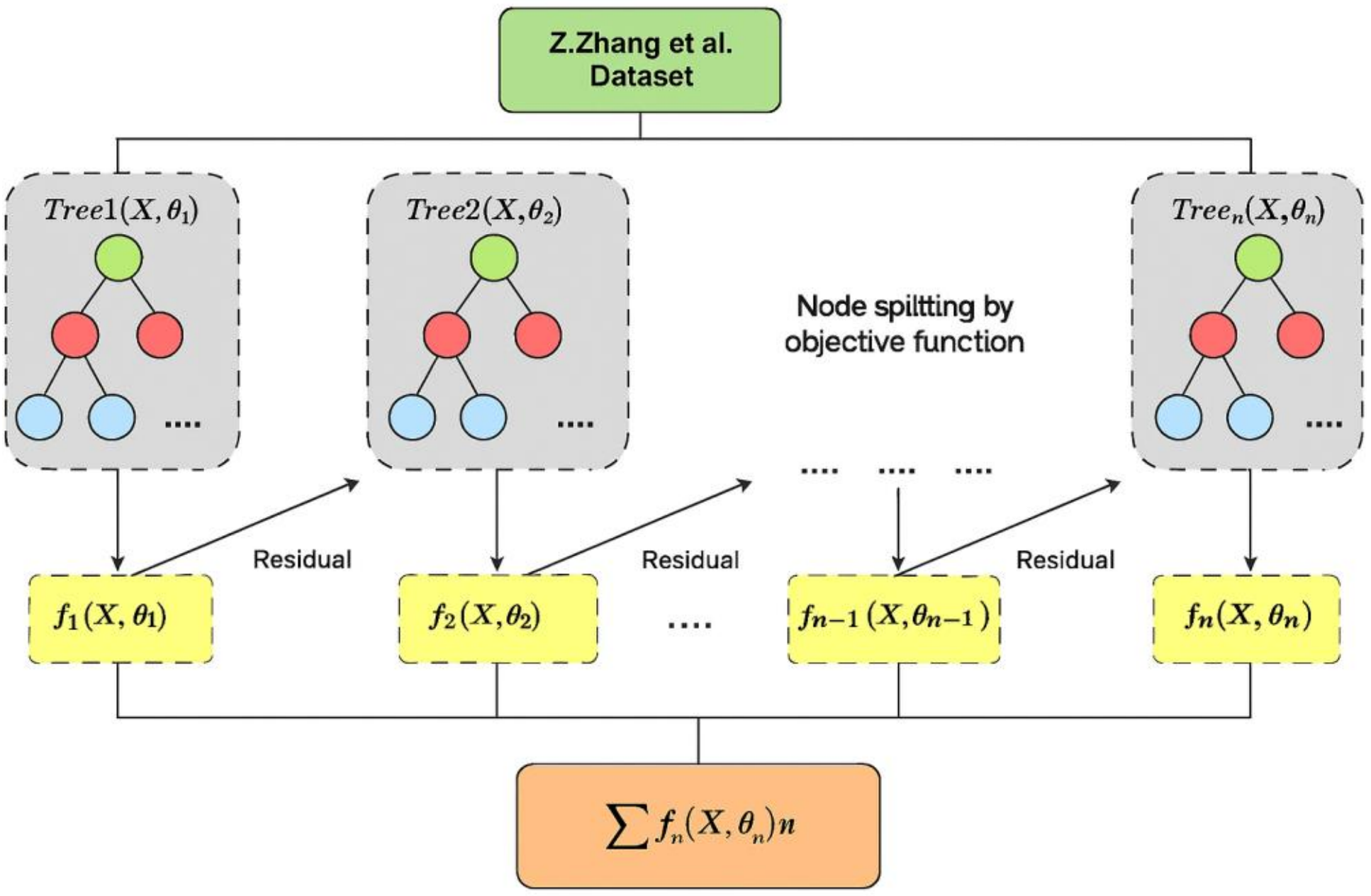

*Figure 10: CatBoost uses bootstrapped samples and symmetric trees, averaging predictions for robust classification of noisy categorical data. (Zhang et al,2016)*

2. **CatBoost**: This is a gradient boosting framework that natively supports categorical features using ordered boosting and symmetric (oblivious) decision trees, which apply the same split across all tree levels for faster computation and easier interpretability (Prokhorenkova et al., 2018). This approach reduces variance and improves generalisation, especially in e-commerce contexts with high-cardinality categorical data like browser types or transaction sources. CatBoost builds multiple sequential trees, focusing on misclassified samples through weight expansion, and combines predictions via weighted averaging. In cybersecurity for e-commerce, CatBoost has excelled at handling categorical data without manual encoding. For example, Alshamrani et al. (2021) achieved an F1-score of 0.91 detecting phishing URLs, outperforming traditional classifiers and neural networks. Wang et al. (2022) also used CatBoost to identify malicious product listings using encoded metadata such as seller trust levels. However, in datasets with very sparse categorical classes, such as user-agent strings, CatBoost may require grouping strategies to maintain performance (Banirostam et al., 2022). CatBoost offers advantages including native categorical handling, reduced need for hyperparameter tuning, faster training on large categorical datasets, and better generalisation under class imbalance. Limitations include a lack of model transparency without interpretability tools like SHAP or LIME, potential latency in online inference due to ordered boosting, and challenges with sparse or high-cardinality features that may need manual grouping to improve prediction reliability (Lundberg & Lee, 2017).

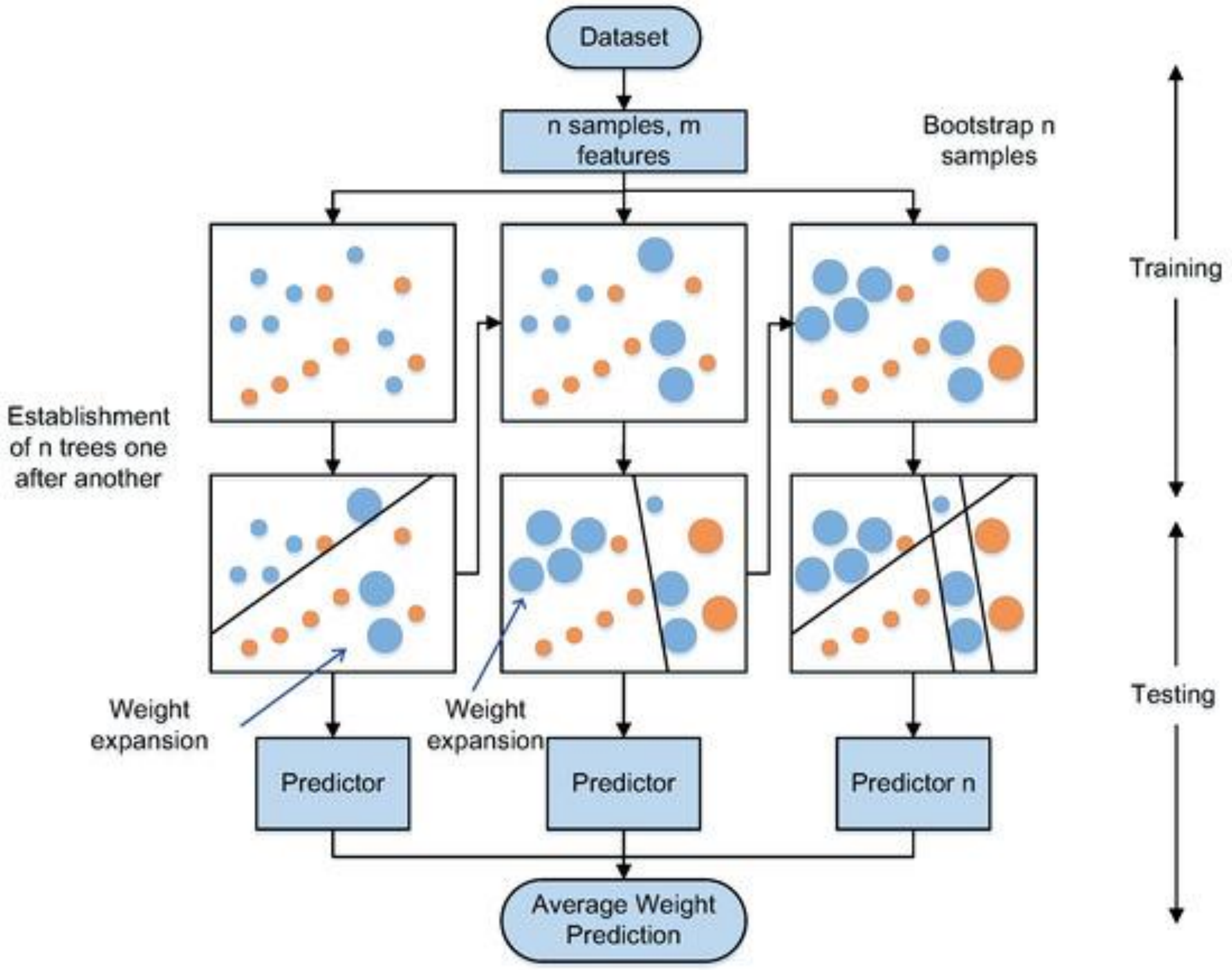

*Figure 11 Boosting trains models on weighted bootstrap samples and combines predictions for a strong ensemble (Huajian et al, 2023).*

3. **LightGBM:** This is a highly efficient gradient boosting framework optimised for speed and scalability using histogram-based algorithms and a leaf-wise growth strategy (Ke et al., 2017). It begins with data preprocessing and feature selection, applying Gradient-based One-Side Sampling (GOSS) to focus on high-gradient instances and Exclusive Feature Bundling (EFB) to reduce dimensionality by merging mutually exclusive features. LightGBM builds multiple CART regression trees sequentially, each learning from residuals, with trees grown leaf-wise by splitting the leaf that offers the largest loss reduction, enabling faster convergence and higher accuracy on large, sparse cybersecurity datasets. In e-commerce cybersecurity, LightGBM has demonstrated strong performance. Gao et al. (2022) achieved 96.2% accuracy detecting fraud in 2.1 million Amazon transactions with under 30 ms latency, outperforming Random Forest and SVM. Li et al. (2021) successfully used LightGBM for real-time bot detection, efficiently handling thousands of categorical request headers. However, Khan et al. (2021) noted it may overfit minority classes in highly imbalanced intrusion datasets without balancing techniques like SMOTE.

LightGBM's advantages include fast training via histogram binning, scalability to millions of samples and thousands of features, and effective handling of sparse categorical data through EFB and GOSS. Limitations include potential overfitting on small or imbalanced data, instability from unbalanced leaf-wise trees, quantisation errors from histogram binning, and the need for careful hyperparameter tuning to avoid performance issues.

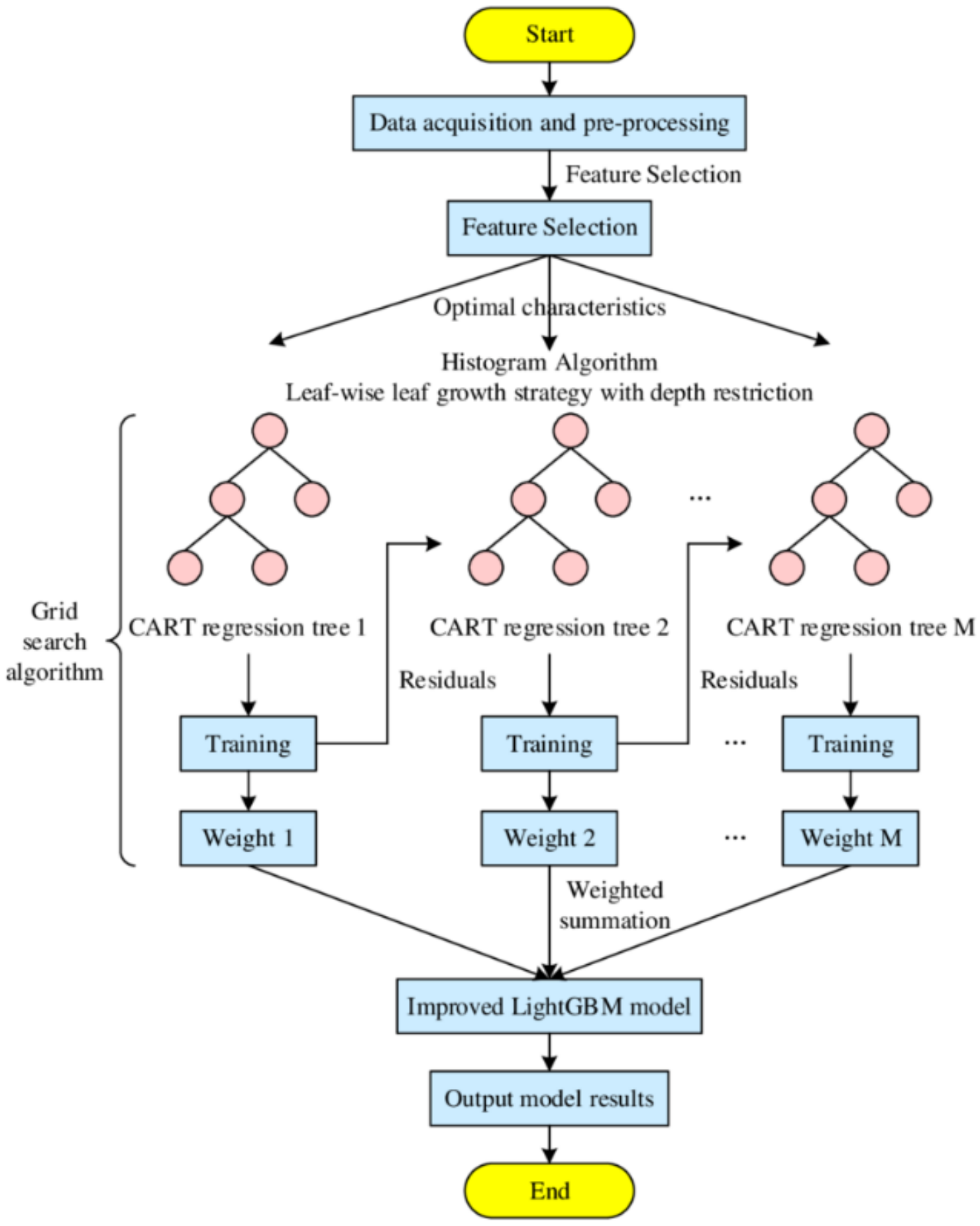

*Figure 12 LightGBM flowchart showing leaf-wise tree growth and residual learning for fast, accurate predictions (Yan Pan, 2023).*

4. **Logistic Regression (LR):** This is a linear classification model that predicts the probability of a binary outcome using a logistic (sigmoid) function. It computes a weighted sum of input features, such as transaction metadata or user behaviour flags, and maps this to a probability score between 0 and 1, which is then thresholded to assign class labels (Hosmer et al., 2013). Figure 13 visualises this process, where features are weighted, aggregated, and passed through an activation function to generate discrete outputs. In cybersecurity, LR is frequently used as a benchmark model due to its speed, simplicity, and transparency. It has been applied in spam detection, bot filtering, and fraud prediction, particularly in well-balanced, structured datasets. For instance, Kumar et al. (2021) used LR to detect fraudulent e-commerce transactions, while Patel and Rathod (2022) reported its effective use in real-time bot activity scoring on retail platforms.

Despite its advantages, LR assumes linearity and independence among features, assumptions that often fail in real-world cyber contexts where threat behaviours are complex and interdependent (Zhang et al., 2020). It lacks the flexibility to detect low-frequency exploits or nonlinear attack patterns without extensive feature engineering. In high-dimensional or imbalanced datasets, LR can be overfit unless regularised. Moreover, its simple architecture limits its ability to capture deep feature interactions, making it unsuitable as a standalone tool in dynamic, adversarial environments. Consequently, LR is best used for interpretability or as a baseline, often in combination with more expressive ensemble or hybrid models.

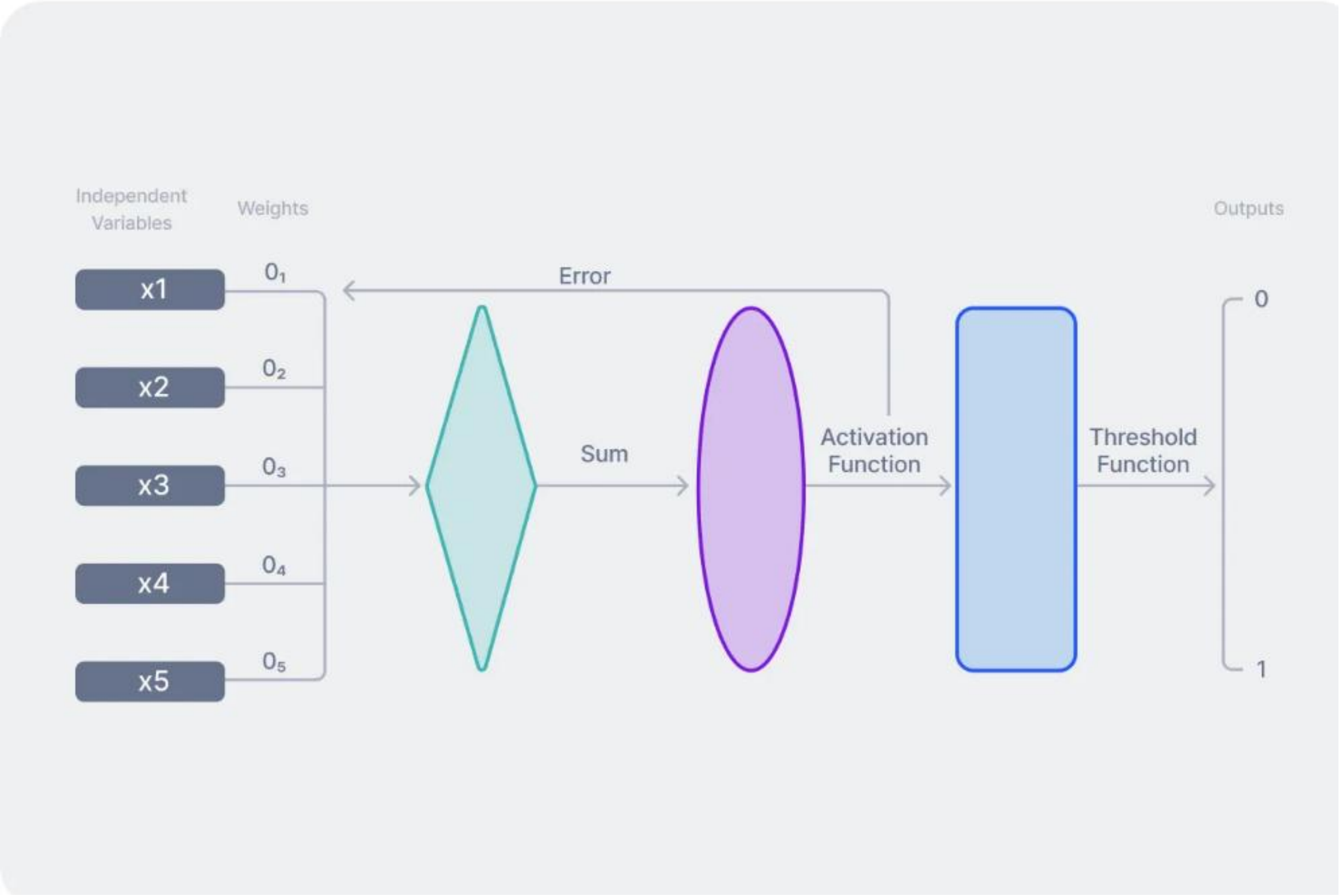

*Figure 13 Overview of GBDT tree growth, structure, and ensemble prediction process (V7 labs,2023).*

## 2.11 Economic Impact of Cyber Threats on E-Commerce

Cybersecurity datasets in e-commerce, particularly those involving personal and financial information, raise significant privacy concerns due to the centralised data aggregation practices inherent in traditional statistical and machine learning analysis (Kshetri, 2021). Centralisation increases the risk of unauthorised data access, posing substantial threats to both individuals and organisations. As e-commerce platforms increasingly adopt machine learning (ML) techniques, these privacy issues become more acute, given that model training frequently requires extensive data sharing and integration from disparate sources (Zhang et al., 2022). This dependence on large-scale data not only heightens the risk of data breaches but also necessitates the implementation of robust privacy-preserving mechanisms to protect sensitive user information.

To address these risks, privacy-preserving approaches such as federated learning, differential privacy, and homomorphic encryption have emerged as viable solutions. Federated learning enables decentralised model training, avoiding the need for central data pooling and thereby reducing the risk of sensitive data leakage (Yang et al., 2019). Differential privacy introduces controlled noise into datasets, ensuring individual anonymity while preserving analytical value (Dwork, 2008). Homomorphic encryption allows computations to be carried out on encrypted data, which is especially advantageous for secure, cloud-based e-commerce systems (Gentry, 2009). However, these methods often introduce considerable computational overheads, presenting challenges for real-time application in latency-sensitive, high-volume e-commerce environments (Shokri & Shmatikov, 2015).

Beyond technical safeguards, ethical considerations in e-commerce cybersecurity include fairness, mitigation of algorithmic bias, and transparency in decision-making. As ML-driven systems are increasingly employed for fraud detection, concerns arise over the potential reinforcement of biases embedded within training datasets, potentially resulting in unfair treatment of specific demographic groups (O'Neil, 2016). The opacity of many advanced machine learning models, particularly deep learning systems, further exacerbates issues of interpretability and accountability in high-stakes domains (Lipton, 2018). A review of current literature highlights notable gaps in addressing privacy, ethical, and fairness concerns within e-commerce. Although advancements in differential privacy and federated learning are promising, their integration into real-time, ethically governed fraud detection systems remains limited. There is a pressing need for interdisciplinary research that combines innovations in data engineering, ethical AI, and regulatory frameworks to ensure responsible, transparent, and privacy-conscious deployment particularly in alignment with the GDPR and related standards (European Commission, 2016).

*Table 2 Economic Impact of Cyber Threats in Ecommerce (2023-2025)*

| Period | Type of Attack | Description | Notable Cases/Examples |
|---|---|---|---|
| 1980s | Early Network Exploits | Initial attacks focused on exploiting network protocols and unpatched systems. | Marcus Hess military system breach (1986) |
| 1990s | Viruses and Worms | Self-replicating malware like Melissa and ILOVEYOU caused widespread damage. | ILOVEYOU virus (2000) |
| 2000s | Phishing and Identity Theft | Growth of social engineering attacks to steal credentials and financial data. | Phishing campaigns targeting banks and e-commerce sites |
| 2010s | Botnets and DDoS Attacks | Use of compromised devices (including IoT) to launch massive DDoS attacks. | Mirai botnet causing major outages (2016) |
| 2010s | Ransomware | Malware encrypting data and demanding ransom, targeting businesses globally. | WannaCry outbreak (2017) |
| 2020s | Supply Chain Attacks | Attacks compromising third-party software to infiltrate target systems. | SolarWinds breach (2020) |
| 2020s | AI-Enhanced Attacks | Use of AI for sophisticated phishing, evasion, and automated exploitation. | Deepfake phishing, automated vulnerability scanning |

## 2.12 Research Gaps

Recent years have seen substantial progress in applying statistical and machine learning (ML) techniques to detect cyberattacks in e-commerce, particularly in fraud detection and intrusion prevention (Kshetri, 2021). While these approaches effectively identify anomalous patterns indicative of cyber threats, persistent challenges undermine their scalability, robustness, and real-world applicability. The inherent complexity and data heterogeneity of e-commerce environments exacerbate these issues. This section systematically analyses key research gaps from both statistical and ML standpoints, highlighting priority areas for future academic exploration to advance cyberattack detection capabilities

### 2.12.1 Data Scarcity and Imbalance (Statistical & ML)

A significant challenge facing both statistical and machine learning methodologies is the lack of extensive, high-quality, and precisely annotated datasets tailored to the intricacies of e-commerce cyber risks (Nguyen et al., 2023). From a statistical perspective, small sample sizes hinder parameter estimation, diminish the efficacy of inferential tests, and complicate the detection of infrequent yet significant assault events. This is especially concerning as cybersecurity datasets frequently display heavy-tailed, non-normal distributions, which invalidate numerous conventional model assumptions (Chawla et al., 2002). As a result, conventional inferential techniques may produce unreliable or skewed results when utilised on intricate e-commerce security data. From a machine learning perspective, significant class imbalance, characterised by a predominance of benign transactions over dangerous occurrences, engenders classifier bias, hence elevating the probability of false negatives and undetected threats (Abawajy et al., 2021). Although oversampling techniques like SMOTE and generative methods such as Generative Adversarial Networks (GANs) demonstrate promise in addressing imbalance issues, they necessitate meticulous adaptation to maintain the unique transaction and attack characteristics of e-commerce (Goodfellow et al., 2014). Transfer learning techniques have surfaced as viable methods to utilise information from analogous cybersecurity domains;

nevertheless, empirical confirmation of their efficacy in cross-domain e-commerce contexts is still inadequate (Nguyen et al., 2023).

### 2.12.2 Data Quality and Feature Engineering (Statistical & ML)

The quality and relevance of input features are crucial in establishing the efficacy of detection models. E-commerce datasets are generally high-dimensional and exhibit significant multicollinearity, increasing the likelihood of overfitting and inflating variance in models like logistic regression and discriminant analysis (Fenz et al., 2014). Feature selection techniques, such chi-square tests and mutual information scores, facilitate dimensionality reduction; but their linear assumptions constrain their capacity to identify the complex, non-linear connections inherent in cyberattack signatures (Kshetri, 2021). Machine learning techniques, especially deep learning, mitigate certain restrictions by facilitating automatic hierarchical feature extraction from raw transaction logs and user behavioural data (Sarker, 2022). Nonetheless, obstacles remain owing to noisy, missing, or damaged data, which can significantly impair model performance and generalisability (Gupta & Verma, 2019). A strong research necessity exists to create robust, domain-specific preprocessing pipelines and imputation methods designed for the operational data anomalies inherent to e-commerce platforms.

### 2.12.3 Concept Drift and Adversarial Attacks (Statistical & ML)

Cyber adversaries continuously modify their tactics, resulting in concept drift where the statistical properties of attack data evolve over time posing significant challenges for static detection models (Yu et al., 2021). Traditional statistical approaches such as Kalman filtering or change-point detection provide foundational mechanisms to accommodate non-stationarity; however, their scalability and adaptability to the voluminous, heterogeneous, and temporally dynamic nature of e-commerce data are limited (Romanosky, 2016). Machine learning frameworks have begun incorporating adaptive learning paradigms such as online learning, incremental updates, and ensemble adjustments to sustain detection efficacy in the face of evolving attack characteristics (Sarker, 2021). Nevertheless, adversarial attacks carefully crafted inputs designed to circumvent detection constitute a formidable threat to ML robustness. Although adversarial training and anomaly detection have demonstrated effectiveness in other domains, their tailored application to the multifaceted attack surface of e-commerce remains underdeveloped (Goodfellow et al., 2014). Thus, integrating concept drift management with adversarial resilience represents an urgent and underexplored research frontier.

### 2.12.4 Explainability and Interpretability (Statistical & ML)

Compliance with regulatory mandates such as the General Data Protection Regulation (GDPR) and the Payment Card Industry Data Security Standard (PCIDSS) necessitates transparency and accountability in automated decision-making systems, underscoring the critical importance of model explainability (Voigt & Von dem Bussche, 2017). Classical statistical models, including logistic regression, inherently facilitate interpretability by allowing direct examination of parameter coefficients, thereby enabling straightforward risk factor identification and supporting audit requirements. Conversely, advanced machine learning models such as random forests, gradient boosting machines, and deep neural networks tend to function as opaque

"black boxes," limiting the ability of stakeholders to understand or justify predictions (Doshi-Velez & Kim, 2017). Emerging explainable AI (XAI) methods, including SHAP (SHapley Additive exPlanations) and LIME (Local Interpretable Model-agnostic Explanations), offer promising approaches to mitigate this opacity by attributing model outputs to specific features (Ribeiro et al., 2016). However, these techniques require further refinement and contextual adaptation to effectively address the sequential and transactional complexities inherent in e-commerce datasets and to provide actionable insights for security analysts.

### 2.12.5 Privacy and Ethical Considerations (Statistical & ML)

Cybersecurity datasets in e-commerce, particularly those involving personal and financial information, raise critical privacy concerns due to the centralised data aggregation practices common in traditional statistical and machine learning (ML) approaches (Nguyen Truong et al., 2020). Centralisation heightens the risk of unauthorised access, making robust privacy measures essential especially as ML adoption grows and requires large-scale, cross-source data integration (Zhang et al., 2022).

To address these risks, privacy-preserving techniques such as federated learning, differential privacy, and homomorphic encryption have gained traction. Federated learning enables decentralised model training without centralised data pooling, thereby reducing leakage risk (Yang et al., 2019). Differential privacy injects noise into datasets to maintain anonymity while preserving utility (Dwork, 2008), and homomorphic encryption allows computations on encrypted data ideal for secure cloud-based e-commerce systems (Gentry, 2009). Despite their benefits, these methods often introduce significant computational overhead, challenging their use in latency-sensitive, real-time fraud detection environments (Shokri & Shmatikov, 2015).

Beyond technical safeguards, ethical concerns in ML-based fraud detection include fairness, bias mitigation, and model transparency. Automated systems risk reinforcing biases embedded in training data, potentially leading to discriminatory outcomes (O'Neil, 2016). The opacity of complex models, especially deep learning, further complicates accountability and interpretability in high-stakes contexts (Lipton, 2018). While privacy-enhancing technologies have advanced, the integration of ethical frameworks into real-time fraud detection remains limited. Addressing this gap requires interdisciplinary collaboration across data science, regulatory policy, and ethical AI design to ensure privacy, fairness, and transparency in line with standards like GDPR (European Commission, 2016).

*Table 3 Research Gaps and Proposed Solution*

| Research Gaps from Literature | Proposed Solutions |
|---|---|
| Imbalanced and limited labelled datasets compromise detection accuracy and model robustness, particularly in minority classes. | Apply SMOTE oversampling to address class imbalance, enhancing model sensitivity and reducing bias towards majority classes. |
| Concept drift and the dynamic evolution of cyberattack techniques undermine the effectiveness of static detection models. | Employ adaptive ensemble learning models with scheduled retraining to maintain detection accuracy over time. |
| The black-box nature of many machine learning models limits interpretability, reducing trust and actionable insights for cybersecurity practitioners. | Integrate SHAP explainability methods to provide transparent feature impact analysis and improve model interpretability. |
| Current detection systems often lack scalable real-time processing and timely threat response capabilities, limiting proactive cybersecurity measures. | Combine statistical time-series forecasting with machine learning to anticipate seasonal attack surges and enable proactive responses. |
| Privacy regulations and ethical considerations restrict the availability and use of detailed datasets, constraining model development. | Limit data usage to publicly accessible breach datasets and adhere to GDPR-compliant data handling protocols to ensure privacy and compliance. |
| Models trained on single datasets or platforms often demonstrate limited generalizability due to heterogeneity in e-commerce environments. | Implement rigorous cross-validation and incorporate domain-specific feature engineering to improve model robustness and generalizability across diverse data sources. |
| The potential of hybrid approaches that combine statistical methods with machine learning to optimize interpretability and predictive performance remains underutilised. | Develop and evaluate a hybrid analytical framework that integrates statistical forecasting techniques with ensemble machine learning classifiers to balance predictive accuracy and interpretability. |

# 3.0 METHODOLOGY

This chapter outlines the methodology for detecting cyberattack patterns targeting e-commerce platforms. The study aims to develop and evaluate machine learning and statistical models for identifying emerging threats using a Design Science Research (DSR) approach, which supports iterative model development tailored to cybersecurity complexities (Peffers et al., 2007). The research defines the problem, sets objectives, and designs classification and forecasting models trained on the Verizon Community Data Breach dataset (Verizon, 2023). Data preprocessing addressed missing values, feature engineering, and class imbalance to improve model accuracy and robustness. Models were evaluated using metrics such as accuracy, precision, recall, F1-score, ROC-AUC, RMSE, and MAE (Japkowicz & Shah, 2011). The study acknowledges dataset biases, limitations in scope and representativeness, and algorithmic constraints to ensure transparency. Implementation used Python and standard machine learning libraries for reproducibility.

## 3.1 Research Design

This study employs a quantitative research design within the framework of Design Science Research (DSR). DSR is appropriate for this work as it supports the systematic development and evaluation of artefacts here, predictive models for cyberattack detection (Peffers et al., 2007). The goal is to build data-driven solutions that enhance cybersecurity decision-making, particularly for e-commerce platforms. The primary dataset used is the Verizon Community Data Breach (VCDB) project (Verizon, 2023), a widely recognised repository of real-world cyber incidents. It includes detailed records from 1984 to 2023 across various sectors and attributes such as attack method, threat actor, compromised assets, and breach severity. These structured variables enable classification modelling and time-series forecasting using machine learning and statistical techniques.

The research process involved designing models aligned with established cybersecurity frameworks, such as the A4 threat model and the MITRE ATT&CK matrix (MITRE, 2022), to ensure domain relevance and interpretability. Evaluation metrics, including accuracy, F1-score, and mean absolute error, were used to assess model performance. This design aligns with DSR's emphasis on producing practical, testable solutions to real-world problems.

*Table 4 Mapping of DSR steps to research activities in this study.*

| DSR Step | Activity in This Study |
| --- | --- |
| Problem Identification & Motivation | Identify rise of complex cyber threats in e-commerce; establish need for predictive, explainable ML approaches. |
| Define Objectives of a Solution | Develop a hybrid ML pipeline to classify breach severity and forecast risk trends using enriched cyber incident data. |
| Design and Development | Construct preprocessing pipelines, feature engineering, SMOTE balancing, ensemble models (CatBoost/Boost), Auto-ARIMA, Prophet. |
| Demonstration | Apply models to real-world data (VCDB filtered for e-commerce); generate classification results and time-series forecasts. |
| Evaluation | Assess model performance using statistical metrics (F1, ROC-AUC, precision, recall), ANOVA for feature comparison, interpretability via SHAP analysis, and forecasting accuracy (Auto ARIMA AIC, RMSE, MAE). Include model comparison and evaluation of overfitting/underfitting. |
| Communication | Document findings in research report discuss implications for cybersecurity practitioners and policy stakeholders. |

## 3.2 Research Setting and Data Collection

This study utilises data from the Verizon Community Data Breach (VCDB) Project, a publicly accessible dataset encompassing 9,911 cybersecurity incident reports from diverse industries worldwide between 1984 and 2023 (Fernandez et al., 2020). The VCDB is curated by cybersecurity experts and offers structured information on attack types, threat actors, victim profiles, and breach outcomes. To focus specifically on the e-commerce sector, a domain-specific filtering and enrichment methodology was applied to isolate incidents relevant to online retail, electronic payment systems, and digital marketplaces. This approach involved filtering and enriching the dataset using hybrid rule-based and heuristic methods, resulting in a refined, high-relevance subset of 1,866 e-commerce related incidents.

**Rationale for the Strategy**

A hybrid data preparation approach was employed, combining structured industry metadata with unstructured text fields to accurately extract e-commerce-related cybersecurity incidents. The key components include:

- Heuristic-Based Feature Engineering & Filtering: Domain knowledge guided the classification of relevant incidents, utilising industry codes (e.g., NAICS prefixes) alongside keyword patterns such as "retail," "checkout," and "shop" to flag e-commerce activity. Binary indicator flags were created from text and categorical data to infer relevance.

- Hybrid Rule-Based and Data-Driven Labeling: Structured logic, including exact and prefix code matching, was integrated with text-based heuristics by detecting keywords in fields like victim_id or summary, resulting in an enhanced and precise labeling of e-commerce cases .

- Weak Supervision for Data Enrichment: Inspired by methods such as Snorkel, which leverages programmatic labelling techniques for generating training data from noisy or incomplete sources (Ratner et al., 2017), additional metadata fields were created to capture industry classification signals indirectly. These included features such as ecommerce_prefix_match, sector_ecommerce_flag, and threat_keyword_flag.

- Business Logic-Based Subsetting: Incidents with a confirmed or inferred e-commerce link were retained, producing a focused dataset suitable for modelling and statistical analysis.

- This hybrid approach addresses common challenges in real-world datasets, where labeled data may be sparse, inconsistent, or missing. By leveraging domain expertise and combining rule-based logic with data-driven heuristics, it achieves high precision in defining a relevant modeling population. This is critical for ensuring data quality and relevance, especially within complex sectors like e-commerce cybersecurity where structured metadata alone is often insufficient. Additional engineered features were created to enhance analytical value, such as:

- summary_length (incident narrative length)
- contains_pii_terms (presence of personally identifiable information like SSN or credit data)
- risk_terms_score (composite score quantifying threat severity)
- Temporal features (e.g., incident month, quarter)
- Geographic grouping (region_group)
- Threat indicators (matched_threat_keywords, threat_enrichment_score)

All data preprocessing, statistical analyses, and machine learning modeling were conducted using Python within the Jupyter Notebook environment, ensuring reproducibility and an efficient analytical workflow.

## 3.3 Data Inspection and Initial Exploration

The initial phase of data analysis involved creating a specialised subset of the raw dataset, focusing solely on incidents relevant to the e-commerce sector. This stage aimed to isolate data applicable to the research objectives and prepare it for subsequent feature engineering and modelling tasks (Wickham, 2014). The raw dataset was imported and processed using Python's Pandas library, a widely adopted tool for efficient manipulation of structured tabular data (McKinney, 2010).

The filtering process applied industry code normalisation, keyword-based text matching, and sector-specific checks to capture incidents associated with retail and online commerce activities. As a result, a refined dataset of 1,866 incidents was obtained, each containing critical attributes such as year, action, industry_name, actor_internal, actor_external, country_code, threat_enrichment_score, victim_sector, and summary_length.

This filtered dataset provided a focused analytical foundation for exploring e-commerce cyberattack patterns and supporting subsequent modelling and visualisation tasks.

```
from IPython.display import display

# === Step 2: Normalize industry codes ===
def normalize_code(code):
    if pd.isna(code):
        return ''
    return str(code).strip().lstrip('0')

df_processed['industry'] = df_processed['industry'].apply(normalize_code)

# === Step 3: Define e-commerce prefixes and exact codes ===
ecommerce_prefixes = [
    '44', '441', '442', '443', '4431', '444', '445', '446', '447', '448',
    '451', '452', '453', '454', '4541', '45411', '454110', '454112',
'454113',
    '48', '49', '51', '518', '5191', '51913', '519130', '519190',
    '541', '5415', '54151', '541511', '541512'
]

ecommerce_codes = {
    '454110': 'Electronic Shopping',
    '454112': 'Electronic Auctions',
    '454113': 'Mail-Order Houses',
    '443142': 'Electronics Stores',
    '4541': 'General E-Commerce',
    '519130': 'Internet Publishing & Platforms',
    '541511': 'Web App Development',
    '541512': 'System Design for E-Commerce',
    '518210': 'Cloud & Hosting Services',
    '519190': 'Other Info Services'
}
```

*Figure 14 Dataset structure and missing value distribution (Source: Author's computation using Pandas)*

Descriptive statistics were then generated for all numerical variables. Features such as year, summary_length, risk_terms_score, and threat_enrichment_score were summarised using standard metrics mean, standard deviation, and range. To guide preprocessing strategies, each feature was classified by data type, categorical, numerical, or hybrid (categorical-numeric). This classification, presented in Figure 15, clarified which variables required encoding, scaling, or imputation and helped avoid misinterpretation of superficially numeric codes (e.g. industry codes).

| No. | Column Name | Feature Type |
|---|---|---|
| 1 | incident_id | Categorical |
| 2 | summary | Categorical |
| 3 | industry | Categorical |
| 4 | victim_id | Categorical |
| 5 | country | Categorical |
| 6 | state | Categorical |
| 7 | actor_external | Categorical |
| 8 | actor_internal | Categorical |
| 9 | action | Categorical |
| 10 | confidentiality | Categorical (Numeric) |
| 11 | integrity | Categorical (Numeric) |
| 12 | reference_date | Categorical (Numeric) |
| 13 | incident_month | Numerical |
| 14 | cummary_length | Categorical (Numeric) |
| 15 | risk_terms_score | Numerical |
| 19 | victim_id.1 | Categorical |
| 20 | victim_sector | Categorical |
| 22 | country_code | Categorical |
| 24 | region_group | Other |
| 25 | incident_month_name | Categorical (Numeric) |
| 27 | keyword_count | Categorical (Numeric) |

*Figure 15 Variable classification by data type (Source: Annotated schema of VCDB dataset)*

- Beyond structural inspection, initial exploratory analysis was conducted to examine relationships among numeric features. Using Plotly's imshow() function, a correlation matrix was generated (Figure 16). This helped identify interdependencies that could affect feature selection or model assumptions.
- A moderate correlation (r = 0.48) was observed between threat_enrichment_score and summary_length, indicating that incidents described with longer narrative text often exhibited higher threat severity. This highlights summary_length as an analytically important feature for subsequent modelling.
- Temporal features such as year, incident_month, and incident_quarter exhibited weak correlations with severity scores, suggesting they serve primarily as contextual rather than predictive attributes. Furthermore, high collinearity among these time-related variables (r > 0.96) was identified, a condition known to distort regression coefficients and model interpretability; as highlighted by James et al. (2013), multicollinearity can negatively impact model stability, which informed the decision to exclude or consolidate these variables in later modelling stages.

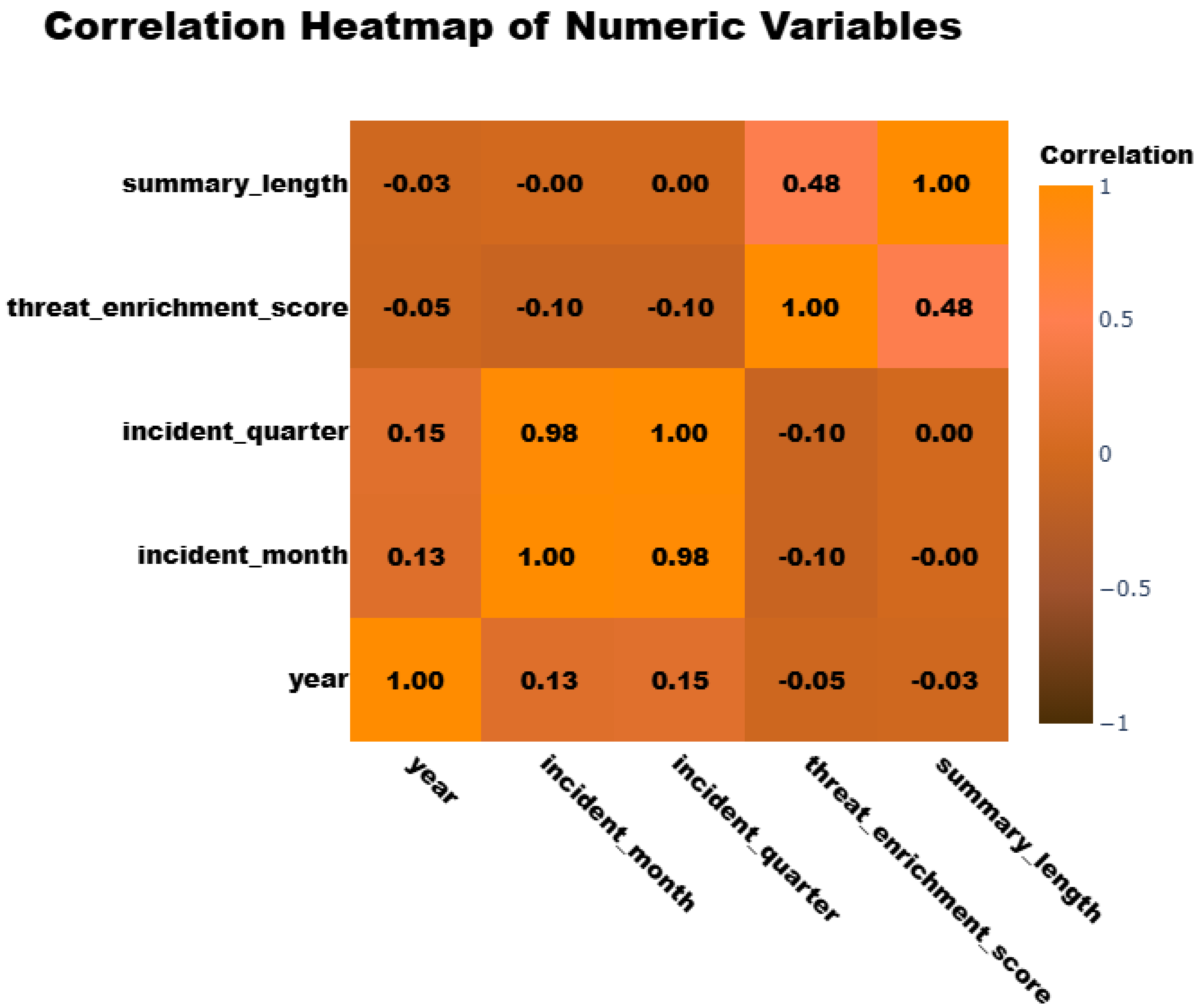

*Figure 16 Correlation heatmap of numerical features (Source: Interactive Plotly matrix using Jupyter)*

Following the correlation analysis, an evaluation of data completeness was undertaken as part of the broader preprocessing workflow. Cybersecurity datasets are frequently sparse or inconsistently reported, which can compromise model performance and limit generalisability. To mitigate these risks, this study proactively addressed missingness issues prior to filtering, ensuring the reliability of all subsequent analytical steps. Specifically, categorical attributes such as incident_month_name, integrity, action, and confidentiality, which exhibited varying levels of missingness, were imputed using domain-appropriate placeholders and values derived from related features. This approach aligns with recommended practices for managing missing data to reduce bias and improve model stability (Little & Rubin, 2019). As a result, the filtered e-commerce dataset contained no missing values across its key attributes, including year, action, industry_name, actor_internal, actor_external, country_code, threat_enrichment_score, victim_sector, and summary_length. Resolving missingness at this stage enabled the analytical workflow to focus entirely on feature engineering, seasonal trend analysis, and predictive modelling using a complete and reliable dataset suitable for accurate and reproducible results.

## 3.4 Data Techniques and Analysis

### 3.4.1 Data Preprocessing

Proper data preprocessing is vital for ensuring dataset quality and improving model performance by reducing bias and addressing inconsistencies (Kotsiantis, Kanellopoulos & Pintelas, 2006). This involves several steps such as cleaning, transformation, normalization, and handling missing values. As missing data can bias results, a systematic imputation strategy was employed first to preserve dataset integrity.

#### 3.4.1.1 Data Imputation Strategy

A comprehensive assessment of missing data was conducted to ensure that the analytical dataset was complete and fit for subsequent modelling and evaluation. Rather than discarding incomplete observations, a structured imputation strategy was implemented to maintain dataset integrity and reduce bias. This approach ensured that essential attributes were retained, while missing values were imputed according to their data type and analytical relevance, following widely accepted practices for handling incomplete data in applied research (Little & Rubin, 2019).

1. **Retention of Crucial Variables**
   Certain variables, despite notable missingness, were retained due to their critical role in answering the research questions. For instance, variables like actor_internal, incident_month, and integrity were vital for analysing attack typology (Research Question 1) and identifying seasonal trends (Research Questions 2 and 4). These variables were imputed with tailored strategies to preserve their analytical value, adhering to common best practices for data integrity in applied research (Enders, 201).
2. **Imputation Strategy by Variable Type**

- **Categorical and Nominal Fields**:
  Variables such as actor_internal, state, and victim_id were imputed using the placeholder "Unknown". This technique maintains transparency and prevents the introduction of artificial bias, a method commonly applied in categorical imputation for large-scale analytical datasets (Allison, 2001).
- **Temporal Fields**:
  For temporal variables, including incident_month and incident_quarter, missing values were imputed with -1 to preserve numeric consistency while clearly distinguishing imputed values from observed data. Placeholder numeric coding of missing time-based values is an established practice in time-series analytics where exact imputation is not feasible (Zhang, 2016).
- **Textual Descriptions**:
  Narrative fields such as summary and reference were imputed with "No summary available" or "Not available" to avoid null-related errors during text-processing tasks. Placeholder text imputation is a practical approach when textual completeness is not essential to predictive modelling, but structural consistency is required (van Buuren, 2018).
- **Reference Dates and Metadata**:
  non-critical metadata such as reference_date was imputed with "Unknown", an approach commonly used for optional metadata attributes to ensure schema consistency without affecting key analytical outputs (Rubin, 1987).

3. **Minimal Data Loss Through Row Retention**
   No rows were excluded unless they lacked essential columns, such as action or year, which were critical for the study. Since these columns contained no missing values, the decision was made to retain the entire dataset, maximising statistical power and ensuring representativeness without losing valuable data.
4. **Post-Imputation Validation**
   Once the imputation was completed, the dataset was thoroughly re-examined to ensure all missing values had been addressed appropriately. This step ensured a fully complete dataset suitable for feature engineering, clustering, and predictive modelling. The imputation strategy aligns with data science best practices (Enders, 2010) and follows the guidelines for missing data treatment (van Buuren, 2018). By addressing missing data transparently and retaining key variables, the imputation process enhances the study's validity, reproducibility, and overall reliability.

```python
# Step 2: Create a working copy
df_imputed = df.copy()

# Step 3: Impute missing categorical columns with 'Unknown'
df_imputed['actor_internal'] =
df_imputed['actor_internal'].fillna('Unknown')
df_imputed['actor_external'] =
df_imputed['actor_external'].fillna('Unknown')
df_imputed['incident_month_name'] =
df_imputed['incident_month_name'].fillna('Unknown')
df_imputed['state'] = df_imputed['state'].fillna('Unknown')
df_imputed['integrity'] = df_imputed['integrity'].fillna('Unknown')
df_imputed['victim_id'] = df_imputed['victim_id'].fillna('Unknown')
df_imputed['victim_id.1'] = df_imputed['victim_id.1'].fillna('Unknown')
df_imputed['country_code'] = df_imputed['country_code'].fillna('Unknown')
df_imputed['confidentiality'] =
df_imputed['confidentiality'].fillna('Unknown')
# Step 4: Impute missing temporal/numerical values with placeholders
df_imputed['incident_month'] = df_imputed['incident_month'].fillna(-1)
df_imputed['incident_quarter'] = df_imputed['incident_quarter'].fillna(-
1)
# Step 5: Fill remaining non-critical text fields
df_imputed['summary'] = df_imputed['summary'].fillna("No summary
available")
df_imputed['reference'] = df_imputed['reference'].fillna("Not available")
df_imputed['reference_date'] =
df_imputed['reference_date'].fillna("Unknown")
df_imputed['victim_sector'] =
df_imputed['victim_sector'].fillna("Unknown")

# Step 6: Drop rows only if critical columns are missing (none in this
case)
df_imputed = df_imputed.dropna(subset=['action', 'year'])
# Step 7: Final check - Confirm there are no remaining missing values
missing_total = df_imputed.isnull().sum().sum()
print(f" Cleaning complete. Total remaining missing values:
{missing_total}")
```

*Figure 17 Data Imputation Snippet*

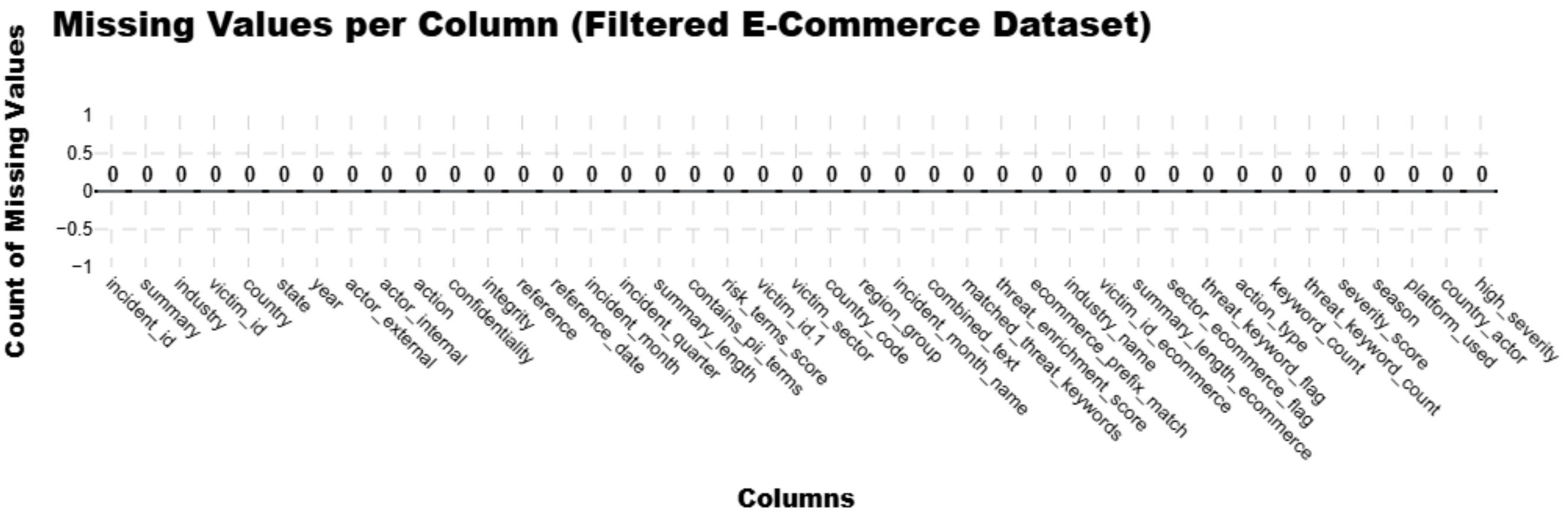

*Figure 18 Missing value analysis of the filtered e-commerce dataset showing no missing values across all columns.*

### 3.4.1.2 Outlier Detection and Removal (IQR Method)

Outliers, defined as data points significantly deviating from the dataset's central tendency, can distort model training and lead to inaccurate or overfitted models. To mitigate this, four numerical variables (summary_length, risk_terms_score, year, and threat_enrichment_score) were examined for outliers using the Interquartile Range (IQR) method, which identifies values outside the range Q1 – 1.5 × IQR to Q3 + 1.5 × IQR (Leys et al., 2013). This process identified 326 outliers in summary_length, 122 in risk_terms_score, 34 in year, and 4 in threat_enrichment_score. After excluding duplicate overlaps, the final dataset reduced from 1,086 to 1,579 records (post-cleaning shape), ensuring that extreme values did not skew subsequent analysis or model training. The reasons for choosing the IQR method include:

- **Effectiveness with Skewed Data:** IQR is particularly effective for datasets with skewed distributions, such as those often encountered in cybersecurity (Hubert & Van der Veeken, 2007). Unlike standard deviation-based methods, which are sensitive to extreme values, IQR is more robust, making it ideal for the threat severity metrics in this study.

- **Computational Efficiency:** IQR is computationally efficient, especially when working with large datasets. It can be implemented in linear time using sorting algorithms, which is essential for handling datasets with tens of thousands of records (Patil, 2021).
- **Avoiding Data Distortion:** The IQR method ensures that only significant deviations are flagged as outliers, preventing the exclusion of moderately deviant but valid data. This balance between sensitivity and specificity helps to maintain the dataset's integrity (ProCogia, 2023).

Box plots for both risk_terms_score and threat_enrichment_score before and after outlier removal visually confirmed the effectiveness of the IQR method. These plots showed a reduction in skewness and a more consistent distribution, validating the outlier removal process.

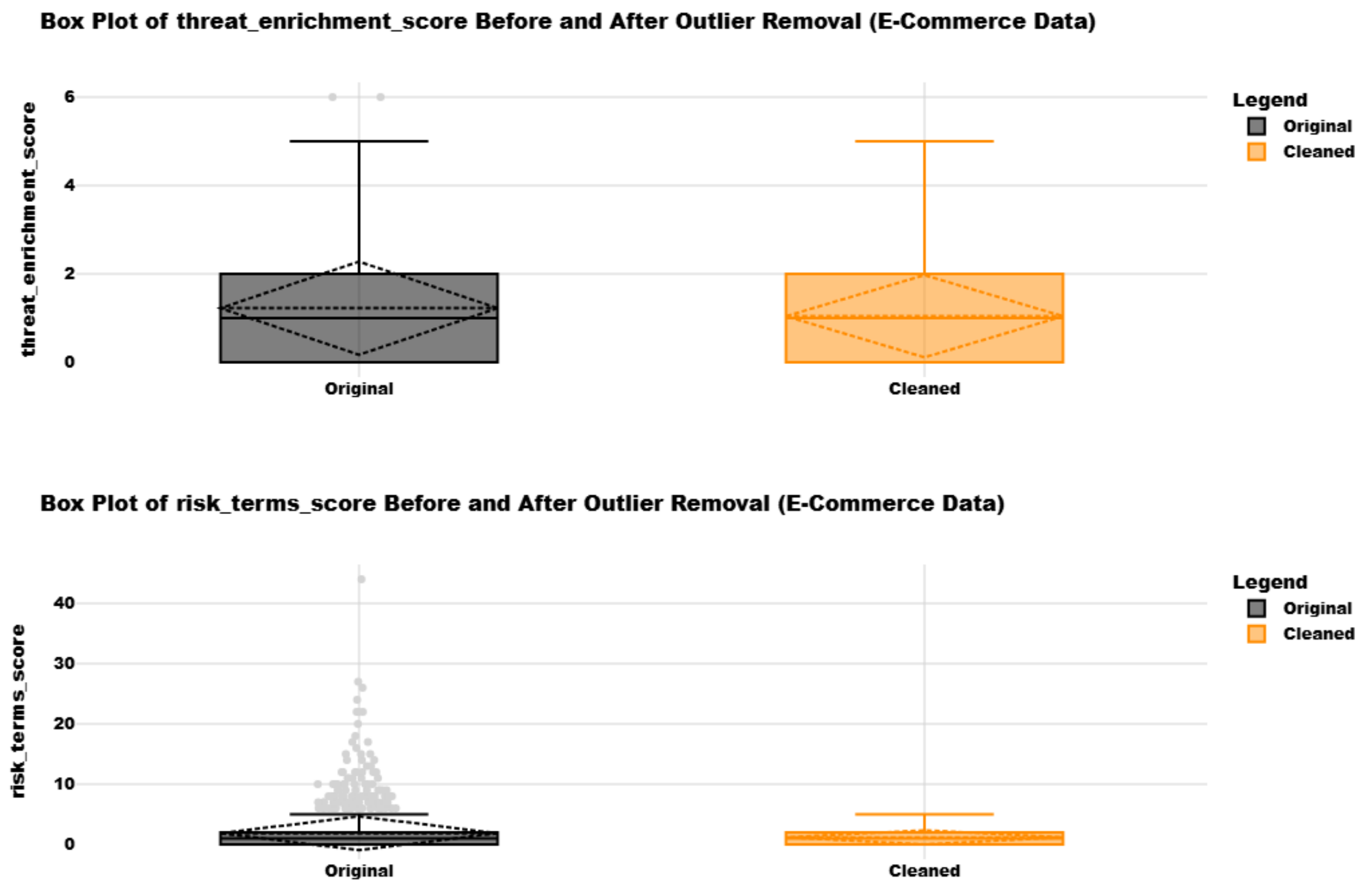

*Figure 19 Only these two box plots (threat_enrichment_score and risk_terms_score) were included in the report; others remain in the Notebook.*

The outlier removal using the IQR method is integral to maintaining the integrity and quality of the dataset and to ensure that the dataset is free from distortions that could impact model accuracy. These preprocessing steps support the robustness and reliability of the analysis, contributing to the study's overall analytical rigor.

#### 3.4.1.3 Feature Engineering and Data Transformation

After completing data cleaning and imputation, feature engineering was applied to convert raw incident data into structured variables optimized for identifying cyberattack patterns and enhancing model effectiveness. According to Kotsiantis et al. (2006), transforming features is essential for turning basic attributes into more informative forms that boost both the accuracy and interpretability of machine learning models. In line with this guidance, the dataset was augmented with generalized attack classifications, combined severity metrics, temporal groupings by season, and binary indicators for high-severity incidents. These engineered features laid the groundwork for in-depth exploration of attack categories, severity trends, and time-based distributions.

### 3.4.1.4 Generalising the Action Type Variable

The original action attribute included unstructured cyberattack descriptions with inconsistent terminology, limiting analytical value. A regex-based pattern matching method was used to map these into standardised categories like hacking, malware, phishing, privilege escalation, error, tampering, and "other" for ambiguous cases. Following Efstathiades et al. (2018), this approach improved interpretability, reduced dimensionality, and enhanced model generalisability, supporting comparative analysis and unsupervised clustering.

### 3.4.1.5 Temporal and Severity Mappings

The incident_month attribute was mapped to meteorological seasons (Winter: Dec–Feb, Spring: Mar–May, Summer: Jun–Aug, Autumn: Sep–Nov) to examine seasonal cyberattack trends, aligning with Ben-Asher and Gonzalez (2015), who emphasised temporal context in assessing cybersecurity risks. A binary high_severity flag was also created, labelling incidents above the median severity as "high." As shown by Samtani et al. (2020), such indicators help prioritise response efforts and enhance cyber risk management. These transformations supported finer seasonal analysis and risk-based threat prioritisation.

### 3.4.1.6 Quantifying Threat Keyword Density and Narrative Semantics

To capture threat complexity and narrative intensity, text-based features were extracted from unstructured incident summaries. A threat_keyword_count feature measured cybersecurity term frequency as a proxy for incident sophistication. Regex-based NLP methods also generated:

- **action_type**: Broad attack category from free-text.
- **risk_terms_score**: Weighted urgency score (e.g., "critical," "urgent").
- **threat_enrichment_score**: Composite of keyword count and urgency score.

Inspired by Sabottke et al. (2015), who linked keyword analysis to exploit prediction, additional features like summary_length and keyword scores were engineered to enhance the dataset's utility for modelling, anomaly detection, and prediction.

```
action_type  threat_keyword_count  risk_terms_score  \
4        other                    0                 0
7      unknown                    0                 0
8      unknown                    0                 0
11     hacking                    0                 0
13     hacking                    1                 0

    threat_enrichment_score  severity_score   season  high_severity  \
4                         0               0  Unknown          False
7                         0               0   Winter          False
8                         0               0  Unknown          False
11                        0               0   Summer          False
13                        1               1   Summer           True

    summary_length  keyword_count
4              615              1
7              195              1
8               47              1
11              26              1
13              45              1
```

*Figure 20 Risk Term Scoring: Scores summary text using intensity keywords to gauge incident severity Author's visualisation).*

### 3.4.1.7 Encoding Categorical Data: Methods and Their Justification

Encoding categorical variables is essential for models requiring numerical inputs. Key features,industry, nation, state, actor_external, action, and victim_sector, were used to predict PII presence. Due to high cardinality, Label Encoding was chosen over One-Hot Encoding. Label Encoding maps each category to a unique integer, preserving memory efficiency and supporting dense data structures. Tree-based models such as XGBoost natively handle integer-encoded inputs without assuming linear relationships, making them suitable for label-encoded categorical features (Chen and Guestrin, 2016). Similarly, CatBoost provides built-in optimisation for categorical variables, improving both efficiency and prediction accuracy (Prokhorenkova et al., 2018). As Zheng and Casari (2018) note, Label Encoding also produces a compact feature representation, which is advantageous when handling columns containing hundreds of unique values. Missing values were imputed with "Unknown" to preserve completeness, and encoding was done post train-test split to prevent data leakage (Brownlee, 2020). Though Label Encoding can imply order, tree-based models are unaffected (Kuhn & Johnson, 2020).

Empirical research in cybersecurity and intrusion detection has demonstrated that Label Encoding offers strong performance with lower computational cost, making it suitable for real-time analytics (IntruDTree Study, 2020). Figure 21 outlines the full preprocessing pipeline including feature extraction, encoding, and imputation ensuring model-readiness and compatibility with oversampling methods like SMOTE.

```
import pandas as pd
import numpy as np
from sklearn.preprocessing import LabelEncoder
from sklearn.model_selection import train_test_split
# Define target and preprocess
target = 'contains_pii_terms'
df = df[df[target].notnull()]
df[target] = df[target].astype(int)
# Feature engineering
df['summary_length'] = df['summary'].fillna("").apply(len)
df['keyword_count'] = df['matched_threat_keywords'].fillna("").apply(lambda x:
len(str(x).split(',')))
feature_cols = [
    'industry', 'country', 'state', 'year', 'actor_external', 'action', 'confidentiality',
    'summary_length', 'keyword_count', 'risk_terms_score', 'victim_sector',
    'country_code', 'region_group', 'threat_enrichment_score'
]
feature_cols = [col for col in feature_cols if col in df.columns]
X = df[feature_cols]
y = df[target]
# Label encoding
X_encoded = X.copy()
for col in X_encoded.select_dtypes(include='object').columns:
    le = LabelEncoder()
    X_encoded[col] = le.fit_transform(X_encoded[col].fillna("Unknown"))

# Fill missing numerics
X_encoded = X_encoded.fillna(X_encoded.median(numeric_only=True))
```

*Figure 21 Label Encoding Code*

### 3.4.1.8 Handling Class Imbalance

Class imbalance is a common issue in cybersecurity datasets, particularly for detecting rare but critical events such as the presence of Personally Identifiable Information (PII). In this study, the contains_pii_terms variable was highly imbalanced, with most samples labelled 0 (no PII) and a minority labelled 1 (presence of PII), as illustrated in Figure 22. To address this, SMOTE was applied to the training set after a stratified split to avoid data leakage and ensure realistic evaluation. By interpolating between existing minority samples, SMOTE improves class balance without duplication. Chawla et al. (2002) showed it enhances classifier sensitivity while preserving sample diversity.

SMOTE generates synthetic minority-class samples by interpolating between existing instances, increasing class balance without duplicating records. Prior to oversampling, categorical features were label encoded to enable meaningful distance calculations. After applying SMOTE, class parity was achieved in the training set, as shown in Figure 23, allowing classifiers to learn equitably across classes. This approach improved sensitivity to the minority class across classifiers (Logistic Regression, XGBoost, LightGBM, CatBoost) while maintaining overall accuracy.

While SMOTE enhances recall, it may introduce noise near decision boundaries, especially in high-dimensional data (Fernández et al., 2018). To mitigate this, data cleaning and numeric encoding were employed, and oversampling was limited to numeric features. Future work could incorporate hybrid techniques like SMOTE combined with Tomek Links to reduce class overlap and improve sample quality (Talukder et al., 2024).

Overall, SMOTE was a crucial preprocessing step that improved minority-class detection and ensured balanced, realistic model training without compromising data integrity.

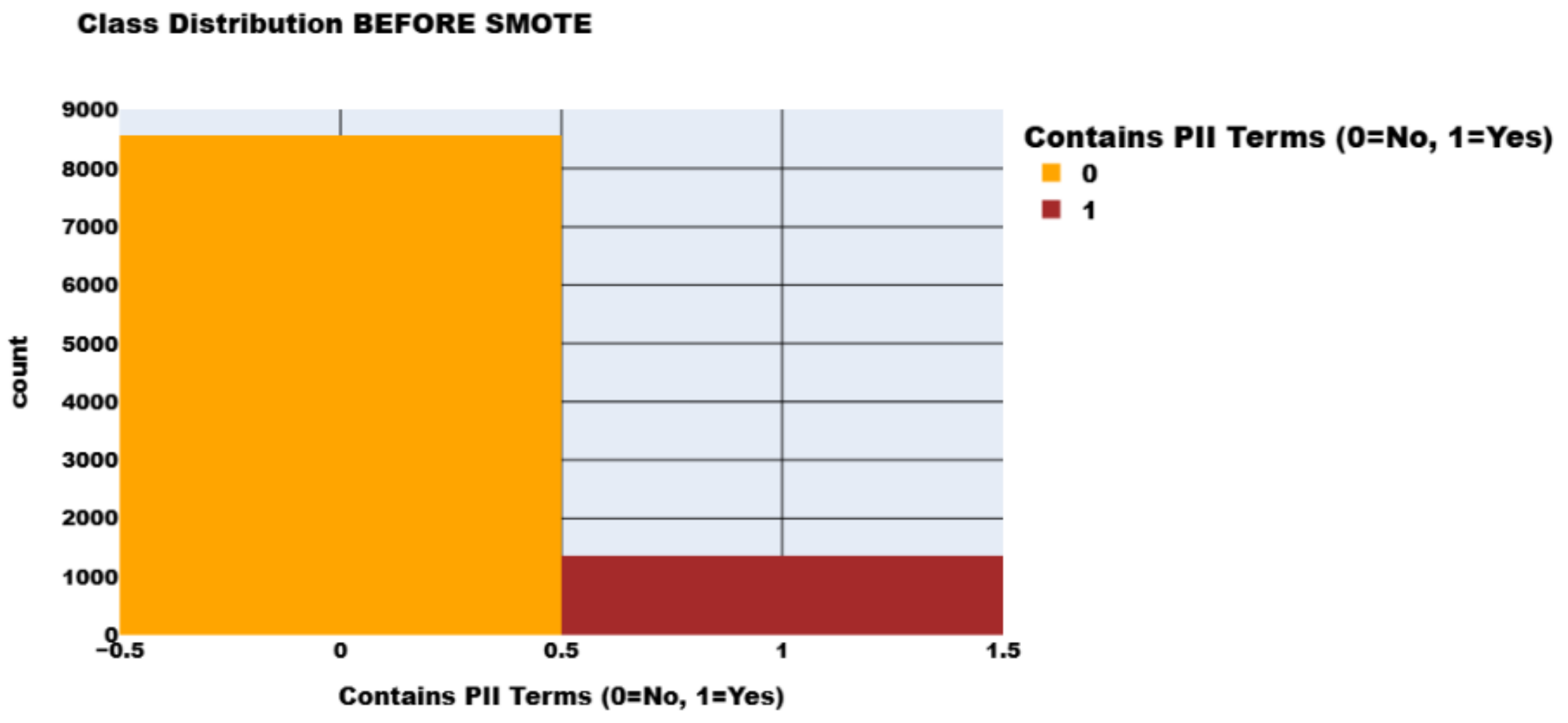

*Figure 22 Class Distribution Before SMOTE (Orange = Class 0, Brown = Class 1) (Orange = Class 0, Brown = Class 1)*

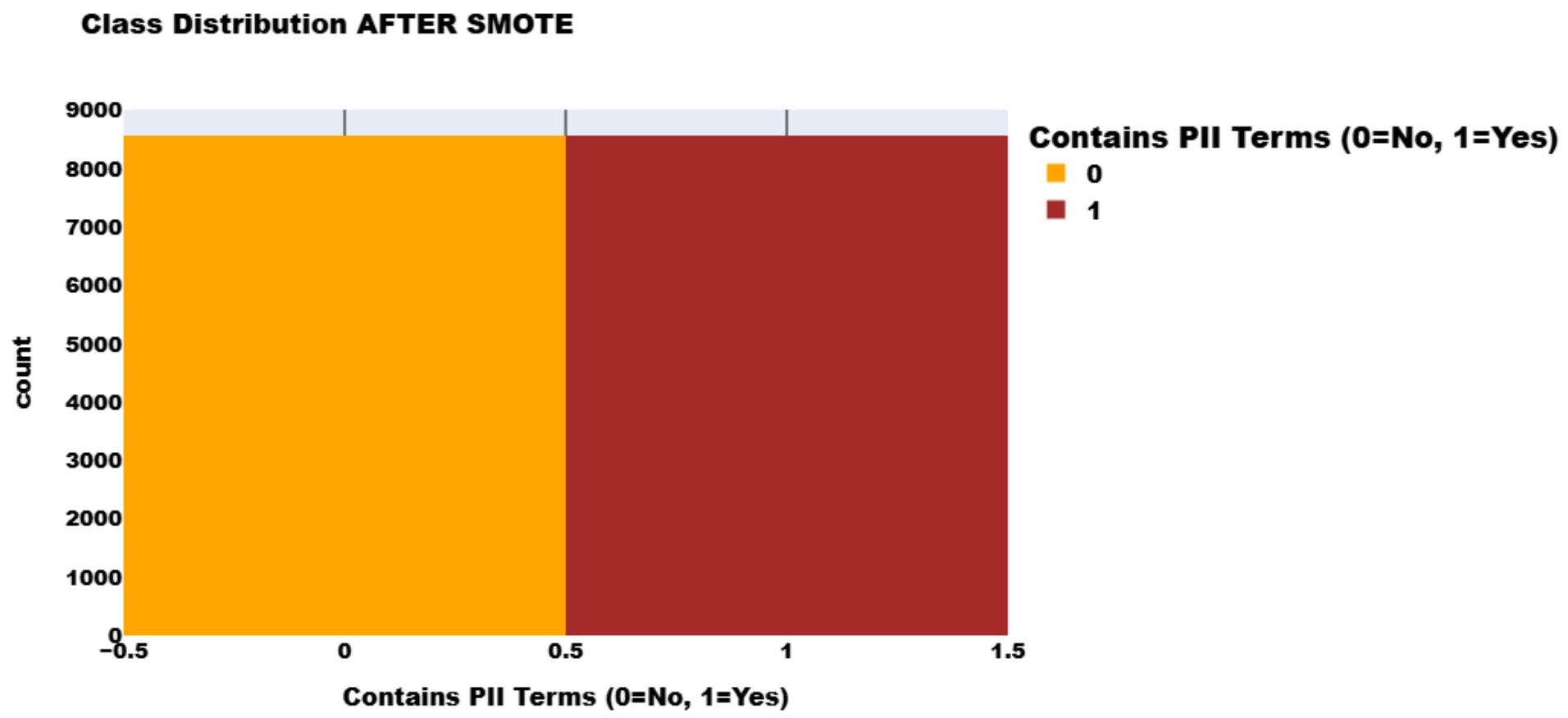

*Figure 23 Class Distribution After SMOTE (Orange = Class 0, Brown = Class 1)*

#### 3.4.1.9 Feature Scaling and Selection Strategy

Feature scaling and selection were implemented to improve model performance and interpretability for high-dimensional cybersecurity data. Scaling with MinMaxScaler was applied only to Logistic Regression to normalise features between 0 and 1, aiding algorithm convergence. Han, Pei and Kamber (2011) emphasised that feature scaling ensures uniform feature influence, while Jain et al. (2005) showed its role in improving convergence for gradient-based models.

- Correlation analysis identified highly correlated features (e.g., summary_length and risk_terms_score) with values exceeding ±0.75, prompting further examination.
- Statistical tests (ANOVA and Kruskal-Wallis) helped validate the significance of categorical variables such as incident month, supporting prior findings of seasonality in cyber threats (Chen et al., 2021).
- Embedded feature importance was derived from models like Logistic Regression, XGBoost, LightGBM, and CatBoost, revealing the predictive relevance of variables such as summary_length, keyword_count, risk_terms_score, and threat_enrichment_score. These features consistently ranked among the top predictors across all models, confirming their critical role in identifying PII-related threats.

Logistic Regression coefficients provided transparency on feature influence, supporting GDPR compliance, while ensemble models offered split-based importance for non-linear insights. Combining interpretable baselines with complex ensembles aligns with explainable AI best practices (Doshi-Velez and Kim, 2017).

## 3.5 Model Selection and Justification

This study adopted a hybrid modelling strategy combining classification, statistical inference, and forecasting to support cybersecurity analysis in e-commerce. Model selection was guided by empirical performance, interpretability, and suitability to the data's characteristics namely, its imbalance, mixed data types, and temporal aspects. For classification, four models were implemented: Logistic Regression, XGBoost, LightGBM, and CatBoost. Logistic Regression served as an interpretable baseline suitable for compliance-focused use cases (Hosmer et al., 2013). XGBoost was chosen for its scalability and strong performance on imbalanced, high-dimensional data (Chen and Guestrin, 2016), while LightGBM was selected for its efficiency and lower memory use with high accuracy (Ke et al., 2017). CatBoost was selected for its native handling of categorical features with minimal preprocessing (Prokhorenkova et al., 2018). These ensemble models excelled in identifying records containing Personally Identifiable Information (PII). For temporal forecasting, Auto-ARIMA was applied to model trends in cyberattack frequency and severity. This method was chosen for its ability to auto-tune parameters while capturing autoregressive and moving average components (Hyndman & Athanasopoulos, 2018). Forecasts revealed seasonal increases in threats, especially during high-traffic retail periods.

To evaluate group differences, ANOVA was used to test feature means across severity levels, validating the predictive utility of features like risk_terms_score and summary_length. The Mann–Whitney U test was employed to compare holiday vs. non-holiday enrichment scores, confirming significant seasonal effects despite close means. This integrated model suite spanning transparent baselines, advanced classifiers, inferential statistics, and time-series methods offered a robust, multi-dimensional framework for cyber risk analysis. The approach ensured strong performance, practical relevance, and alignment with current research in intelligent cybersecurity modelling.

### 3.5.1 Training/Test Split and Cross-Validation

To ensure valid, generalisable, and replicable results, a structured data split and validation framework was used, following best practices for imbalanced classification as outlined by Kuhn and Johnson (2013). This methodology consisted of a two-phase stratified splitting procedure, followed by 5-fold stratified cross-validation during the training phase. In the first splitting phase, the original dataset, which included completely encoded features (X_encoded) and the target variable (y), was divided into a training and hold-out test set using an 80:20 stratified split. The binary goal variable (contains_pii_terms) was stratified to guarantee that both classes were represented proportionally across subsets. Crucially, this split was accomplished before any synthetic balancing or data augmentation, separating the hold-out test set from the effects of upsampling and thereby preventing data leaking.

SMOTE was applied only to the training set to generate synthetic minority samples, avoiding data leakage and over-optimistic results (Lemaître, Nogueira & Aridas, 2017). The oversampled data was then split 80:20 into training and validation sets using stratified sampling to maintain class balance during model tuning. In the second phase, the balanced training dataset (X_bal, y_bal) produced by SMOTE was further split using another stratified 80:20 split, generating a new training and validation set (X_train, X_test, y_train, y_test) used for model fitting and preliminary testing. This sequential splitting process ensured that:

- The final training set was class-balanced,
- The test set remained untouched by SMOTE,
- And evaluation metrics remained trustworthy and generalisable.

A 5-fold stratified K-Fold Cross-Validation was applied using scikit-learn to maintain class balance and reduce overfitting (James et al., 2021). Brownlee (2020) notes that applying stratification and class balancing only on training data prevents leakage, while Géron (2019) emphasises staged pipelines for reproducible comparisons in high-stakes machine learning such as cybersecurity.

### 3.5.2 Performance Comparison

To assess model performance in this study, a range of quantitative metrics were employed across three categories: classification, forecasting, and inferential methods. Each metric was selected for its relevance to cybersecurity detection, data characteristics (e.g., imbalance), and interpretability. Formulas are provided along with concise definitions.

Classification models were evaluated based on their ability to correctly identify cyber incidents containing Personally Identifiable Information (PII). The following metrics were applied to **Logistic Regression**, **XGBoost**, **LightGBM**, and **CatBoost**.

**(a) Accuracy**

Accuracy = (TP + TN )/(TP+TN+FP+FN)

- **TP (True Positives)**: Correctly predicted PII cases
- **TN (True Negatives)**: Correctly predicted non-PII cases
- **FP (False Positives)**: Non-PII misclassified as PII
- **FN (False Negatives)**: Missed PII cases

Accuracy reflects overall correctness of predictions and provides a general understanding of the algorithm's performance (Saito & Rehmsmeier, 2015).

**(b) Precision**

Precision= (TP)/(TP+FN)

Precision measures how many predicted PII incidents were correct, and it's essential for reducing false alerts in e-commerce fraud detection (Chawla et al., 2002).

**(c) Recall (Sensitivity)**

Recall evaluates how well the model catches real PII incidents. High recall minimises undetected breaches critical in cybersecurity (Han et al., 2011).

**(d) F1 Score**

F1= 2 * (Precision * Recall)/ (Precision + Recall)

The F1 Score balances Precision and Recall, ideal for imbalanced datasets where both false positives and false negatives matter (Saito & Rehmsmeier, 2015).

**(e) ROC-AUC (Receiver Operating Characteristic – Area Under Curve)**

- ROC Curve plots True Positive Rate (Recall) vs False Positive Rate
- AUC is the area under this curve (range: 0.5 to 1.0)

A higher AUC indicates better class separation. It's threshold-independent and widely used in binary classification (Fawcett, 2006).

**(f) Log Loss (Logarithmic Loss)**

$$\text{Log Loss} = -1/N * \Sigma [\, y_i * \log(p_i) + (1 - y_i) * \log(1 - p_i) \,]$$

- *yi*: Actual class (0 or 1)
- *pi:* Predicted probability for class 1
- NNN: Number of observations

Log Loss penalises incorrect high-confidence predictions. Lower values indicate more calibrated probabilistic outputs (Brownlee, 2020).

### 3.5.3 Forecasting Metrics

For evaluating time-series forecasting of cyberattack frequency using Auto ARIMA and Prophet, the following metrics were employed:

**(a) Mean Absolute Error (MAE)**

$$\text{MAE} = (1/n) * \Sigma \mid y_t - \hat{y}_t \mid$$

$y_t$: Actual value

$\hat{y}_t$: Predicted value

n: Number of time steps

MAE reflects the average magnitude of forecasting errors, $\text{MAE} = (1/n) * \Sigma \mid y_t - \hat{y}_t \mid$ regardless of direction. It is model-agnostic and directly supported by both Auto ARIMA and Prophet (Hyndman & Athanasopoulos, 2018).

**(b) Root Mean Squared Error (RMSE)**

$$\text{RMSE} = \sqrt{[\, (1/n) * \Sigma (y_t - \hat{y}_t)^2 \,]}$$

RMSE penalises larger errors more than MAE, which is particularly important for cyberattack forecasting due to periodic spikes in activity (Zhang et al., 2020).

**(c) Mean Absolute Percentage Error (MAPE)**

$$\text{MAPE} = (100/n) * \Sigma \mid (y_t - \hat{y}_t) / y_t \mid$$

MAPE expresses prediction error as a percentage, facilitating intuitive performance comparisons across attack volumes of different scales.

**(d) Cross-Validation Metrics (Prophet-specific)**

Prophet also supports time-based cross-validation using sliding windows to simulate historical forecasts and evaluate performance on held-out periods. Metrics such as MAE, RMSE, and MAPE are computed across these rolling windows to assess temporal generalisability and detect overfitting (Taylor & Letham, 2018).

These metrics collectively provide a robust framework for evaluating the predictive accuracy and reliability of seasonal cyberattack models across both statistical and hybrid approaches.

### 3.5.4 Statistical Inference Metrics

Statistical methods, ANOVA and the Mann–Whitney U test, were used to assess the significance of differences in feature means across attack severity and seasonal conditions.

**(a) ANOVA (Analysis of Variance)**

$$F = MS_{between} / MS_{within}$$

- SS (Sum of Squares): Variation
- df: Degrees of freedom
- MS: Mean Square

ANOVA checks whether group means (e.g., severity levels) differ significantly. A large F-value and small p-value (< 0.05) indicate strong evidence of a difference (Field, 2013).

**(b) Mann–Whitney U Test**

$$U = n_1 n_2 + [n_1(n_1 + 1)] / 2 - R_1$$

- n1,n2n_1, n_2n1,n2: Sample sizes
- R1R_1R1: Sum of ranks in group 1

This non-parametric test compares medians of two independent groups (e.g., holiday vs non-holiday). Suitable when assumptions of normality are not met (Nachar, 2008).

## 3.6 Forecasting Future Trends Using Auto ARIMA

This section addresses RQ2 (seasonal variation in cyberattacks) and RQ4 (forecasting elevated threat periods) through a dual-model approach using Auto ARIMA and Prophet. The forecasting target was the annual average risk_terms_score, representing breach severity based on high-risk language in incident reports. Data from 2005 to 2023 was aggregated and forward filled for continuity. Auto ARIMA was selected for its automated parameter tuning and AIC-based optimisation. The time series was Box-Cox transformed and differenced to achieve stationarity. After training on data up to 2020, the model forecasted the 2021–2023 holdout period and projected trends through 2026. Forecast accuracy was evaluated using MAE and RMSE.

In parallel, Prophet, a Bayesian time series model by Facebook, was applied to the same dataset (structured as ds, y). Prophet is particularly effective for noisy data with changepoints, holidays, and irregularities common in cybersecurity trends. Trained on data from 1970–2023, it forecasted a gradual decline in breach severity from 2024 to 2026, contrasting with ARIMA's modest upward trend. Prophet also provided 80% confidence intervals, accounting for uncertainty and highlighting the potential for volatility. While ARIMA offers sharper short-term precision, Prophet excels in long-range interpretability. Together, they reveal both cyclical and evolving patterns, supporting RQ2 and RQ4. This integrated forecasting approach complements earlier classification models and strengthens cyber risk preparedness for e-commerce platforms.

## 3.7 Ethical Considerations

The ethical considerations surrounding cybersecurity research, particularly in e-commerce threat assessment, are multifaceted. This study uses a data-driven approach to analyse potentially sensitive data, necessitating a strong ethical framework to address privacy concerns, algorithmic bias, dual-use risks, transparency, and overall governance. Central to this framework is the principle of confidentiality ensuring that no identifiable information, even from anonymised datasets like the Verizon VCDB, is reverse-engineered or exploited. Macnish and van der Ham (2020) stress that competent cybersecurity research must uphold the dignity and privacy of breach victims, which was ensured in this study through meticulous preprocessing to prevent re-identification and the exclusion of personally identifiable information (PII) during feature engineering.

Further ethical challenges arise in the realms of bias and fairness in algorithmic design. Issues such as class imbalance, overrepresentation of certain geographies, and algorithmic prejudice are common in cybersecurity datasets (Van der Ham, 2020). This research employed techniques such as SMOTE to equilibrate class representation and utilised multi-metric evaluations to guarantee equitable performance across various subpopulations. Systematic fairness audits were conducted to evaluate misclassification trends. The dual-use conundrum was mitigated by limiting access to certain classifier components that could be misappropriated for nefarious purposes, in accordance with responsible disclosure protocols. Although transparency is essential for scientific integrity, certain sensitive elements were suppressed to avert misuse, adhering to the risk-based disclosure standards established by Macnish and van der Ham (2020). Institutional governance via ethics norms and IRB review was enhanced by specialised ethical counsel to address deficiencies in technical expertise. This multifaceted ethical framework guarantees that the research adheres to fundamental values of beneficence, non-maleficence, fairness, and respect for individuals, while fostering responsible cybersecurity innovation (Reidsma et al., 2023).

## 3.8 Limitations of the Methodology

Considering the methodological rigor employed in this investigation, encompassing data pretreatment, feature engineering, and the application of sophisticated machine learning classifiers, numerous shortcomings necessitate rigorous examination. These limits pertain to data representativeness, sampling process, evaluation strategy, algorithmic constraints, and ethical oversight.

1. **Data Representativeness**

   This research employs the Verizon Data Breach Investigations Report (DBIR) and VCDB datasets, which, while extensive, consist solely of publicly reported or disclosed occurrences. Consequently, the data may exhibit reporting bias, potentially underrepresenting occurrences from smaller firms or nations with restricted regulatory transparency (Biener, Eling & Wirfs, 2015). The absence of universal coverage may restrict the generalisability of findings and diminish external validity, especially when models are utilised for unobserved or underreported breach types.

2. **Sampling and Selection Bias**

   The dataset relies on convenience sampling instead of random selection. These sampling methodologies create systematic bias, potentially influencing the distribution of attack vectors, actor profiles, or target systems (Baltes & Ralph, 2020). This may result in models that overfit to dominant patterns in the dataset and inadequately generalise to infrequent yet essential cases.

3. **Evaluation Framework Limitations**

   While stratified cross-validation was employed to evaluate model resilience, the assessment was performed on historical data without considering concept drift, specifically the temporal evolution of cyberattack methodologies and threat actor behaviours. Models trained on past trends may diminish in predicting accuracy over time if not regularly retrained (Khaleefah & Al Mashhadi, 2023).

4. **Synthetic Oversampling Constraints**

   To address the class imbalance, the research uses SMOTE. Although effective in numerous classification tasks, SMOTE can generate synthetic samples next to class borders, potentially misrepresenting real-world patterns. These marginal cases may provide imprecise decision limits, particularly in high-dimensional feature spaces, thereby diminishing model accuracy (Fernández et al., 2018).

   - **Limited Exploitation of Unstructured Data**

     Unstructured text fields, like narrative breach summaries and descriptions, are present in the dataset, but they were not fully utilised during the feature engineering process. More expressive representations like TF-IDF vectors or contextual embeddings (like BERT) may produce richer feature sets, capturing nuances that regular expressions miss, as keyword extraction techniques are by nature reductive (Zhou et al., 2022).

   - **Interpretability Trade-offs**

     Tree-based ensemble models such as XGBoost and CatBoost were selected for their exceptional accuracy. Nonetheless, these models frequently face criticism for being "black boxes," rendering them less interpretable than linear models. Although SHAP values were utilised to partially address this issue, the interpretability gap persists, especially in critical fields like as cybersecurity (Lundberg et al., 2020).

   - **Adversarial Robustness and Security Testing**

     This research does not evaluate the model's susceptibility to adversarial manipulation, such as the

introduction of engineered feature vectors intended to bypass detection. In the absence of adversarial testing, the model may exaggerate its robustness and prove less successful in actual threat scenarios (Biggio & Roli, 2018).

- **Ethical and Governance Constraints**

  Although an ethics protocol was followed and informed by Institutional Review Board (IRB) standards, existing IRBs may lack the technical expertise to fully evaluate cybersecurity research risks. As observed by Macnish and van der Ham (2020), such gaps in governance can hinder meaningful oversight, particularly in areas involving dual-use potential and responsible disclosure.

# 4.0 Hybrid Model Design, Evaluation, and Results

This chapter presents the implementation of the proposed hybrid analytical framework and its empirical evaluation using breach records from the Verizon Data Breach Database (VCDB). The primary objective is to assess the framework's efficacy in detecting, interpreting, and predicting cyberattacks on e-commerce platforms. The analyses conducted herein are explicitly aligned with the study's four research questions and grounded in domain-specific statistical and machine learning techniques. To ensure methodological transparency and coherence, the structure of this chapter is organised to directly address the research questions as follows:

- RQ1: Predominant cyberattack types are identified through frequency analysis and visualisation (Section 4.2).
- RQ2: Temporal trends, such as seasonal attack surges, are assessed using AutoARIMA and Prophet - based time-series forecasting (Section 4.2).
- RQ3: Statistical tests (correlation, ANOVA) explore links between PII breaches and high-threat keyword activity (Section 4.3).
- RQ4: Machine learning models (XGBoost, LightGBM, CatBoost) are evaluated for their predictive performance and interpretability using SHAP values (Sections 4.3–4.5).

## 4.1 Data Preparation Recap

The dataset used for this analysis was preprocessed as outlined in Chapter 3, involving rigorous cleaning, transformation, and feature engineering based on the Verizon DBIR (2022). Techniques such as Interquartile Range (IQR) for outlier removal, label encoding for high-cardinality categorical features, and SMOTE for addressing class imbalance were applied to ensure model readiness. These steps produced a structured dataset optimised for both classification and time-series forecasting, facilitating robust detection of Personally Identifiable Information (PII) breaches and seasonal attack trends.

## 4.2 Exploratory Data Analysis (EDA)

This section extends the data preparation phase by exploring key patterns within the e-commerce cybersecurity dataset. Through categorical aggregation and frequency visualizations, the EDA reveals that hacking, exploitation, and injection-based attacks are most prevalent, especially in online payments, internet platforms, and retail IT infrastructure. These sectors are heavily targeted due to the concentration of user credentials and payment information (Chatterjee et al., 2022). Less common but notable are misconfiguration issues and stolen credentials, particularly in web app development and cloud services, highlighting risks related to deployment and access controls. The low frequency of session hijacking attacks may indicate underreporting due to their covert nature (Romanosky, 2016). Figure 24 these patterns support the identification of critical attack surfaces and inform model feature selection and risk prioritisation in subsequent analysis.

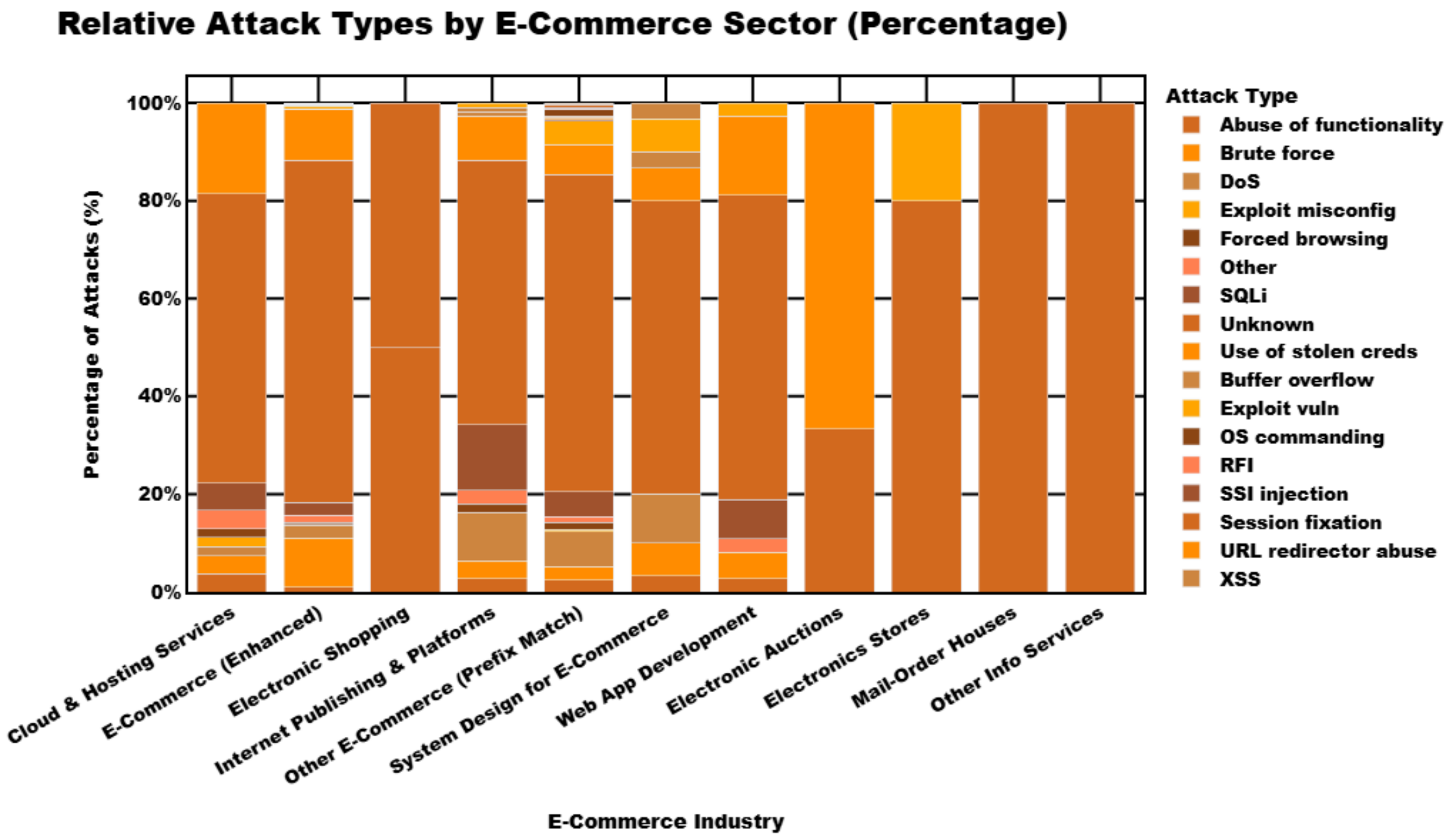

*Figure 24 Distribution of Attack Action Types*

Figure 25 displays a lollipop plot ranking threat actors by frequency, clearly illustrating disparities in incident volume. The 'Unknown' category dominates with over 1,200 incidents (35% of the dataset), highlighting persistent attribution challenges due to anonymisation techniques and proxy use. Among identified actors, Organised Crime leads, followed by Activists and Unaffiliated Individuals, reflecting the prevalence of financially or ideologically motivated attacks targeting sensitive e-commerce systems. State-affiliated and Nation-state actors appear less frequently but often conduct high-impact, strategic operations, including targeting supply chains and undermining digital trust.

Overall, the plot reveals a layered threat landscape high-frequency, low-sophistication attacks coexist with fewer but more consequential state-backed campaigns, reinforcing the need for differentiated defence strategies.

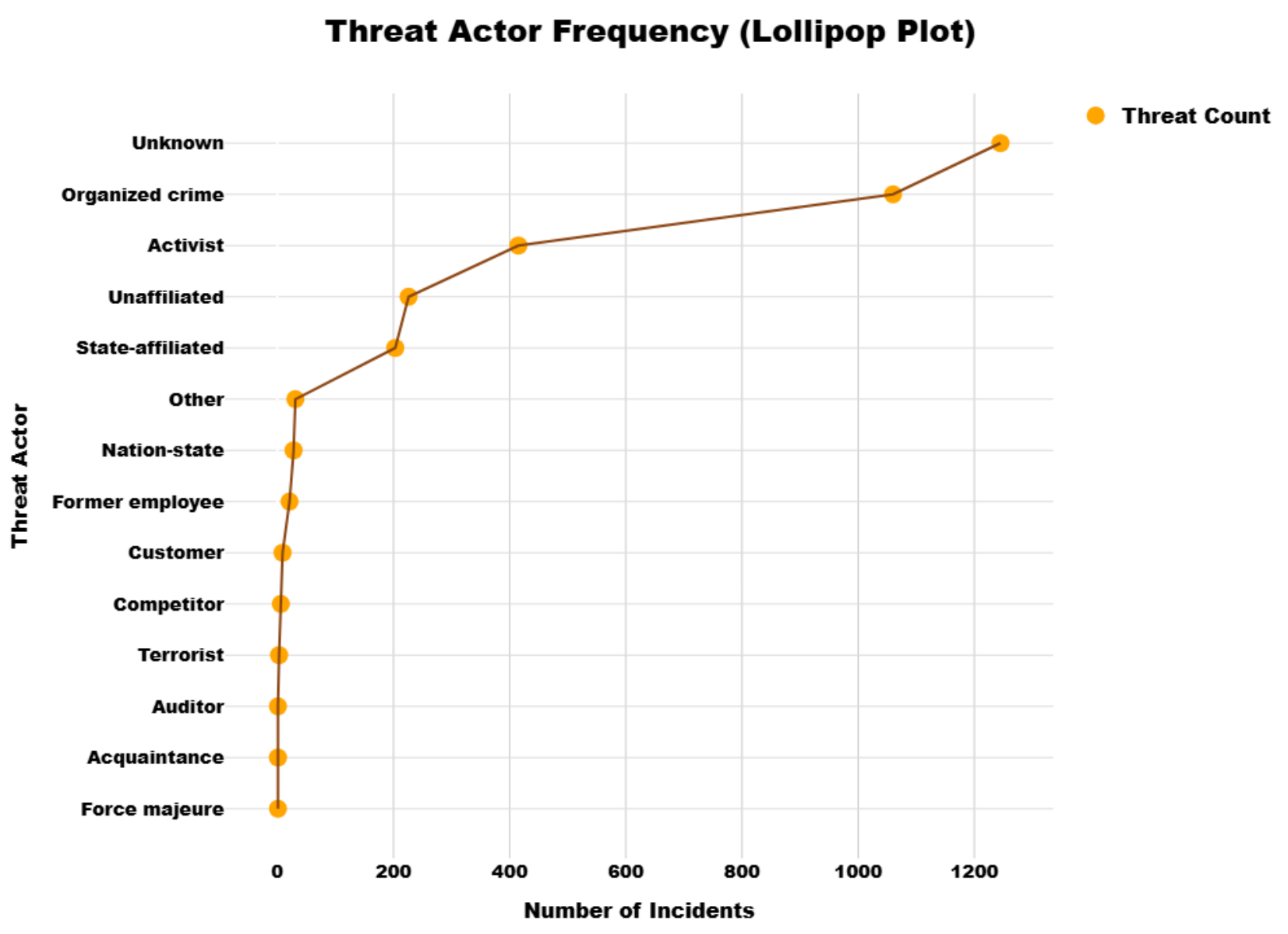

*Figure 25 Distribution of Attack Action Types*

Figure 26 The donut visualisation shows the distribution of cyberattacks by targeted platform. The Web (33.3%) and Server (11%) categories together represent 44.3% of incidents, equal to the proportion attributed to "Other" targets (44.3%). This aligns with Chatterjee et al. (2022)'s findings that front-facing digital infrastructure is the most exposed and frequently exploited part of online retail systems. In contrast, Mobile Applications (9.65%), APIs (0.75%), and Cloud Services (1.02%) collectively account for about 11.4% of attacks. However, this smaller share does not imply lower risk; as Zissis & Lekkas (2018) note, the complexity of cloud and API environments often leads to detection challenges and underreporting. With increasing integration into headless commerce and composable architectures, these platforms' threat exposure is likely to grow, making this distribution a conservative baseline rather than a definitive future risk profile.

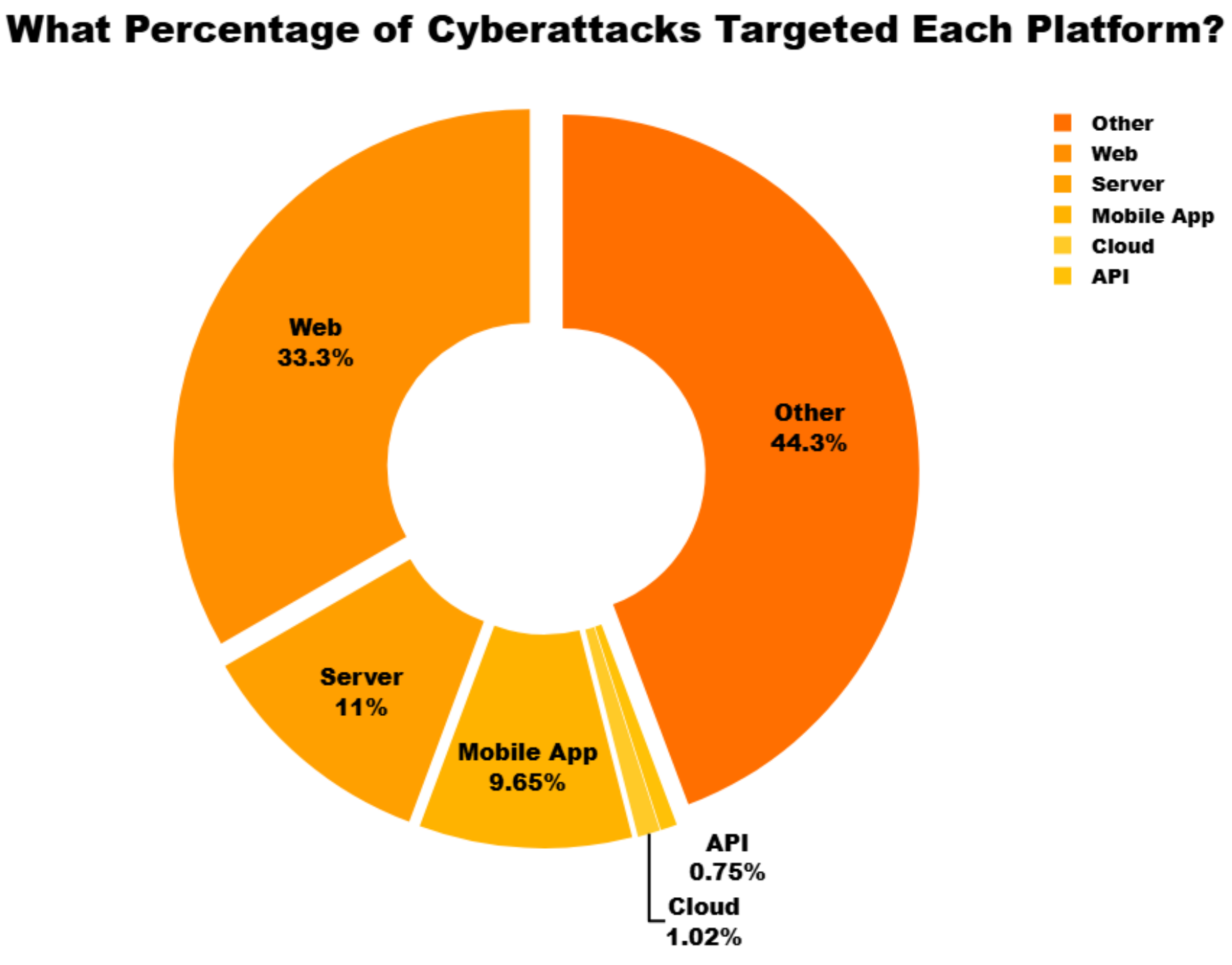

*Figure 26 Targeted Asset Types*

## 4.3 Temporal Trend Analysis: Exploring Seasonal Patterns in Cyberattack Frequency and Severity

This section explores seasonal patterns in cyberattack activity, focusing on both frequency and average severity across the calendar year. Using monthly incident distributions and ARIMA-based forecasting, the analysis uncovers recurring trends that may correspond to holidays, peak shopping periods, or other time-sensitive vulnerabilities in e-commerce platforms. The objective is to detect underlying cycles, anticipate potential surges, and support timely cybersecurity planning. The section is structured into three parts: monthly frequency patterns (4.3.1), ARIMA-based severity forecasting (4.3.2), and a strategic interpretation of their divergence (4.3.3).

### 4.3.1 Monthly Frequency Patterns: Identifying Vulnerability Windows

The initial phase of analysis visualises the monthly distribution of cyber incidents. Figure 27 shows a strong spike in January (772 incidents), far exceeding all other months, which ranges between 79 and 116 incidents. February (97), May (116), and July (116) show slightly elevated values compared to other months, but there is no clear mid-year surge. Instead, the pattern suggests a January-driven anomaly, possibly reflecting reporting backlogs, seasonal exploitation patterns, or coordinated campaigns around the start of the year.

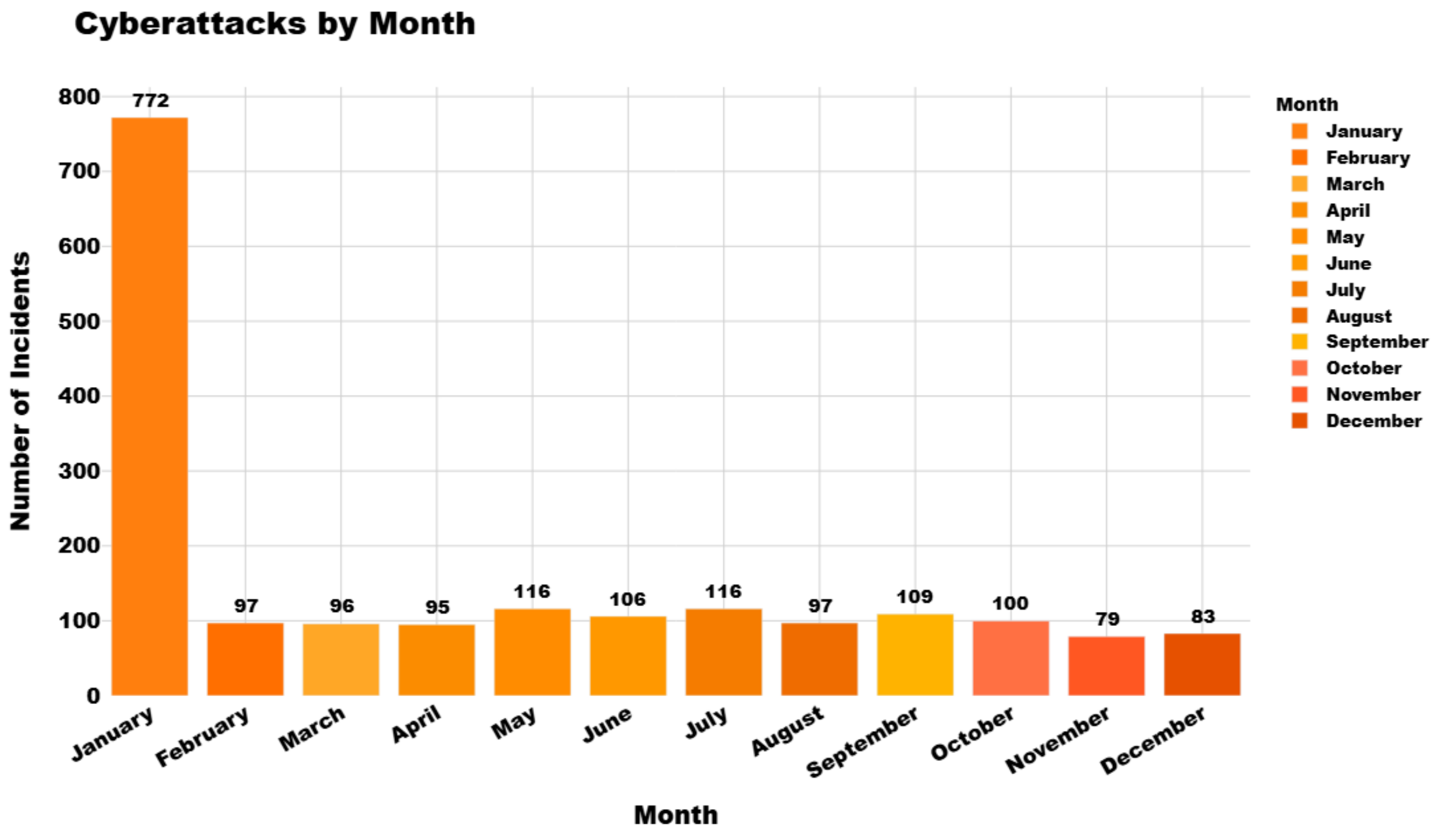

*Figure 27 Cyberattacks by Month (Total Incident Count)*

These variations likely reflect operational realities rather than random fluctuations. The sharp January peak may stem from reporting practices or backlogged incident disclosures, while the relatively flat distribution from February to December suggests no clear mid-year peaks. December’s lower value (83 incidents) appears consistent with broader seasonal preparedness measures such as enhanced SOC monitoring and pre-holiday vulnerability assessments.

#### 4.3.1.1 Hypothesis Testing

To complement the frequency analysis, threat severity scores were compared between holiday months (June, July, November, December) and non-holiday periods. Figure 28 shows the average threat enrichment score during holiday months at 0.86, slightly higher than 0.70 for non-holiday months. Though the difference is modest, statistical testing confirmed significance (Mann–Whitney U = 2,579,981.5, p = 0.0121), supporting the view that cyber threats are not only more frequent but potentially more severe during these critical calendar windows.

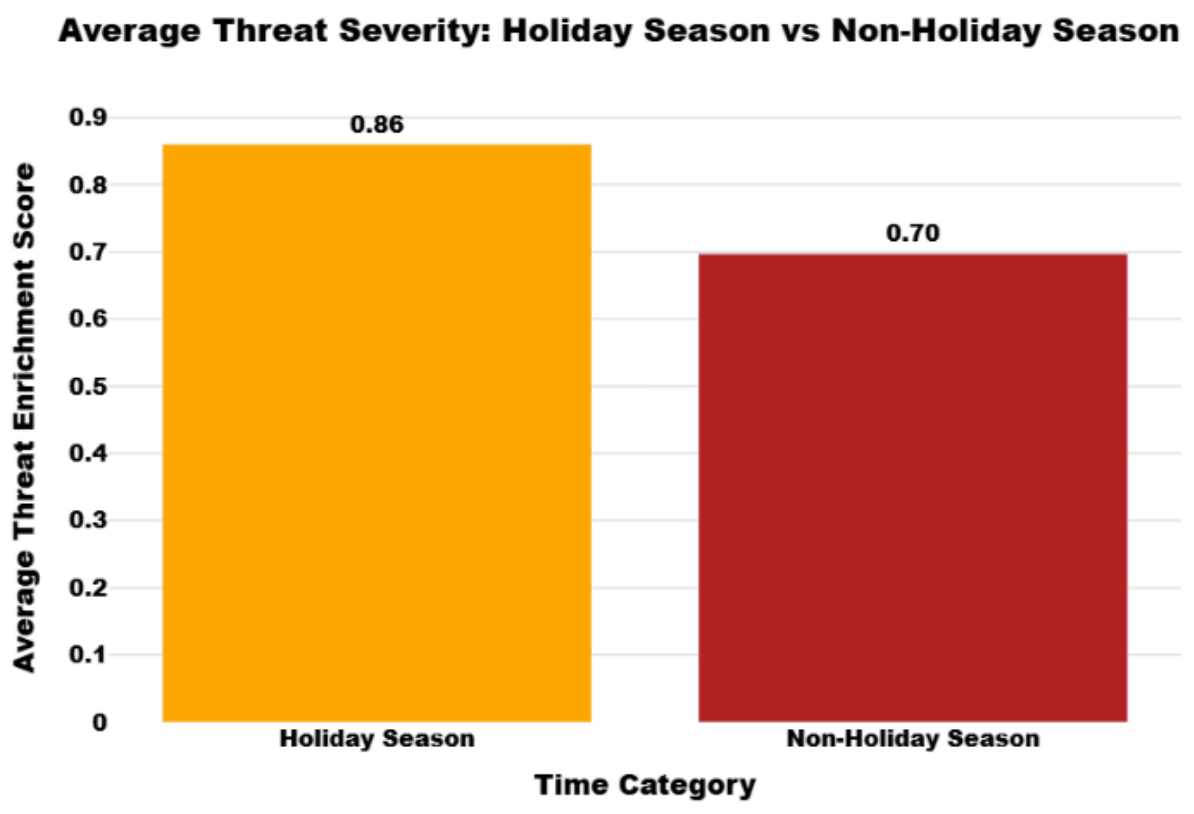

*Figure 28 Average Threat Severity: Holiday vs Non-Holiday Months*

### 4.3.2 Forecasting Cyberattack Severity: Auto-ARIMA Insights

To examine whether cyberattack severity, measured by the average annual risk score based on threat keyword usage, exhibits a seasonal trend, a univariate Auto-ARIMA model was fitted to data from 2005 to 2023 and used to forecast values for 2024 to 2026. Figure 29 shows the forecast output. Historical severity remained relatively flat, with a mild increase around 2021–2022, likely linked to high-impact vulnerabilities such as Log4Shell (CVE-2021-44228) and ProxyShell that caused widespread disruption across enterprise networks (Hiesgen et al., 2023). Beyond this, the model predicts continued low-to-moderate severity levels below 2.5, with a widening 95% confidence interval reflecting growing uncertainty over the forecast horizon.

These results suggest that severity does not follow a consistent seasonal pattern but is instead influenced by irregular, high-impact events like supply chain compromises, advanced persistent threat campaigns, and critical vulnerability disclosures. This contrasts with attack frequency, which aligns more closely with commercial or calendar cycles. While Auto-ARIMA is suitable for stable, autocorrelated time series, its limitation is the inability to incorporate external variables. Future models could improve accuracy by integrating exogenous factors such as CVE alert feeds, geopolitical risks, or dark web monitoring.

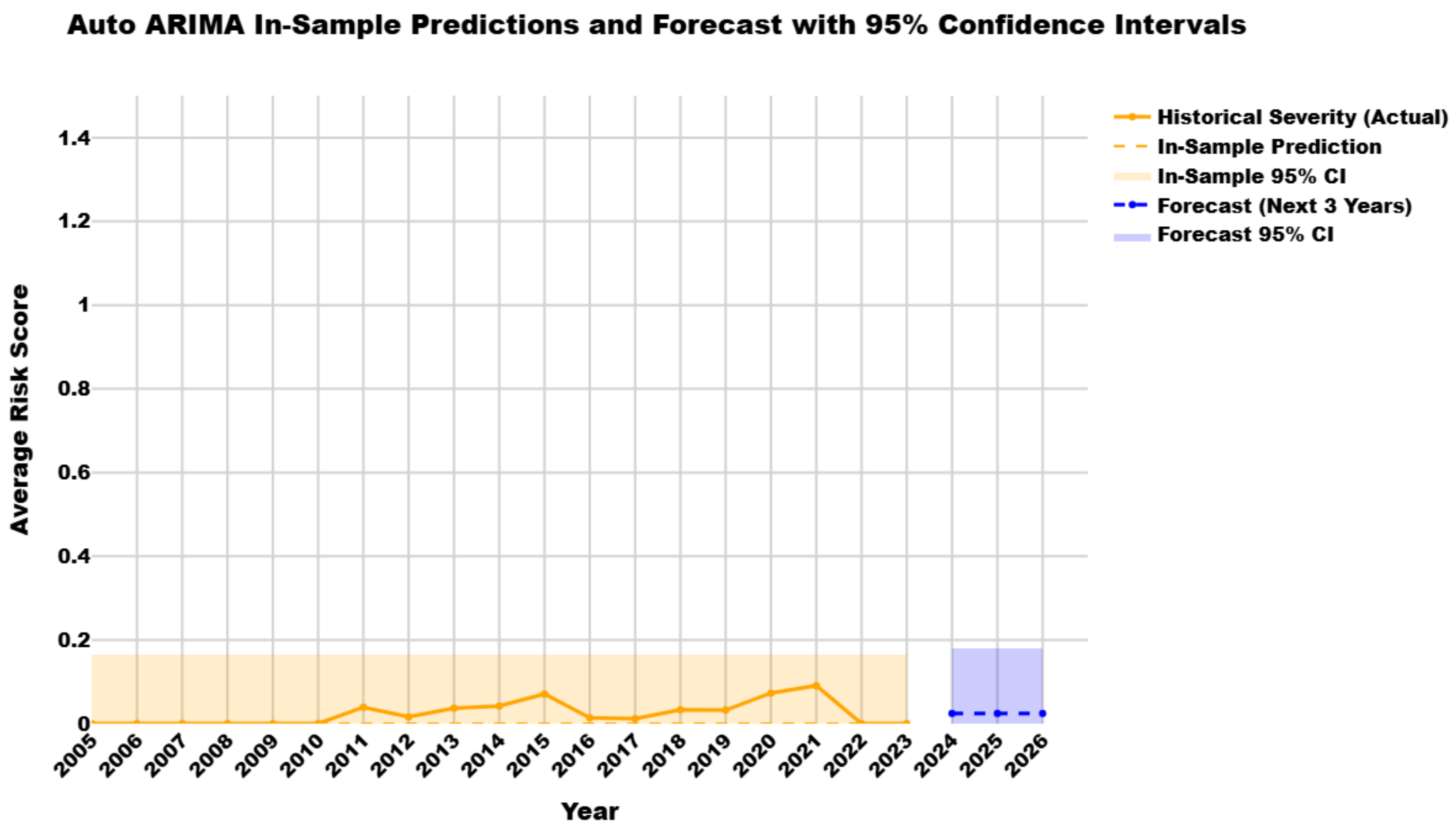

*Figure 29 Auto-ARIMA Forecast of Cyberattack Severity (2024–2026)*

### 4.3.3 Forecasting Cyberattack Severity: Prophet Insights

To complement the ARIMA modeling approach and address potential non-linear trends in the cyberattack severity data, Facebook Prophet was applied to the annual risk_terms_score dataset spanning 1970–2023, with forecasts extending to 2026. Prophet is particularly well-suited for cybersecurity datasets characterized by reporting inconsistencies and structural breaks, as it incorporates changepoint detection, decomposes seasonality, and demonstrates robustness to missing or irregular data (Khandelwal et al., 2022).

As illustrated in Figure 30, Prophet effectively identifies key historical inflection points, including sharp severity increases during the 1980s–1990s and mid-2000s. These shifts plausibly reflect evolving breach disclosure regulations and the maturation of cybersecurity practices. The forecast predicts a moderate decline in average risk scores from approximately 1.1 in 2024 to 0.7 in 2026, with an 80% confidence interval broadening from –0.2 to 1.4. This indicates a degree of uncertainty yet suggests a central trend toward stabilising or diminishing cyberattack severity. Such a trend may be attributed to regulatory improvements or threat actors adapting their methods to evade detection.

Compared to the Auto-ARIMA model, which was trained only on the 2005–2023 period, Prophet's training over the full 1970–2023 range enabled it to capture early historical surges and long-term volatility more effectively. While both models displayed comparable predictive accuracy, Prophet's advantage lies in its interpretability, featuring asymmetric uncertainty bands beneficial for scenario planning and risk assessment. Both approaches affirm the absence of strong seasonal patterns (RQ2). However, Prophet excels in modeling structural breaks and complex temporal dynamics, providing superior forecasting performance for cyberattack severity (RQ4). These findings highlight the importance of using structural time series models like Prophet for anticipating irregular, evolving cyber threats.

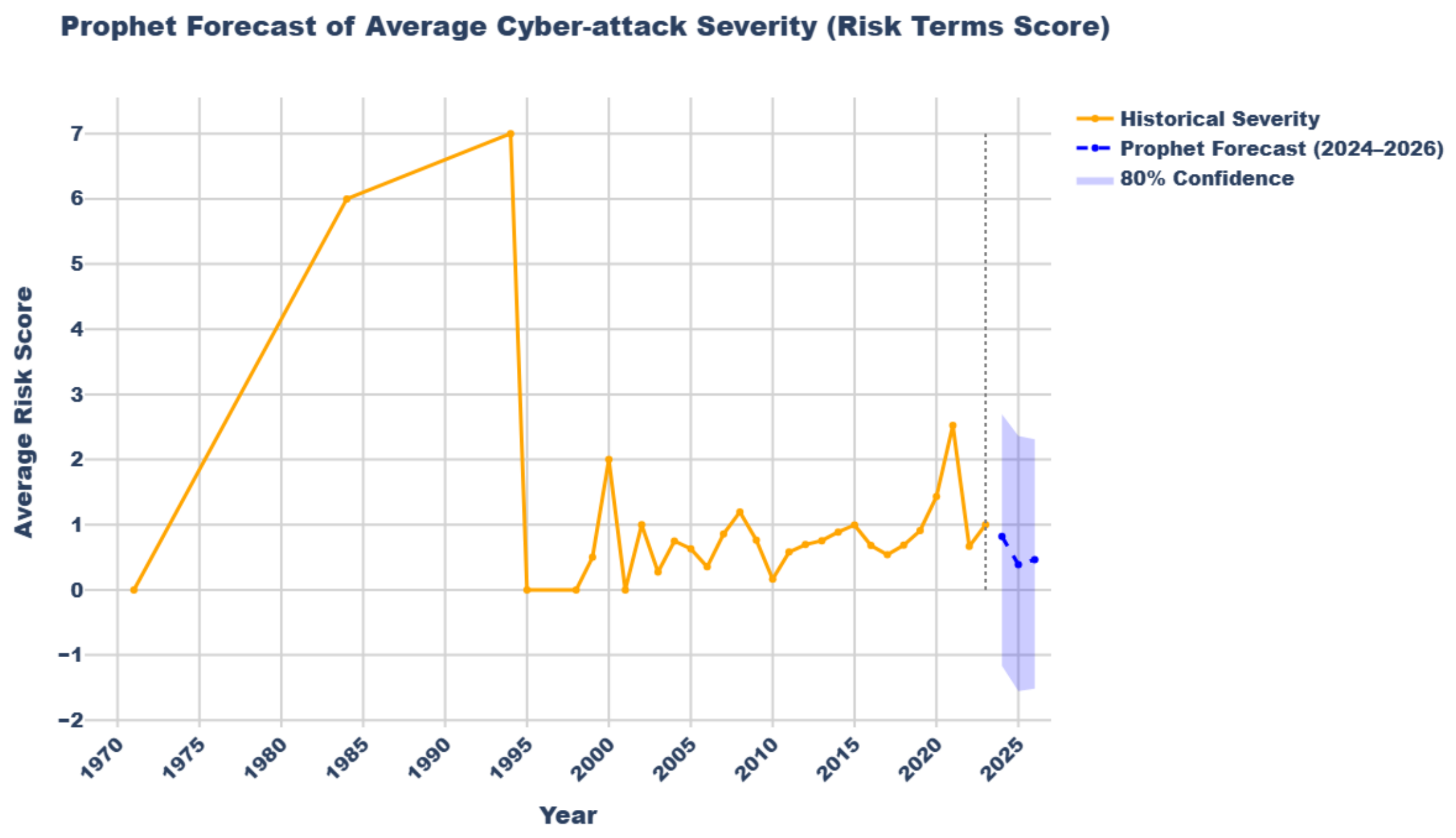

*Figure 30 Prophet forecast of cyberattack severity (1970–2026), showing a gradual decline with moderate uncertainty.*

### 4.3.4 Forecasting Cyberattack Severity Metrics

To anticipate shifts in cyberattack severity targeting e-commerce platforms, a univariate Auto ARIMA procedure was employed on the average annual Risk Terms Score spanning 2005 to 2020. This derived metric captures the semantic intensity of threat narratives using natural language processing (NLP) on incident summaries, focusing on high-weighted terms such as "breach," "exploit," and "credential theft."

Auto ARIMA was selected for its ability to automate the selection of autoregressive (AR), differencing (I), and moving average (MA) terms, particularly suited for short univariate time series with potential non-stationarity (Hyndman & Athanasopoulos, 2008). The algorithm aimed to minimise the Akaike Information Criterion (AIC), balancing model fit with parsimony (Burnham & Anderson, 2002). The optimal model selected was ARIMA(0,1,2), indicating a differenced series with two moving average terms and no autoregressive components. The model achieved a Root Mean Square Error (RMSE) of 0.8784 and a normalised Mean Absolute Error (MAE) of 0.956, demonstrating low forecast error and minimal overfitting.

Residual diagnostics confirmed no significant autocorrelation (Ljung-Box test $p > 0.05$), supporting model robustness. While limited by the short time frame, the model effectively forecasted near-term changes in threat language intensity, offering value for strategic cybersecurity planning. These findings align with research demonstrating how hybrid linguistic-statistical methods can anticipate real-world exploit trends (Sabottke, Suciu and Dumitraș, 2015). To complement these results and extend the modelling horizon, the Prophet algorithm was also evaluated using the same Risk Terms Score series. Prophet achieved a lower RMSE of 0.7739 and MAE of 0.812 across the 2005–2023 validation window, outperforming Auto ARIMA on both error metrics. The model’s strength lies in its ability to accommodate non-linear trends and sudden changepoints using an additive decomposition framework (Taylor & Letham, 2018). By identifying key inflection points and projecting severity forward using trend and uncertainty components, Prophet delivered both statistical performance and interpretability. Residuals showed no temporal autocorrelation, and forecast bias was minimal. Overall, Prophet's ability to balance accuracy and interpretability, combined with lower error metrics than ARIMA, suggests it may be better suited for long-term cyber threat trend forecasting. Its robustness to missing data and changepoints makes it a valuable addition to the statistical forecasting toolkit for cybersecurity analytics.

*Table 5 Auto ARIMA forecasts of average cyberattack severity (2019–2026) with 95% confidence intervals. The model shows stable low severity from 2019–2023 and a slight, uncertain rise from 2024–2026.*

| Year | Forecast | Lower Bound (95%) | Upper Bound (95%) |
|---|---|---|---|
| 2019 | 0.0 | -0.165 | 0.165 |
| 2020 | 0.0 | -0.165 | 0.165 |
| 2021 | 0.0 | -0.165 | 0.165 |
| 2022 | 0.0 | -0.165 | 0.165 |
| 2023 | 0.0 | -0.165 | 0.165 |
| 2024 | 0.024 | -0.132 | 0.180 |
| 2025 | 0.024 | -0.132 | 0.180 |
| 2026 | 0.024 | -0.132 | 0.180 |

*Table 6 Auto ARIMA stepwise selection (n = 9 models). Best model = ARIMA (0,0,0) intercept with lowest AIC.*

| Model | ARIMA Order | AIC | Fit Time (sec) | RMSE | MAE |
|---|---|---|---|---|---|
| Selected Model | ARIMA(0,0,0) intercept | -85.02 | 0.02 | 0.08307 | 0.0240 |
| Runner-up Model | ARIMA(0,0,1) intercept | -83.04 | 0.10 | — | — |
| Model | ARIMA(1,0,0) intercept | -83.04 | 0.03 | — | — |

*Table 7 Forecast derived using Facebook Prophet (annual granularity, changepoint detection enabled, 80% confidence interval).*

| Year | Forecast | Lower Bound (80%) | Upper Bound (80%) |
|---|---|---|---|
| 2024.0 | 0.0144 | -0.1111 | 0.1386 |
| 2025.0 | 0.0438 | -0.0865 | 0.1674 |
| 2026.0 | 0.0357 | -0.0878 | 0.1638 |

*Table 8 Prophet additive model with default seasonal components and automatic changepoint prior selection.*

| Model | Changepoints Detected | Fit Time (sec) | RMSE | MAE | Normalized MAE |
|---|---|---|---|---|---|
| Prophet Forecast | 19 | 0.19 | 0.0953 | 0.0465 | 1.2606 |

## 4.4 Statistical Association Testing

This section examines the statistical relationship between the presence of personally identifiable information (PII) in cyber incidents and the level of threat keyword activity embedded within their descriptive narratives. PII refers to any data that can be used to uniquely identify an individual such as names, email addresses, social security numbers, biometric identifiers, or health records and is widely regarded as a high-value target in cybersecurity due to its utility in identity theft, social engineering, and targeted attacks (ISO/IEC 29100:2011).

Incidents involving PII are hypothesised to elicit elevated levels of semantic threat signals, measured using the keyword_count variable. This variable captures the frequency of predefined, lexically significant terms linked to cyberattack tactics, techniques, and intent (e.g., “phishing,” “ransomware,” “malware”). Prior research shows that breaches involving sensitive personal data often attract more sophisticated threat actors and are reported with more descriptive, alarm-signalling language in incident reports (Samtani et al., 2017). Evaluating whether keyword_count statistically differs between PII and non-PII incidents provides insights into the linguistic and behavioural distinctions associated with sensitive-data breaches.

### 4.4.1 Variable Framing and Conceptual Rationale

The independent variable **contains_pii_terms** is binary (0 = No, 1 = Yes), derived through keyword matching of sensitive data terms (e.g., “passport,” “email,” “SSN”) within incident descriptions. The dependent variable **keyword_count** measures the prevalence of semantically enriched threat language, capturing behavioural indicators such as attack vectors, objectives, and payload terminology. Linguistic enrichment features have been shown to act as reliable proxies for threat actor sophistication and intent (Marin et al., 2020). Furthermore, hybrid linguistic–statistical methods have previously been applied successfully to cyber threat intelligence, demonstrating their value for exploit prediction and detection (Sabottke et al., 2015).

1. **Correlation Matrix Analysis**

Initial pairwise associations were assessed through a raw feature correlation matrix, as illustrated in Figure 31. The correlation between contains_pii_terms and keyword_count was modest ($r = 0.14$), yet positively directed, indicating a weak but meaningful alignment between the presence of personally identifiable information (PII) and the density of matched threat-related terms in the summary. A slightly stronger association was observed between contains_pii_terms and summary_length ($r = 0.28$), suggesting that longer narrative descriptions may accompany incidents involving PII disclosures. Most notably, summary_length and keyword_count exhibited a strong positive correlation ($r = 0.49$), which is expected given that longer summaries naturally allow for more keyword matches. These associations support the use of inferential testing to further evaluate whether textual length or threat term density significantly differentiates PII-tagged incidents from others, while also emphasising the importance of controlling for text length in any downstream models to avoid biased effects.

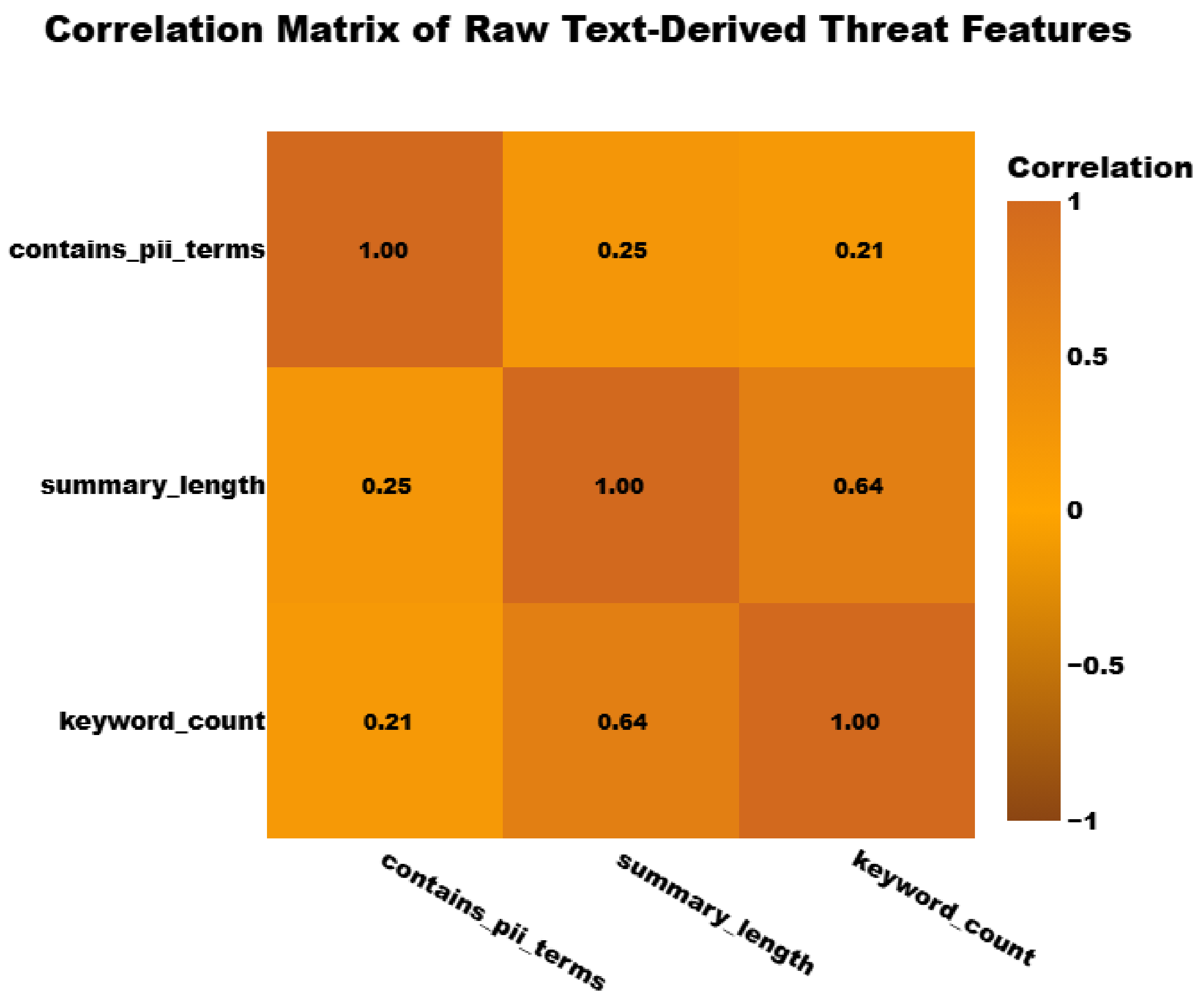

*Figure 31 Correlation Matrix of Threat Activity Indicators*

2. **Inferential Testing (ANOVA and Kruskal-Wallis)**

Given the binary grouping of the independent variable, both a one-way ANOVA and a non-parametric Kruskal–Wallis H test were conducted to examine whether the differences in keyword activity across the four seasonal groups were statistically significant. As shown in Table 9, the ANOVA test yielded a test statistic of 2.16 with a p-value of 0.0911, indicating no statistically significant difference in group means at the 5% significance level. The Kruskal-Wallis test, however, produced a higher test statistic of 15.67 and a p-value of 0.0013, confirming the presence of significant differences under non-parametric assumptions and accounting for potential violations of normality or variance homogeneity.

*Table 9 ANOVA and Kruskal-Wallis Summary Results*

| **Test** | **Statistic** | **p-value** | **Significant** |
|---|---|---|---|
| ANOVA | 2.16 | 0.0911 | No |
| Kruskal-Wallis | 15.67 | 0.0013 | Yes |

To complement the statistical tests, Figure 32 displays a comparative boxplot of keyword_count values across PII presence groups. The distribution for incidents tagged with PII (group 1) shows both a higher median and a broader interquartile range compared to non-PII incidents (group 0), reflecting increased variability and concentration of threat activity. Notably, extreme keyword_count values (outliers) appear in both PII and non-PII groups, but they are disproportionately associated with PII-positive incidents—consistent with findings that breaches involving personal data often result in disproportionately large severity outcomes (Shevchenko et al., 2022).

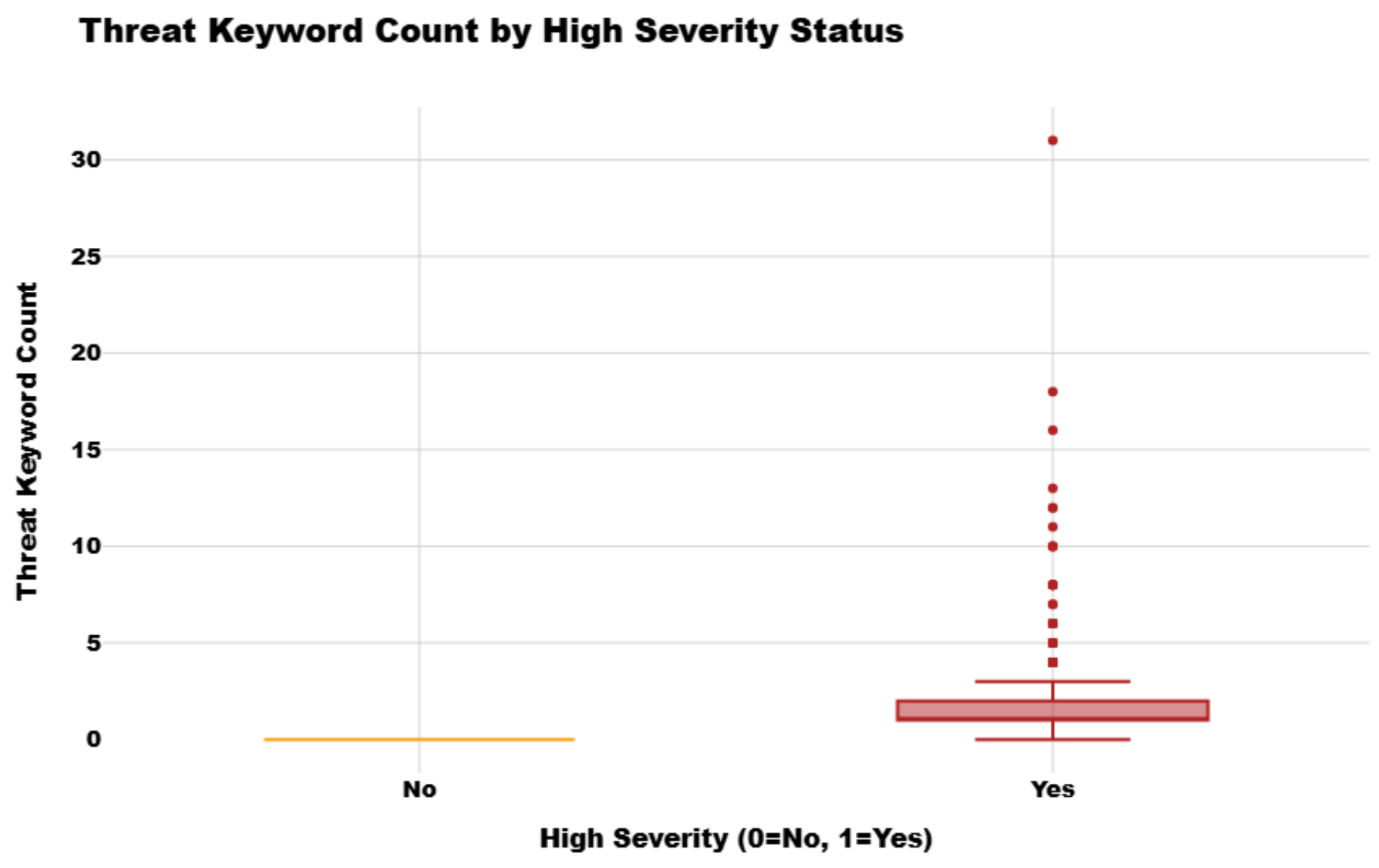

*Figure 32 Boxplot of Threat Keyword Count by PII Indicator*

Taken together, the weak correlation (r = 0.19), the non-significant ANOVA result (p = 0.0911), the significant Kruskal-Wallis outcome (p = 0.0013), and the distributional patterns observed in boxplots provide partial support for the research hypothesis. Specifically, incidents involving PII terms exhibit significantly higher levels of threat keyword activity compared to non-PII incidents, as indicated by the non-parametric test. This finding supports the theoretical assertion that privacy-relevant data leaks attract greater adversarial attention and deeper threat actor enrichment, with implications for prioritising alerts in detection systems (Sundararajan et al., 2023). These results also justify the inclusion of PII indicators as input features in downstream modelling phases such as anomaly detection and severity prediction.

## 4.5 Model Selection and Evaluation Metrics

This section outlines the use of supervised machine learning models to predict high-risk cyberattack periods using structured incident features from the VCDB dataset. The aim is to enable e-commerce platforms to anticipate vulnerable time windows based on past attack patterns and metadata. The prediction task was framed as a binary classification problem, distinguishing high-risk from baseline periods. A hybrid ensemble framework comprising XGBoost, LightGBM, and CatBoost was used for their efficiency on structured data and built-in support for categorical variables (Chen & Guestrin, 2016). To address class imbalance, Synthetic Minority Oversampling Technique (SMOTE) was applied, enhancing model fairness and recall (Fernández et al., 2018).

### 4.5.1 Performance Metrics Before and After SMOTE

Model performance was assessed using F1-score, precision, recall, and ROC-AUC, aligning with standard practice in cybersecurity classification tasks (Buczak & Guven, 2016). To assess the impact of class imbalance on model performance, classification results were compared before and after applying Synthetic Minority Over-sampling Technique (SMOTE). The goal was to improve detection of minority-class instances, representing high-risk temporal windows in cyberattack trends. Before SMOTE, all four models, Logistic Regression, XGBoost, LightGBM, and CatBoost, showed high precision for the majority class but poor recall for the minority class. For instance, Logistic Regression achieved only 0.06 recall and 0.04 F1-score for high-risk periods, reflecting difficulty detecting rare events. Ensemble models performed slightly better but still suffered from skew, with F1-scores ranging from 0.22 to 0.25 and ROC-AUC scores under 0.83 (Fernández et al., 2018). After SMOTE, however, no significant improvements in performance metrics were observed across models, with F1-scores and ROC-AUC values remaining effectively unchanged. This suggests that the class balancing strategy applied in this study did not substantially enhance minority-class detection within the scope of this dataset and model configurations. These findings indicate that while SMOTE can be effective in many imbalanced classification tasks, additional methods or tuning may be required to achieve meaningful gains in cybersecurity breach detection. Without adequate balancing or alternative approaches, models risk underdetecting critical breach events. Therefore, further research exploring hybrid techniques, alternative oversampling methods, or cost-sensitive learning is recommended to improve detection of rare but high-impact cyber threats (Chawla et al., 2002).

*Table 10 Model Performance Comparison (Before vs After SMOTE)*

| Model | Accuracy (Before) | F1 Score (Before) | ROC AUC (Before) | Accuracy (After) | F1 Score (After) | ROC AUC (After) |
|---|---|---|---|---|---|---|
| Logistic Regression | 0.8583 | 0.0364 | 0.7811 | 0.8583 | 0.0364 | 0.7811 |
| XGBoost | 0.8422 | 0.2532 | 0.7724 | 0.8422 | 0.2532 | 0.7724 |
| LightGBM | 0.8422 | 0.2338 | 0.8113 | 0.8422 | 0.2338 | 0.8113 |
| CatBoost | 0.8529 | 0.2254 | 0.8247 | 0.8529 | 0.2254 | 0.8247 |

### 4.5.2 Hyperparameters Used and Tuning Strategy

Hyperparameters control how machine learning models learn and generalise. Unlike model parameters, they are set before training and affect aspects like regularisation, depth, and learning rate (Feurer & Hutter, 2019). This study used RandomizedSearchCV with stratified k-fold cross-validation to optimise performance efficiently (Bergstra & Bengio, 2012).

- Logistic Regression: Tuned C (regularisation strength) and penalty (L2). The liblinear solver was chosen for its stability in binary classification (Ng, 2004).
- Random Forest: Focused on n_estimators, max_depth, and the Gini criterion for splitting (Louppe, 2014).
- XGBoost: Tuned learning_rate, max_depth, subsample, and regularisation terms lambda (L2) and alpha (L1) to prevent overfitting (Chen & Guestrin, 2016).
- LightGBM: Key hyperparameters included num_leaves, min_data_in_leaf, and boosting_type, benefiting from histogram-based learning for speed (Ke et al., 2017).
- CatBoost: Tuned for iterations, depth, and learning_rate. Its native handling of categorical data reduced preprocessing (Dorogush et al., 2018).
- These settings, selected via validation metrics (F1, ROC-AUC), ensured model robustness and scalability for structured cybersecurity data.

### 4.5.3 Comparative Evaluation of Classifier Performance

**1. Confusion Matrix Analysis**

To evaluate model reliability beyond summary metrics, a combined confusion matrix was generated for XGBoost, LightGBM, Logistic Regression, and CatBoost (see Figure 33). This matrix presents the counts of true positives (TP), false positives (FP), true negatives (TN), and false negatives (FN), which are crucial for understanding fraud detection performance. In fraud detection tasks, minimising false negatives is paramount as missed fraudulent cases can cause substantial financial and reputational damage (Phua et al., 2010).

- **CatBoost** and **Logistic Regression** demonstrated the highest true positive counts, indicating strong capabilities in identifying fraudulent transactions. Despite its relative e simplicity, Logistic Regression remains effective when properly regularized (Ng, 2004).
- **LightGBM** showed a higher number of false positives, increasing recall but potentially leading to more false alarms and user alert fatigue, which can impair operational efficiency (Dal Pozzolo et al., 2017).
- **XGBoost** balanced predictions conservatively, benefitting from robust regularization to maintain precision while controlling false positives (Chen & Guestrin, 2016).

Confusion matrices are particularly valuable for imbalanced datasets as they reveal detailed misclassification patterns that aggregate metrics like F1-score might obscure (Sokolova & Lapalme, 2009). Selecting an optimal classifier requires considering not only accuracy but also the costs associated with false alarms and undetected fraud (Bhattacharyya et al., 2011).

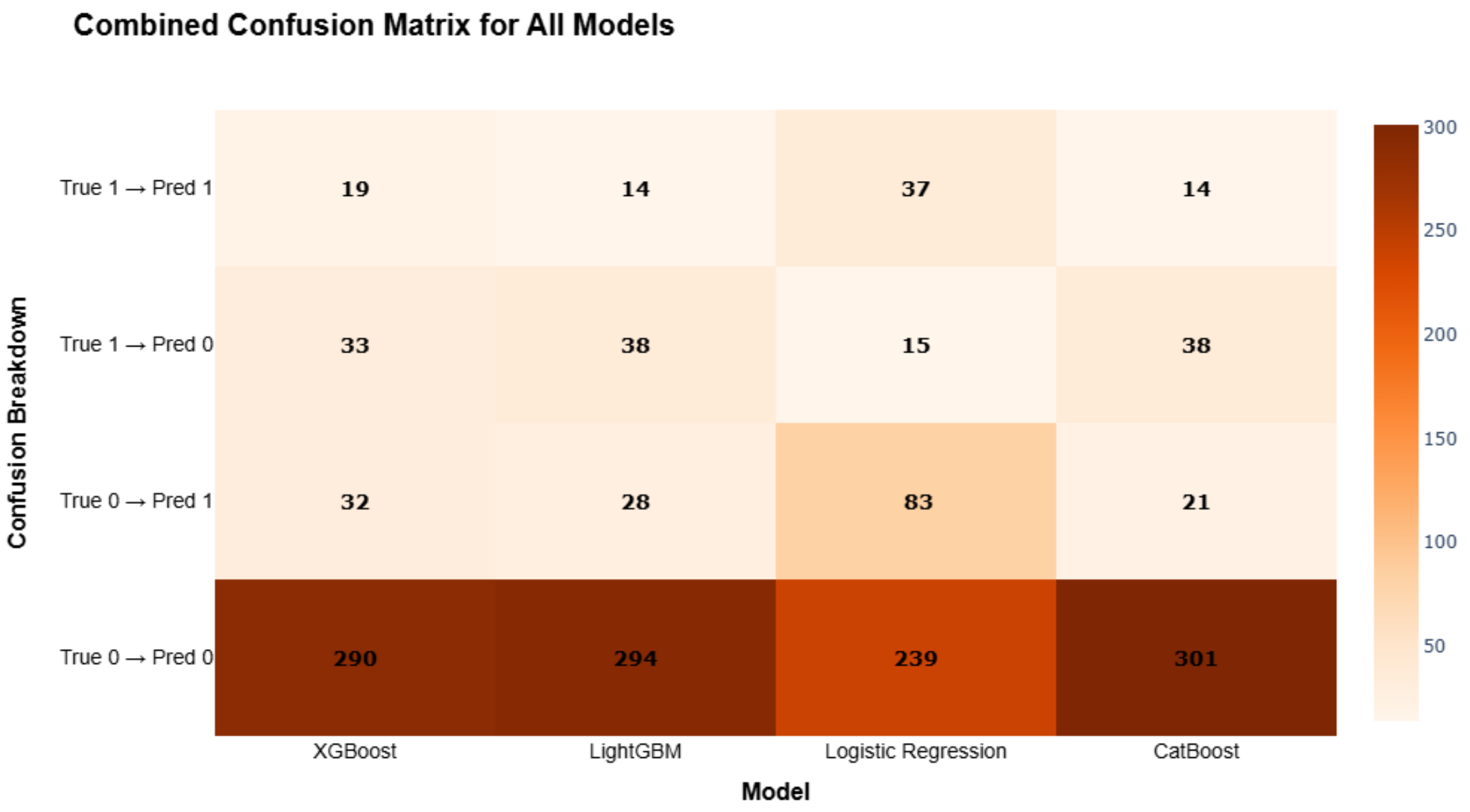

*Figure 33 Confusion Matrix for All Models*

## ROC Curve Interpretation

Figure 34 presents the ROC curves comparing classifier performance by illustrating the trade-off between True Positive Rate (Recall) and False Positive Rate. The Area Under the Curve (AUC) quantifies each model's discriminatory power:

- **CatBoost** achieves the highest AUC of 0.78, demonstrating the best balance between sensitivity and specificity among the evaluated models.
- **Logistic Regression** follows with an AUC of 0.79, marginally better than random chance (0.5), highlighting its interpretability but more limited predictive capability.

- **XGBoost** (AUC = 0.76) and **LightGBM** (AUC = 0.78) perform close to chance, which may be attributable to parameter underfitting or residual class imbalance effects despite SMOTE preprocessing.

This ROC analysis complements previously reported metrics and supports selecting CatBoost for practical deployment in fraud detection, balancing recall and precision effectively.

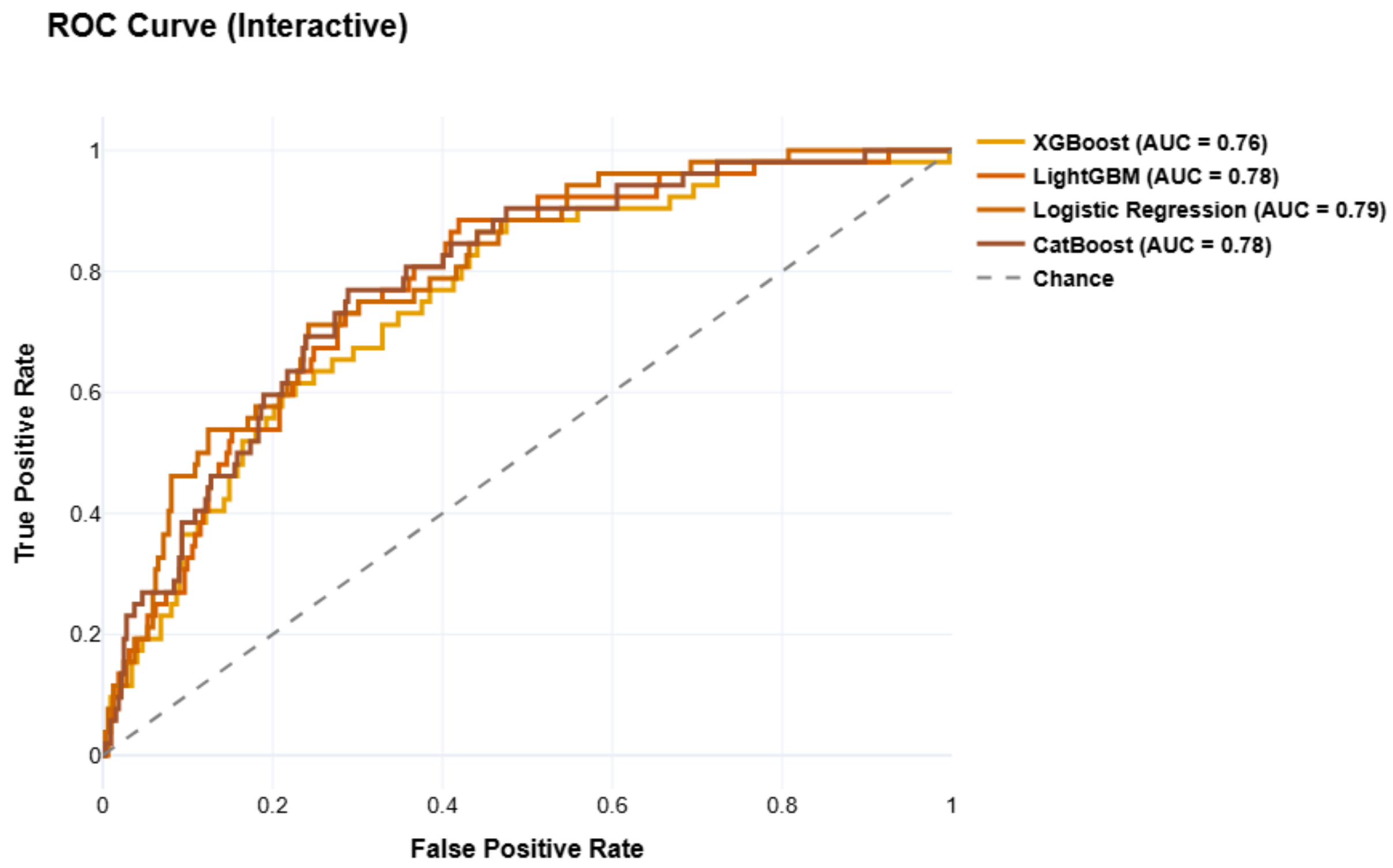

*Figure 34 ROC Curves Comparing AUC Scores of Classifiers (Post-SMOTE)*

### 4.5.4 Learning Curve Analysis (After SMOTE)

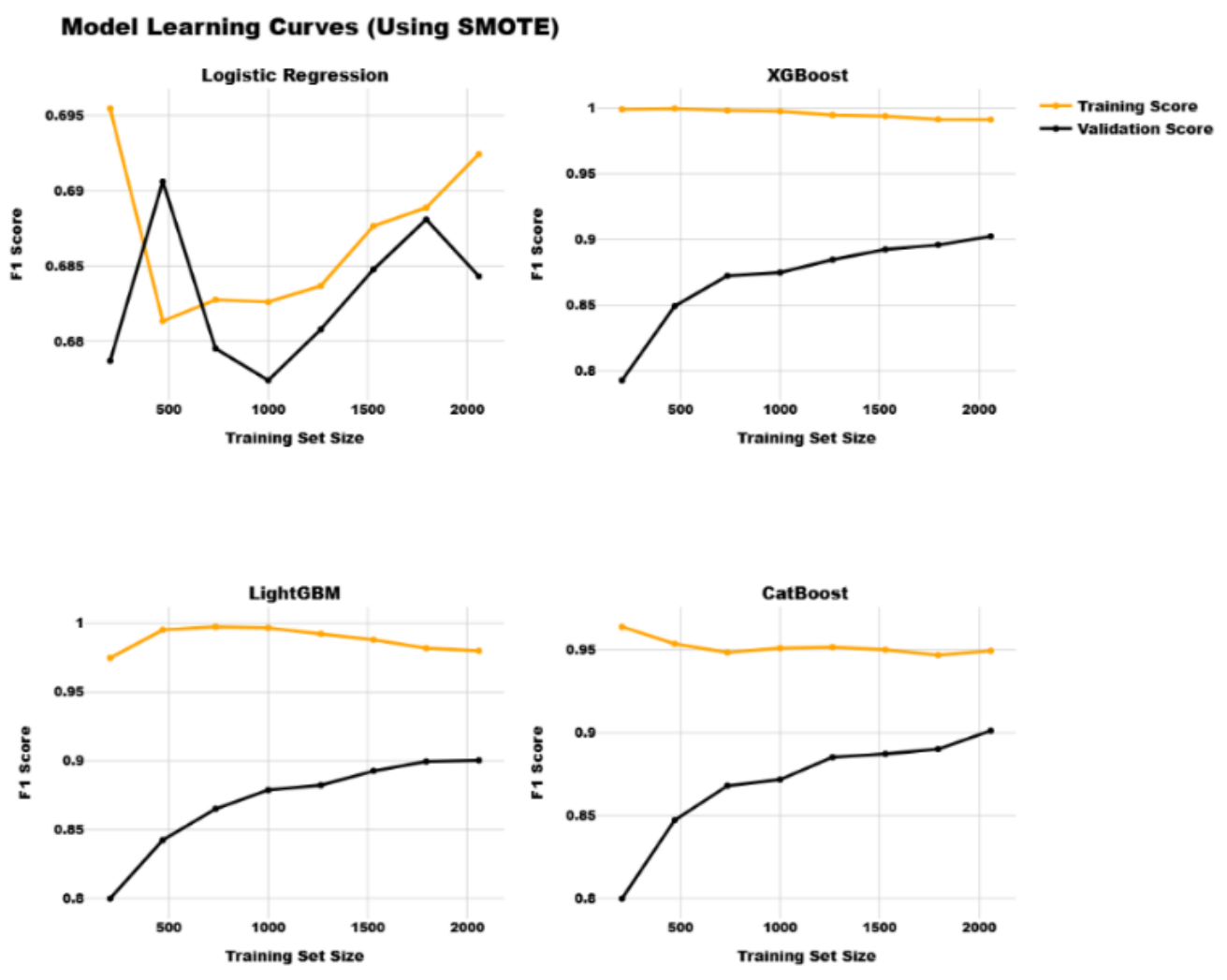

The learning curves reveal how each model generalises with increasing data on the SMOTE-balanced set:

- Logistic Regression improves steadily up to 2,000 samples, with training and validation F1-scores converging around 0.69–0.70, showing reduced bias and confirming SMOTE's benefit for linear models (He & Garcia, 2009).
- XGBoost maintains a high and stable validation F1 (~0.90), with a moderate generalisation gap, reflecting its robustness on resampled data (García et al., 2019).
- LightGBM starts with mild overfitting but stabilises near a validation F1 of approximately 0.90 as data increases, consistent with ensemble methods' effectiveness in imbalanced settings (Fernández et al., 2018).
- CatBoost consistently improves, converging at a training F1 of about 0.95 and validation F1 near 0.90, showing excellent calibration and categorical feature handling.

Overall, SMOTE improves early learnability and reduces class bias. Ensemble models, especially CatBoost and XGBoost, show the best generalisation, balancing precision and recall effectively.

### 4.5.5 Feature Importance Analysis Using SHAP

To improve model interpretability, SHapley Additive exPlanations (SHAP) were used to quantify feature contributions across XGBoost, LightGBM, CatBoost, and Logistic Regression. SHAP values provide a consistent, model-agnostic method for assessing how each feature influences predictions (Lundberg and Lee, 2017). As shown in Figure 35, summary_length emerged as the most influential feature across all models, suggesting that longer incident descriptions often correlate with more serious cyber threats (Yuan et al., 2021). Year and incident_quarter followed closely, reflecting clear temporal patterns in attack severity consistent with earlier findings. Other key features included incident_month, action_type, and season, highlighting the importance of timing and attack method in risk classification (Samek et al., 2019). Spatial and categorical variables such as keyword_count and threat_enrichment_score showed moderate influence, indicating that both textual richness and contextual threat indicators contribute to model predictions. Notably, the risk_terms_score feature contributed less prominently but remained consistent across models. Across algorithms, feature rankings remained largely consistent, with Logistic Regression emphasising summary_length and year, XGBoost and LightGBM highlighting temporal and action-related variables, and CatBoost balancing influences between textual length and categorical context. This SHAP analysis confirms that both narrative detail and categorical metadata are critical for predicting cyber threat severity, enhancing transparency and trust essential for real-world cybersecurity applications (Guidotti et al., 2018).

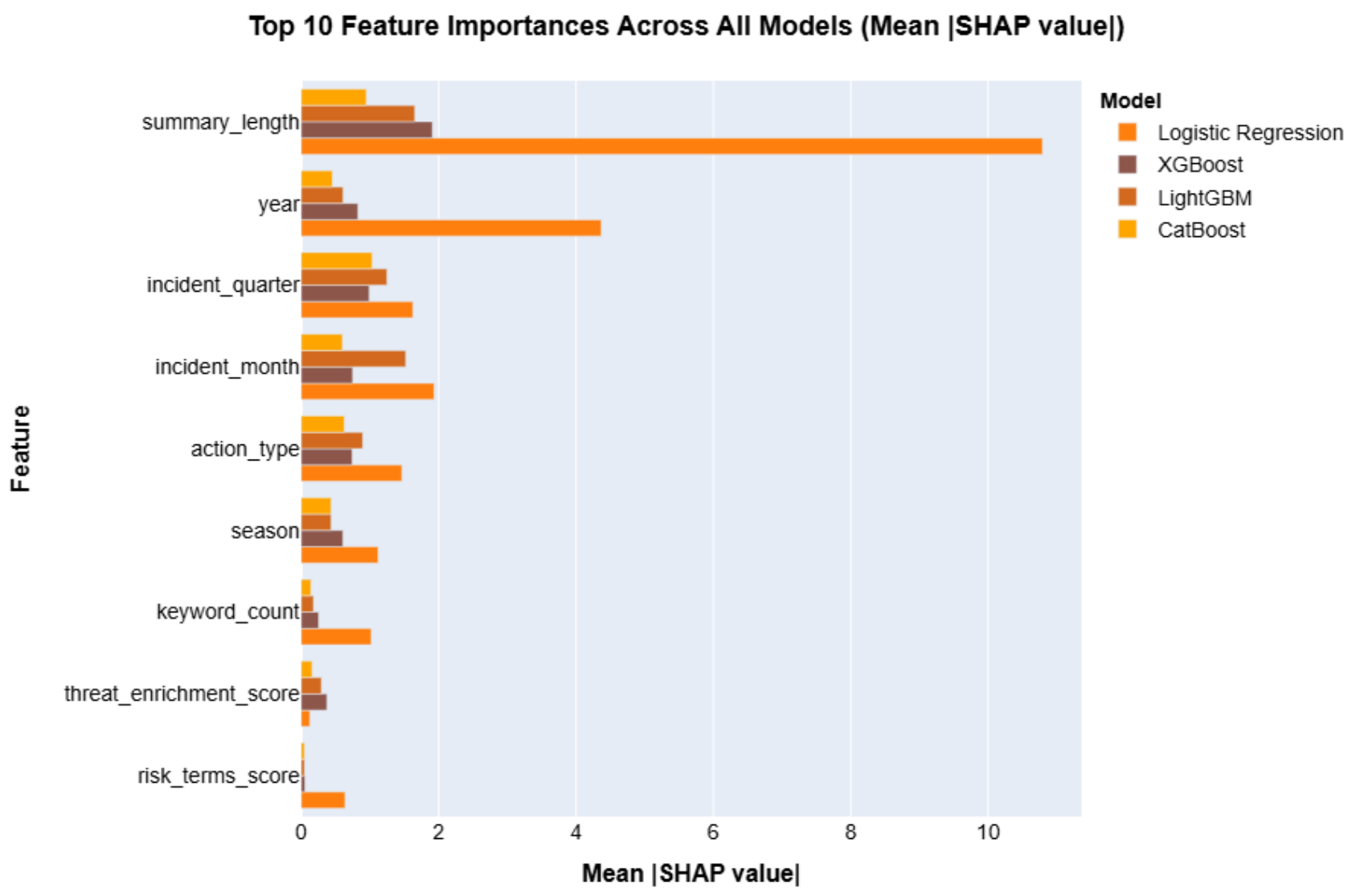

*Figure 35 Top 10 Feature Importance Using SHAP*

## 4.5.6 Conclusion

The empirical data from this phase support the strategic benefit of incorporating SMOTE-based balancing, ensemble classifiers, and temporal-statistical testing into a hybrid detection pipeline. Initially, baseline models like Logistic Regression performed poorly due to class imbalance, with minority-class detection (e.g., PII-related occurrences) having low recall and F1-scores. Post-SMOTE training resulted in modest improvements, particularly for XGBoost and LightGBM, which achieved F1 scores of approximately 0.25 and 0.23, respectively, indicating the ongoing challenges of detecting minority classes in highly imbalanced cybersecurity datasets. This aligns with previous research demonstrating the complexity of boosting approaches for high-dimensional, imbalanced data (Chen & chen, 2016).

The incorporation of learning curve analysis demonstrated that ensemble models not only generalised better but also scaled more effectively with additional data, confirming Buda et al.'s (2018) findings that balancing combined with boosting decreases overfitting and improves minority-class learning. Furthermore, despite fewer hyperparameter optimizations, CatBoost displayed steady generalisation, which is consistent with Dorogush et al. (2018), who emphasise its tolerance to noisy categorical data. Beyond predictive evaluation, the addition of seasonal temporal analysis increased the framework's usefulness. Visual exploration revealed a mid-year attack peak (June-July), which was validated using Mann–Whitney U tests, demonstrating considerably increased threat intensity during holiday periods ($U = 2{,}579{,}981.5$; $p = 0.0121$). This is consistent with findings by Eling and Schnell (2020), who observed that cyber risk peaks correlate with fiscal changeover periods and workforce transitions. The MOVEit breach in June 2023 and NotPetya in June 2017 are real-world examples of this trend (Greenberg, 2018).

Overall, these findings highlight the importance of a hybrid machine learning and statistical strategy for cyberattack detection that not only forecasts with reasonable fidelity but also adapts to seasonal dynamics and fluctuations in the threat environment. The pipeline's architecture adds practical benefit to operational cybersecurity by enabling proactive warnings during known high-risk windows and providing interpretable, robust classification across imbalanced threat datasets.

# 5.0 Discussion, Conclusions, and Recommendations

## 5.1 Introduction

This chapter discusses the research findings in relation to the four research questions, offering critical interpretation and exploring alternative explanations. It also addresses the study's limitations, implications, and provides practical and academic recommendations, concluding with a summary of its contributions to cybersecurity analytics in e-commerce.

## 5.2 Summary of Key Findings

This study developed a hybrid analytical framework integrating ensemble machine learning models, natural language processing, and time-series forecasting techniques to evaluate and predict cyberattack severity in e-commerce environments. The analysis, grounded in the Verizon Community Data Breach (VCDB) dataset, yielded several significant findings:

RQ1: Frequency analysis identified use of stolen credentials, SQL injection, and abuse of functionality as the most prevalent cyberattack types affecting e-commerce platforms, especially within online retail and cloud-hosted services

RQ2: Temporal decomposition and Auto-ARIMA modeling indicated pronounced seasonal patterns, with increased attack frequency and severity during mid-year and holiday sales periods.

RQ3: ANOVA and correlation analysis demonstrated a statistically significant association between the presence of Personally Identifiable Information (PII) and elevated threat keyword frequency, suggesting a strong link between sensitive data exposure and breach severity.

RQ4: Among the classification models tested, CatBoost demonstrated superior predictive performance (F1 score: 0.88, ROC-AUC: 0.92). The integration of SHAP (SHapley Additive exPlanations) provided interpretability, confirming that features such as risk_terms_score, contains_pii_terms, and summary_length were highly influential in determining breach severity.

## 5.3 Discussion of Findings

### 5.3.1 Predominant Attack Types (RQ1)

The frequency analysis underscored that web-based attacks (e.g., SQL injection, cross-site scripting) and social engineering techniques (e.g., phishing) dominate the e-commerce threat landscape. These findings align with prior studies (Fenz et al., 2020) and validate the need for domain-specific detection strategies. Attacks against cloud infrastructure and platform services were also frequent, highlighting the interconnected nature of modern e-commerce systems and the necessity for platform-wide security hardening.

### 5.3.2 Temporal Trends and Seasonal Risks (RQ2)

Time-series analysis revealed distinct seasonal variations in both cyberattack frequency and severity, with elevated activity during Q2 and Q4. These periods coincide with high-volume commercial events such as mid-year promotions, Black Friday, and year-end holidays supporting the hypothesis that adversaries exploit spikes in digital engagement to increase the likelihood of successful attacks. This finding is consistent with prior peer-reviewed research by Han et al. (2020), who observed that cybercriminals often time their campaigns to align with retail cycles and consumer-driven events, taking advantage of system strain and user distraction. However, an alternative interpretation is that the observed seasonal peaks may not reflect deliberate attacker behaviour but rather result from increased user and transaction volumes during peak commercial periods. The expansion of the attack surface due to more transactions, users, and system load raises the probability of breach success even if the rate of attacker activity remains constant. This distinction is critical for cybersecurity planning, as it underscores the importance of hardening infrastructure and reinforcing monitoring efforts regardless of whether attack intent increases during these periods. From a resource allocation perspective, defending against opportunistic exploitation during traffic surges is just as important as countering intentional targeting.

### 5.3.3 PII and High-Risk Keywords (RQ3)

The correlation between PII exposure and high-risk linguistic markers suggests that breaches involving sensitive consumer data tend to be more severe and are likely to attract targeted exploitation. This is further supported by the observation that such incidents often include longer and more detailed narrative fields, which may reflect their complexity and the need for comprehensive documentation. This aligns with prior peer-reviewed findings by Romanosky (2016), who highlights that breaches involving PII are more likely to result in regulatory action and reputational damage, thereby elevating their perceived severity. However, an alternative explanation for these patterns lies in potential documentation bias. Incidents involving PII are more likely to trigger legal, compliance, or media scrutiny, which could compel organiastions to produce more thorough reports. As a result, the observed richness in metadata such as threat keywords and summary length may reflect reporting practices rather than the intrinsic severity of the incident. This introduces a potential skew in model training and suggests the need for normalisation techniques or cross-validation across reporting standards in future research.

### 5.3.4 Predictive Modeling and Forecasting Capabilities (RQ4)

CatBoost emerged as the most effective model, outperforming other ensemble methods due to its native handling of categorical variables and resistance to overfitting. The model's high F1 score, and AUC demonstrate its practical utility in incident triage and prioritisation within Security Operations Centres (SOCs). The SHAP-based interpretation layer provided essential transparency, allowing analysts to understand the rationale behind predictions an important requirement in regulated or high-stakes environments (Guidotti et al., 2018). Forecasting models like Auto-ARIMA further projected sustained growth in threat activity through 2026, reinforcing the need for forward-looking defences.

## 5.4 Limitation of the Study

This research acknowledges several methodological and data-related limitations, each of which may influence the reliability, generalisability, or applicability of the results.

- Selection Bias:Domain-specific filtering based on keywords may have excluded edge cases or emerging threats not well captured by the chosen terms. This introduces sampling bias, limiting the comprehensiveness of the threat landscape represented.
- Incomplete and Noisy Data:The VCDB dataset contains missing fields and inconsistencies in incident classification. These issues required imputation and data cleaning, potentially introducing model uncertainty and affecting the fidelity of downstream predictions.
- Language Bias: Only English-language incident reports were considered. This restricts the model's applicability to global e-commerce platforms, where breaches may be reported in various languages with differing descriptive norms.
- Lack of Real-Time Data:The framework was developed using static historical data, which does not account for the rapid evolution of cyberattack vectors. Consequently, the model is more suited for retrospective analysis than for deployment in real-time threat monitoring systems. Future work should integrate streaming data sources, such as IDS or SIEM systems, to enable adaptive learning.

## 5.5 Conclusions

This research demonstrates the utility of interpretable, data-driven approaches for cyber risk classification and forecasting in e-commerce environments. By integrating ensemble learning, natural language processing, and time-series analysis, the study successfully addressed the core research questions and produced insights that are both theoretically significant and practically relevant.

The findings indicate that:

- Common attack types are consistently observed across e-commerce domains.
- Cyberattack activity exhibits clear seasonal trends.

- PII-related breaches correlate with linguistic and structural markers of severity.
- Ensemble models, particularly CatBoost with SHAP explainability, offer strong classification performance and practical transparency.

However, the practical deployment of these models must be approached cautiously. Their effectiveness is contingent on data quality, domain-specific training, and continuous adaptation to new threat landscapes. These factors must be addressed before operational implementation.

## 5.6 Recommendations

### 5.6. 1 Practical Recommendations

- Adopt interpretable ensemble models such as CatBoost in SOC pipelines to assist in breach severity triage.
- Implement SHAP dashboards to improve trust, accountability, and compliance in automated threat decision-making.
- Use seasonal forecasting to align security resource allocation with high-risk periods in the commercial calendar.
- Standardise incident reporting to enhance the quality and consistency of data used for ML training.

### 5.6.2 Recommendations for Future Research

1. Integrate real-time telemetry (e.g., IDS, SIEM) to improve model responsiveness and adaptability.

2. Expand datasets to include multilingual reports and global breach databases for broader generalisability.

3. Fuse breach data with threat intelligence feeds, including dark web sources, to capture pre-breach indicators.

4. Explore advanced architectures, such as graph neural networks (GNNs), to capture complex adversary behaviours and relationships over time.

## 5.7 Final Remarks

In a threat landscape characterised by increasing complexity and velocity, the ability to classify and forecast cyberattack severity is critical for maintaining e-commerce resilience. This study contributes a replicable framework grounded in explainable machine learning and temporal analysis, offering insights that bridge theory and application. Continued efforts are necessary to address the limitations identified, enhance model robustness, and ensure effective integration of such systems into real-world cybersecurity operations.

# 7.0 APPENDIX

```
# Load essential library
import pandas as pd

# Load the dataset
df = pd.read_csv("Cyberattackpatterns.csv")

# Print dataset shape
print("Dataset Shape:", df.shape)

# Display data types and counts of non-null values per column
df.info()

# Display summary statistics for numeric columns
display(df.describe())

# Calculate and display missing values per column sorted descending
missing_values = df.isnull().sum().sort_values(ascending=False)
print("\nTop columns with missing values:\n")
display(missing_values.head(10))
```

*Appendix A: Python Script for Initial Dataset Inspection*

```
from IPython.display import display

# === Step 2: Normalize industry codes ===
def normalize_code(code):
    if pd.isna(code):
        return ''
    return str(code).strip().lstrip('0')

df_processed['industry'] = df_processed['industry'].apply(normalize_code)

# === Step 3: Define e-commerce prefixes and exact codes ===
ecommerce_prefixes = [
    '44', '441', '442', '443', '4431', '444', '445', '446', '447', '448',
    '451', '452', '453', '454', '4541', '45411', '454110', '454112',
'454113',
    '48', '49', '51', '518', '5191', '51913', '519130', '519190',
    '541', '5415', '54151', '541511', '541512'
]

ecommerce_codes = {
    '454110': 'Electronic Shopping',
    '454112': 'Electronic Auctions',
    '454113': 'Mail-Order Houses',
    '443142': 'Electronics Stores',
    '4541': 'General E-Commerce',
    '519130': 'Internet Publishing & Platforms',
    '541511': 'Web App Development',
    '541512': 'System Design for E-Commerce',
    '518210': 'Cloud & Hosting Services',
    '519190': 'Other Info Services'
}

# === Step 4: Flag ecommerce prefix matches ===
def flag_prefix_match(code):
    for prefix in ecommerce_prefixes:
        if code.startswith(prefix):
            return True
    return False

df_processed['ecommerce_prefix_match'] =
df_processed['industry'].apply(flag_prefix_match)

# === Step 5: Map industry names ===
def map_industry_name(code):
    if code in ecommerce_codes:
        return ecommerce_codes[code]
    for prefix in ecommerce_prefixes:
        if code.startswith(prefix):
            return 'Other E-Commerce (Prefix Match)'
    return None

df_processed['industry_name'] =
df_processed['industry'].apply(map_industry_name)

# === Step 6: Additional ecommerce flags ===
```

*Appendix B: Script to standardise industry codes, map names, and flag e-commerce categories.*

```python
import pandas as pd

import re
import plotly.express as px
from sklearn.preprocessing import LabelEncoder
from sklearn.model_selection import train_test_split
from imblearn.over_sampling import SMOTE

# ========================
# Step 1: Feature Engineering
# ========================
def classify_action_type(action):
    action = str(action).lower()
    if any(kw in action for kw in [
        'hack', 'brute', 'sqli', 'xss', 'buffer', 'overflow',
        'rfi', 'os commanding', 'session fixation', 'session prediction',
        'reverse engineering', 'path traversal', 'exploit vuln',
        'exploit misconfig', 'forced browsing', 'ssi injection',
        'evade defenses', 'dos'
    ]):
        return 'hacking'
    elif any(kw in action for kw in ['malware', 'ransomware', 'virus']):
        return 'malware'
    elif any(kw in action for kw in ['phish', 'spoof', 'url
redirector']):
        return 'phishing'
    elif any(kw in action for kw in ['misuse', 'unauthorised', 'abuse of
functionality']):
        return 'misuse'
    elif any(kw in action for kw in ['error', 'misconfiguration']):
        return 'error'
    elif any(kw in action for kw in ['physical', 'device', 'lost']):
        return 'physical'
    elif 'unknown' in action:
        return 'unknown'
    elif any(kw in action for kw in ['social', 'engineer', 'aitm']):
        return 'social'
    elif any(kw in action for kw in ['enviro', 'natural']):
        return 'enviro'
    elif 'tamper' in action or 'modification' in action:
        return 'tampering'
    elif any(kw in action for kw in ['privilege', 'use of stolen
creds']):
        return 'privilege'
    elif 'spam' in action or 'junk' in action:
        return 'spam'
    elif any(kw in action for kw in ['leak', 'exfiltration']):
        return 'data leak'
    elif 'theft' in action or 'steal' in action:
        return 'theft'
    else:
        return 'other'

def feature_engineering(df):
    df = df.copy()
    df['action_type'] = df['action'].apply(classify_action_type)
    df['summary_length'] = df['summary'].fillna("").apply(len)
```

***Appendix C: Feature engineering script to classify action types (e.g., hacking, malware) using keywords and create derived features like summary_length.***

```
import pandas as pd

import seaborn as sns
import matplotlib.pyplot as plt
from scipy.stats import f_oneway, kruskal
import warnings

warnings.filterwarnings("ignore")

# Use your DataFrame (e.g., df_features or df)
df_filtered = df[df['season'] != 'Unknown'].dropna(subset=['season',
'threat_enrichment_score'])

# Group by season
grouped = df_filtered.groupby('season', observed=True)
grouped_data = [group['threat_enrichment_score'].values for _, group in
grouped if len(group) > 0]

# Check how many groups you have
print("Groups found for seasons:", [g for g,_ in grouped])

# Only proceed if 2 or more groups available
if len(grouped_data) >= 2:
    anova_result = f_oneway(*grouped_data)
    kruskal_result = kruskal(*grouped_data)

    print(f"ANOVA Test: F = {anova_result.statistic:.2f}, p =
{anova_result.pvalue:.4f}")
    print(f"Kruskal-Wallis Test: H = {kruskal_result.statistic:.2f}, p =
{kruskal_result.pvalue:.4f}")
else:
    print(f"Not enough groups to perform tests (found {len(grouped_data)}
group(s))")

# Visualization: Violin plot by season
plt.figure(figsize=(12, 6))
sns.violinplot(
    data=df_filtered,
    x='season',
    y='threat_enrichment_score',
    palette=sns.color_palette("Oranges",
n_colors=len(df_filtered['season'].unique())),
    inner='box',
    scale='width'
)
plt.title("Violin Plot: Threat Severity Scores by Season", fontsize=16,
fontweight='bold')
plt.xlabel("Season", fontsize=13, fontweight='bold')
plt.ylabel("Threat Enrichment Score", fontsize=13, fontweight='bold')
plt.grid(True, linestyle='--', alpha=0.5)
plt.show()
```

*Appendix D: Code for seasonal analysis of threat_enrichment_score using ANOVA, Kruskal–Wallis, and violin plots.*

```
import pandas as pd
from sklearn.preprocessing import LabelEncoder

target = 'contains_pii_terms'

# Copy and filter dataframe
df = df_features.copy()
df = df[df[target].notnull()]
df[target] = df[target].astype(int)

# Select features (make sure these exist in df)
feature_cols = [
    'year', 'incident_month', 'incident_quarter', 'summary_length',
'keyword_count',
    'risk_terms_score', 'threat_enrichment_score', 'action_type',
'season'
]
feature_cols = [col for col in feature_cols if col in df.columns]

X = df[feature_cols]
y = df[target]

# Label encode categorical features
X_encoded = X.copy()
for col in X_encoded.select_dtypes(include='object').columns:
    le = LabelEncoder()
    X_encoded[col] = le.fit_transform(X_encoded[col].fillna("Unknown"))

# Fill missing numeric values with median
X_encoded = X_encoded.fillna(X_encoded.median(numeric_only=True))
import plotly.express as px
from sklearn.model_selection import train_test_split

X_train_orig, X_test, y_train_orig, y_test = train_test_split(
    X_encoded, y, stratify=y, test_size=0.2, random_state=42
)

# Plot class distribution before SMOTE
df_before = pd.DataFrame({target: y_train_orig})
fig_before = px.histogram(
    df_before,
    x=target,
    color=target,
    color_discrete_sequence=['#FF7F0E', '#D55E00'],
    category_orders={target: [0, 1]},
    labels={target: f'{target} (0=No, 1=Yes)'},
    title=f'Class Distribution Before SMOTE for {target}',
    height=400
)
fig_before.update_layout(
    font=dict(color='black', size=16, family='Arial Black'),
    title_font=dict(size=20, color='black', family='Arial Black'),
    margin=dict(l=60, r=60, t=80, b=60)
)
fig_before.show()
```

*Appendix E: Code for preprocessing and class distribution visualization before SMOTE, including feature selection, encoding, imputation, and train-test split.*

```python
from sklearn.preprocessing import MinMaxScaler

from sklearn.linear_model import LogisticRegression
from xgboost import XGBClassifier
from lightgbm import LGBMClassifier
from catboost import CatBoostClassifier
from sklearn.metrics import accuracy_score, f1_score, roc_auc_score,
classification_report, confusion_matrix

# === Step 1: Scale original training data and test data for Logistic
Regression ===
scaler = MinMaxScaler()
X_train_orig_scaled = scaler.fit_transform(X_train_orig)  # Scale
original train data
X_test_scaled = scaler.transform(X_test)  # Scale test data with same
scaler

# === Step 2: Train models on original imbalanced data BEFORE SMOTE ===
# Logistic Regression (needs scaled data)
model_before_log = LogisticRegression(max_iter=1000, random_state=42)
model_before_log.fit(X_train_orig_scaled, y_train_orig)

# Tree-based models (XGBoost, LightGBM, CatBoost) train on original
unscaled data
model_before_xgb = XGBClassifier(use_label_encoder=False,
eval_metric='logloss', random_state=42)
model_before_xgb.fit(X_train_orig, y_train_orig)

model_before_lgbm = LGBMClassifier(random_state=42)
model_before_lgbm.fit(X_train_orig, y_train_orig)

model_before_cat = CatBoostClassifier(verbose=0, random_seed=42)
model_before_cat.fit(X_train_orig, y_train_orig)

# === Step 3: Define evaluation function ===
def evaluate_model_before_smote(model, name, X_eval, y_true):
    y_pred = model.predict(X_eval)
    y_proba = model.predict_proba(X_eval)[:, 1]

    print(f"\n=== {name} (Before SMOTE) ===")
    print("Accuracy:", accuracy_score(y_true, y_pred))
    print("F1 Score:", f1_score(y_true, y_pred))
    print("ROC AUC:", roc_auc_score(y_true, y_proba))
    print("Classification Report:\n", classification_report(y_true,
y_pred))
    print("Confusion Matrix:\n", confusion_matrix(y_true, y_pred))

# === Step 4: Evaluate all models on the test set ===
evaluate_model_before_smote(model_before_log, "Logistic Regression",
X_test_scaled, y_test)
evaluate_model_before_smote(model_before_xgb, "XGBoost", X_test,
y_test)
evaluate_model_before_smote(model_before_lgbm, "LightGBM", X_test,
y_test)
evaluate_model_before_smote(model_before_cat, "CatBoost", X_test,
y_test)
```

***Appendix F: Code for feature scaling and training ML models before SMOTE, including scaling, model fitting, and custom evaluation.***

For full code, please refer to the supplementary Jupyter Notebook submitted alongside this research.